\documentclass[12pt]{article}
\usepackage{amsmath}
\usepackage{graphicx}
\usepackage{xr-hyper}
\usepackage{natbib}
\usepackage{url} % not crucial - just used below for the URL 
\usepackage{subfig}
\usepackage{comment}
\usepackage{nicematrix}

\newcommand{\blind}{1}

\makeatletter
\newcommand*{\addFileDependency}[1]{
  \typeout{(#1)}
  \@addtofilelist{#1}
  \IfFileExists{#1}{}{\typeout{No file #1.}}
}
\makeatother

\usepackage{amssymb,amsfonts,mathrsfs,amsmath,amsthm,mathtools,bm,dsfont, thmtools,bbm}\allowdisplaybreaks
\usepackage[dvipsnames]{xcolor}
\usepackage{array, graphicx, graphics,float,multirow,tabularx,tikz,pgfplots, longtable,threeparttable}
\usepackage{setspace}
\usepackage[colorlinks,citecolor=blue,urlcolor=blue, runcolor=blue, filecolor=cyan]{hyperref}
\usepackage{footnotebackref}
\usepackage{bibentry, natbib} % reference
\usepackage{paralist}  % intermize and enumerate as paragraph
\usepackage{appendix} % appendix
\usepackage{mathrsfs}
\usepackage{enumitem}
\usepackage{authblk}
\usepackage{caption}
\usepackage{titlesec}
\usepackage{textcase,relsize}
\usepackage{booktabs}

\declaretheoremstyle[notefont=\bfseries,notebraces={}{},%
    headpunct={},postheadspace=1em]{mystyle}
\declaretheorem[style=mystyle,numbered=no,name=Assumption]{asmp-hand}
\declaretheorem[style=mystyle,numbered=no,name=Condition]{cond-hand}
\declaretheorem[style=mystyle,numbered=no,name=Example]{exmp-hand}

	\def \calA {\mathcal{A}}		
	\def \calB {\mathcal{B}}		
			
	\def \calD {\mathcal{D}}		
\def \bbE {\mathbb{E}}			
	\def \calF {\mathcal{F}}		
			
	\def \calH {\mathcal{H}}

	\def \calN {\mathcal{N}}		
			
\def \bbP {\mathbb{P}}

	\def \calW {\mathcal{W}}

\def \Cov {\text{Cov}}

\def \conP {\overset{\bbP}\longrightarrow}
\def \conD {\overset{D}\longrightarrow}

\def \vect {\textrm{vec}}

\def \diag {\textrm{diag}}

\newcommand{\norm}[1]{\left\Vert#1\right\Vert}
\newcommand{\abs}[1]{\left\vert#1\right\vert}

\DeclareMathOperator*{\argmin}{arg\,min}

\usepackage{subfig}

\theoremstyle{definition}

\newtheorem{assumption}{Assumption}[section]
\newtheorem*{assumption*}{Assumption}

\theoremstyle{plain}
\newtheorem{theorem}{Theorem}[section]

\newtheorem{lemma}[theorem]{Lemma}

\numberwithin{equation}{section}
\numberwithin{table}{section}
\numberwithin{figure}{section}

\def \conP {\to_p}
\def \conD {\to_d}
\def \bbE {\mathbb{E}}

\def \calW {\mathcal{W}}
\def \RR {{\mathbb R}}

\allowdisplaybreaks

\usepackage{algorithm}
\usepackage{algpseudocode}

\makeatletter
\newenvironment{breakablealgorithm}
{% \begin{breakablealgorithm}
	\begin{center}
		\refstepcounter{algorithm}% New algorithm
		\hrule height.8pt depth0pt \kern2pt% \@fs@pre for \@fs@ruled
		\renewcommand{\caption}[2][\relax]{% Make a new \caption
			{\raggedright\textbf{\fname@algorithm~\thealgorithm} ##2\par}%
			\ifx\relax##1\relax % #1 is \relax
			\addcontentsline{loa}{algorithm}{\protect\numberline{\thealgorithm}##2}%
			\else % #1 is not \relax
			\addcontentsline{loa}{algorithm}{\protect\numberline{\thealgorithm}##1}%
			\fi
			\kern2pt\hrule\kern2pt
		}
	}{% \end{breakablealgorithm}
		\kern2pt\hrule\relax% \@fs@post for \@fs@ruled
	\end{center}
}
\makeatother

\begin{document}

%\doparttoc % Tell to minitoc to generate a toc for the parts
%\faketableofcontents % Run a fake tableofcontents command for the partocs

%\bibliographystyle{natbib}

\def\spacingset#1{\renewcommand{\baselinestretch}%
{#1}\small\normalsize} \spacingset{1}

%%%%%%%%%%%%%%%%%%%%%%%%%%%%%%%%%%%%%%%%%%%%%%%%%%%%%%%%%%%%%%%%%%%%%%%%%%%%%%

\if1\blind
{
	\title{Inferential Theory for Pricing Errors with Latent Factors and Firm Characteristics\footnote{This research was supported by NSF Grant DMS-2052955.}
	}
	\author{Jungjun Choi\footnote{Address for Correspondence: Department of Computer Science and Statistics, University of Rhode Island, Tyler Hall,
9 Greenhouse Road, Kingston, RI 02881. Email: jungjun.choi@uri.edu.}\ \ and Ming Yuan\\
            Department of CS \& Statistics, University of Rhode Island\\
		Department of Statistics, Columbia University}
	\maketitle
} \fi

\if0\blind
{
  \title{   }
	\maketitle
} \fi

\bigskip

\begin{abstract}
We study factor models that combine latent factors with firm characteristics and propose a new framework for modeling, estimating, and inferring pricing errors. Following \cite{zhang2024testing}, our approach decomposes mispricing into two distinct components: inside alpha, explained by firm characteristics but orthogonal to factor exposures, and outside alpha, orthogonal to both factors and characteristics. Our model generalizes those developed recently such as \cite{kelly2019characteristics} and \cite{zhang2024testing}, resolving issues of orthogonality, basis dependence, and unit sensitivity. Methodologically, we develop estimators grounded in low-rank methods with explicit debiasing, providing closed-form solutions and a rigorous inferential theory that accommodates a growing number of characteristics and relaxes standard assumptions on sample dimensions. Empirically, using U.S. stock returns from 2000–2019, we document strong evidence of both inside and outside alphas, with the former showing industry-level co-movements and the latter reflecting idiosyncratic shocks beyond firm fundamentals. Our framework thus unifies statistical and characteristic-based approaches to factor modeling, offering both theoretical advances and new insights into the structure of pricing errors.
\end{abstract}

%\noindent{\it Keywords: IPCA}
\vfill

%\part{} % Start the document part

\newpage 

\spacingset{1.4} 

\setlength{\abovedisplayskip}{8pt}
\setlength{\belowdisplayskip}{8pt}
\setlength\intextsep{8pt}
\setlength{\abovecaptionskip}{4pt}

\section{Introduction}\label{sec:intro}

The search for a parsimonious yet interpretable representation of asset returns lies at the heart of modern asset pricing. Since the seminal works of \cite{sharpe1964capital,ross1976arbitrage,fama1973risk}, researchers have studied linear factor models where excess returns are driven by a small number of systematic risk factors. A dominant empirical approach to uncover these factors has been statistical, relying on principal component analysis (PCA) to extract latent sources of common variation \citep[e.g.,][]{chamberlain1982arbitrage,connor1986performance, connor1988risk}. While such latent-factor models effectively capture the covariance structure of returns, they often lack clear economic interpretation and are static in nature, making them ill-suited for conditional or time-varying risk exposures.

In parallel, a large literature in empirical finance has emphasized firm characteristics as the basis for factor construction, most prominently through the portfolio-sorting tradition that culminated in the Fama--French family of factor models \citep{fama1993common}. 
By anchoring factors in observable firm fundamentals, these models yield interpretable risk premia and direct economic meaning. 
However, ad hoc portfolio sorts can sacrifice statistical efficiency, discarding variation that is captured by latent statistical factors. 
Consequently, two lines of research, statistical factor extraction and characteristic-based portfolio construction, have developed largely in parallel, each offering distinct advantages but limited integration.

Recent advances in conditional and high-dimensional asset pricing have sought to bridge these approaches by allowing latent factor structures to depend explicitly on firm characteristics. \citet{fan2016projected} introduced projected PCA; \citet{kelly2019characteristics} proposed Instrumented PCA (IPCA), in which factor loadings and pricing errors are modeled as functions of firm characteristics; and \citet{kim2021arbitrage} and \citet{zhang2024testing} further refined this framework by relaxing identification restrictions and improving estimation. A complementary literature has incorporated nonlinear and machine-learning-based representations of characteristics, including deep factor and autoencoder models \citep[e.g.,][]{bryzgalova2019forest, gu2021autoencoder, feng2024deep}, which demonstrate that firm fundamentals can efficiently span the space of risk exposures.  At the same time, econometric work on high-dimensional factor models has developed a rigorous asymptotic theory for latent-factor estimation and inference \citep[e.g.,][]{bai2003inferential, fan2016projected, chernozhukov2023inference, chen2023semiparametric}. Yet despite this progress, a unified framework that combines the interpretability of characteristic-based models with the inferential rigor of modern econometrics remains elusive.

Two methodological gaps are particularly salient. First, the IPCA model of \citet{kelly2019characteristics} assumes that pricing errors (alphas) are fully explained by characteristics, violating the orthogonality condition between alphas and factor loadings required by the Arbitrage Pricing Theory (APT).  
This undermines the economic interpretation of estimated ``pricing errors'', as they may inadvertently load on systematic factors. \cite{zhang2024testing} highlighted this issue and proposed a decomposition of alphas into components inside and outside the span of characteristics. However, Zhang’s formulation depends on arbitrary choices of orthonormal bases and is not invariant to the rescaling of characteristics, raising concerns about robustness and interpretability. Moreover, the approach remains algorithmic: estimation relies on iterative numerical procedures with bootstrap-based inference but without accompanying asymptotic theory, leaving the econometric underpinnings incomplete.

This paper develops a general econometric framework that addresses these limitations and formally unifies latent-factor and characteristic-based approaches.  
Building on advances in low-rank and debiased estimation, we propose a model that decomposes pricing errors into two orthogonal components: inside alpha, the portion of mispricing attributable to firm characteristics but orthogonal to factor exposures; and outside alpha, the residual component orthogonal to both factors and characteristics. This decomposition restores theoretical consistency with APT while allowing a richer economic interpretation of both components.  
By deriving closed-form estimators and explicit bias corrections, we obtain tractable estimators that admit Gaussian inference even as the number of characteristics grows with the sample size. Specifically, our contributions are fourfold:

\paragraph{Modeling.} 
We provide a new decomposition of pricing errors that is basis-free, unit-invariant, and consistent with the orthogonality implied by APT.  
The decomposition generalizes \citet{zhang2024testing} and extends the IPCA framework of \citet{kelly2019characteristics} to accommodate both characteristic-driven and residual mispricing components, allowing for richer dynamics and greater interpretability of both components.

\paragraph{Methodology.} 
Using recent developments in low-rank and debiased estimation \citep[e.g.,][]{fan2022structural, chernozhukov2023inference}, we derive closed-form estimators that are computationally efficient and theoretically grounded, and well suited for high-dimensional panels. Unlike previous iterative procedures, our estimators ensure valid orthogonality between pricing errors and factor betas and incorporate debiasing steps that are essential for inference.

\paragraph{Theoretical Contributions.} 
We establish a full inferential theory for characteristic loadings, inside alphas, and outside alphas. We relax the conventional assumption on the relative size of the cross-sectional dimension ($N$) and time-series length ($T$), and introduce bias-correction techniques that allow inference without requiring the restrictive assumption that $T/N\to \infty$ and the number of characteristics is finite, extending the asymptotic theory of high-dimensional factor models \citep{bai2003inferential, fan2016projected, chen2023semiparametric}. These results place our framework on a firmer statistical footing than previous approaches and make it applicable to a wide range of empirical settings.

\paragraph{Empirical Findings.} 
Applying our methodology to U.S. stock returns and the same 36 firm characteristics considered by \cite{kelly2019characteristics} and \cite{zhang2024testing} from 2000 to 2019, we uncover new insights into the structure of pricing errors. We find strong evidence of both inside and outside alphas. Inside alphas exhibit persistent industry-level co-movements associated with fundamental drivers such as technology or finance sector shocks, while outside alphas capture transitory, firm-specific deviations consistent with behavioral or liquidity-based anomalies.

\medskip
\medskip
In summary, our framework unifies statistical and characteristic-based approaches, yielding both methodological innovations and substantive insights into the nature of pricing errors. It connects recent econometric innovations in high-dimensional inference with ongoing efforts in finance to rationalize the vast number of empirical return predictors \citep[e.g.,][]{harvey2016and, hou2020replicating}, offering a richer and more interpretable decomposition of pricing errors, grounding estimation in modern econometric methods with rigorous inferential guarantees, and providing new empirical evidence on the structure of mispricings in equity markets.

The remainder of this paper is organized as follows. Section \ref{sec:model} introduces the model of our paper and Section \ref{sec:estimation} discusses the estimation and debiasing procedure. Section \ref{sec:inferential_theory} provides the inferential theory of our estimators. Section \ref{sec:empirical} shows how our inferential theory can be applied to infer the US stock market and presents the empirical findings of our analysis. Finally, we conclude with a few remarks in Section \ref{sec:conclusion}. All proofs and simulation studies are relegated to the supplement due to the space limit.

In what follows, we use $\|\cdot\|_{\rm F}$ and $\|\cdot\|$ to denote the matrix Frobenius norm and the spectral norm, respectively. For any vector $a$, $\|a\|$ denotes its $\ell_2$ norm. For any set $\calA$, $|\calA|$ is the number of elements in $\calA$. We use $\otimes$ to denote the Kronecker product. $a \lesssim b$ means $|a|/|b| \leq C_1$ for some constant $C_1>0$ and $a \gtrsim b$ means $|a|/|b| \geq C_2$ for some constant $C_2>0$. $c \asymp d$ means that both $c/d $ and $d/c$ are bounded. $a \ll b$ indicates $|a| / |b| \rightarrow 0$ and $a \gg b$ indicates $|b| / |a| \rightarrow 0$. In addition, $I_n$ denotes the $n \times n$ identity matrix, $\textbf{1}_n$ denotes the $n \times 1$ vector of $1$, and $\boldsymbol{0}_{n\times m}$ denotes the $n \times m$ matrix consisting of zeros. In addition, $e_l$ is the $l$-th column of the identity matrix.

\section{Modeling Two Types of Mispricing}\label{sec:model}

Let $R_{t+1}$ the vector of excess returns on $N$ assets from period $t$ to $t+1$. A general factor pricing model posits that
$$
R_{t+1}  = \alpha_t + B_t f_{t+1} + E_{t+1},
$$
where $f_{t}$ is a $K \times 1$ vector of $K$ systematic factors, $B_t$ is the $N\times K$ matrix of factor loadings, and $E_{t+1}$ is an  idiosyncratic noise vector. The vector $\alpha_t$ captures pricing errors (or ``alphas'') and plays a critical role: under the Arbitrage Pricing Theory (APT), alphas should be orthogonal to factor exposures, i.e., $\alpha_t^\top B_t=0$. Otherwise, what appears as mispricing could simply reflect unmodeled factor risk.

\subsection{The KPS Model and Its Limitations}
\cite{kelly2019characteristics}, henceforth KPS, proposed an influential specification in which both factor loadings and pricing errors are modeled as linear functions of firm characteristics. Specifically, let $X_t$ denote the $N \times L$ matrix of firm characteristics observed at time $t$. The KPS model imposes:
$$
\alpha_t = X_{t} \eta, \qquad {\rm and}\qquad B_t = X_{t} \Gamma,
$$
for parameter matrix $\Gamma\in \RR^{L\times K}$ and $\eta\in \RR^L$. This setup blends the strengths of statistical factor analysis with characteristic-based portfolio construction, allowing latent factors to be systematically linked to observable firm-level information.

While elegant, as pointed out in \cite{zhang2024testing}, the KPS specification suffers from two major drawbacks. First, it does not enforce the orthogonality condition $\alpha_t^\top B_t = 0$. As a result, the so-called “pricing error” may in fact load on systematic factors, undermining its interpretation as pure mispricing. Second, by constraining $\alpha_t$ to lie in the span of $X_t$, the model rules out the possibility that some pricing errors are unrelated to the chosen set of characteristics. This restriction may omit economically meaningful forms of mispricing.

\subsection{A Decomposition into Inside and Outside Alphas}
To address these shortcomings, we propose decomposing the pricing error into two orthogonal components: 
$$\alpha_t=\alpha_{I,t}+\alpha_{O,t},$$
where 
\paragraph{Inside Alpha ($\alpha_{I,t}$):} the component of mispricing that is both orthogonal to the factor loadings and spanned by firm characteristics. This represents pricing errors that can be systematically related to observable fundamentals. Formally,
$$\alpha_{I,t}=(I_N - P_{B_t}) X_{t} \eta,$$
where $P_{B_t} = B_t \left( B_t^\top B_t \right)^{-1} B_t^\top$ is the projection matrix onto the linear space spanned by $B_t$. It is clear that for any $\eta\in \RR^L$, there exists $\eta_\perp\in \RR^L$ such that $\eta_\perp^\top \Gamma=0$ and
$$
(I_N-P_{B_t})X_t\eta=(I_N-P_{B_t})X_t\eta_\perp.
$$
Thus, without loss of generality, we shall assume in what follows that
$$\alpha_{I,t} = (I_N-P_{B_t})X_t\eta, \qquad {\rm and}\qquad \eta^\top \Gamma=0.$$

\paragraph{Outside Alpha ($\alpha_{O,t}$):} the residual mispricing component orthogonal to both $B_t$ and the span of $X_t$. This captures idiosyncratic pricing errors not explained by firm characteristics. We represent it as
$$\alpha_{O,t}=B_t^o\delta_{o,t},$$
where $B_t^o$ is a basis for the subspace orthogonal to $X_t$, defined by
\begin{equation}\label{eq:basic_basis}
	B^{o}_{t} =  X^{o}_{t} (X^{o \top}_{t} X^{o}_{t}/N)^{-1/2}, \quad X^{o}_{t} = \left[I_N - P_{X,t} \right] \Omega_{N \times (N-L)}
	,
\end{equation}
where $P_{X_t} = X_{t} (X_{t}^\top X_{t})^{-1} X_{t}^\top$ and $\Omega_{N \times (N-L)}$ is some full column rank matrix like $\left[ I_{N-L} \ \
		\boldsymbol{0}_{(N-L) \times L}
	\right]^{\top}$.

\medskip
\medskip
This decomposition preserves the crucial orthogonality $\alpha_{I,t}^\top B_t=\alpha_{O,t}^\top B_t=0$ for both types of alphas by construction. Economically, it disentangles mispricing attributable to observable fundamentals (inside alpha) from residual, potentially behavioral or market-friction-driven anomalies (outside alpha).

The decomposition into inside and outside alphas has important economic implications. Inside alphas capture systematic mispricing tied to firm characteristics, which may reflect persistent risk premia omitted from standard factor models or inefficiencies linked to observable fundamentals. Outside alphas, in contrast, capture residual idiosyncratic deviations that cannot be traced back to known characteristics, and may be driven by liquidity frictions, behavioral biases, or institutional trading pressures. By separating the two, our framework provides both a sharper theoretical alignment with APT and a more flexible empirical tool for studying the sources of mispricing.

\subsection{Comparison with \cite{zhang2024testing}}
Our decomposition is inspired by the approach of \cite{zhang2024testing}, who also distinguishes between pricing errors within and outside the span of firm characteristics. However, there are important differences:
\paragraph{Unit Invariance.} Zhang’s model can be sensitive to the scaling of firm characteristics, meaning that changing measurement units (e.g., dollars vs. millions) can alter the representation of alphas. Our formulation is invariant to such rescaling, making it more robust for empirical implementation as noted in Appendix \ref{sec:units}.
\paragraph{Basis Dependence.} Zhang defines inside alpha as $\alpha_{I,t} = B_t^{I} \delta_{I}$ where $B_t^{I}$ is an orthonormal basis for the subspace orthogonal to $B_t$ but within the span of $X_t$, and $\delta_I$ is time-invariant. This construction depends critically on the choice of basis, which can change over time and affect the stability of estimation. In contrast, our specification $(I_N - P_{B_t}) X_{t} \eta$ avoids this indeterminacy and ensures that inside alphas are basis-free.
\paragraph{Outside Alpha Dynamics.} Zhang assumes the outside pricing error $\alpha_{O,t}=B_t^{o}\delta_{o}$ for a time-invariant $\delta_{o}$, which is restrictive and may bias inference. We allow for more flexible dynamics by modeling
$$
\delta_{o,t}=\zeta + \xi_{t}
$$
where $\zeta$ captures a persistent component and $\xi_t$ is a sparse, time-varying shock. This assumption balances flexibility with tractability and reflects the plausible view that idiosyncratic mispricings may occasionally shift due to market conditions or firm-specific events.

\section{Estimation and Debiasing}\label{sec:estimation}

In this section, we describe how to estimate the parameters of the model introduced above -- namely, the characteristic-loading matrix $\Gamma$, the latent factors $f_t$, and the pricing error components $\alpha_{I,t}$ and $\alpha_{O,t}$. Our procedure builds on low-rank estimation methods but is carefully modified to ensure identification, orthogonality, and valid inference even when the number of characteristics $L$ is large relative to the number of assets $N$.

\subsection{Estimation of $\Gamma$ and Latent Factors}

\subsubsection{Model Transformation and Motivation}

Starting from our model
\begin{equation*}
	R_{t+1} = \alpha_{O,t} + \alpha_{I,t} + B_t f_{t+1} + E_{t+1},
\end{equation*}
and substituting $\alpha_{I,t} = (I_N - P_{B_t}) X_t \eta$, $\alpha_{O,t} = B^o_t \delta_{o,t}$, and $B_t = X_t \Gamma$, we obtain
\begin{equation}\label{eq:model2}
	R_{t+1} = B^o_t \delta_{o,t} + X_t \eta + X_t \Gamma \breve f_{t+1} + E_{t+1}, 
\end{equation}
where
\begin{equation*}
	\breve f_{t+1} = f_{t+1} - (B_t^\top B_t)^{-1} B_t^\top X_t \eta 
	= f_{t+1} - (\Gamma^\top X_t^\top X_t \Gamma)^{-1} \Gamma^\top X_t^\top X_t \eta.
\end{equation*}

Equation~\eqref{eq:model2} shows that once we account for the part of the pricing error captured by firm characteristics, the transformed return dynamics are effectively governed by a \emph{low-rank structure}: $R_{t+1}$ depends linearly on $X_t \Gamma$ through a small number of latent factors $\breve f_{t+1}$.

To exploit this structure, we pre-multiply both sides of \eqref{eq:model2} by $(X_t^\top X_t)^{-1} X_t^\top$. This step removes the cross-sectional dependence induced by $X_t$ and yields
\begin{equation*}
	\ddot R_{t+1} = \eta + \Gamma \breve f_{t+1} + \ddot E_{t+1},
\end{equation*}
where $\ddot R_{t+1} = (X_t^\top X_t)^{-1} X_t^\top R_{t+1}$ and $\ddot E_{t+1} = (X_t^\top X_t)^{-1} X_t^\top E_{t+1}$.  
Averaging over time and centering give
\begin{equation}\label{eq:lowrank}
	\ddot R^{d}_{t+1} = \Gamma f^{d}_{t+1} + \ddot E^{d}_{t+1},
\end{equation}
where $f^{d}_{t+1} = \breve f_{t+1} - T^{-1}\sum_t \breve f_{t+1}$, and the superscript $d$ denotes de-meaned quantities.  
Equation~\eqref{eq:lowrank} reveals that $\ddot R^d = [\ddot R_2^d, \ldots, \ddot R_{T+1}^d]$ admits a low-rank factor structure, $\ddot R^d = \Gamma F^d + \ddot E^d$, with $\mathrm{rank}(\Gamma F^d) = K$. Here $ F^d = [f^d_{2},\cdots, f^d_{T+1}] $ and $\ddot{E}^d = [ \ddot{E}^d_{2}, \cdots , \ddot{E}^d_{T + 1}]$.

\subsubsection{Initial Estimator via Low-Rank Approximation}

We obtain an initial estimator $\tilde \Gamma$ as the top $K$ left singular vectors of $\ddot R^d$.  
This spectral estimator parallels the principal components estimator in classical factor analysis but operates in the transformed ``characteristics space,'' ensuring that the estimated factors are conditionally orthogonal given $X_t$.

This estimator is $\sqrt{NT}$-unbaised when $T \ll N$, but as $T$ grows relative to $N$, it can suffer from bias due to the finite-sample correlation between estimated factors and residuals. We next correct this bias using a debiasing step grounded in recent developments in low-rank inference.

\subsubsection{Bias and Debiasing of $\Gamma$}

Given $\tilde \Gamma$, we estimate the de-meaned factor matrix as
$$
\tilde{F}^d =\argmin_A\|\ddot{R}^d-\tilde{\Gamma}A\|_{\rm F}^2= \left(\tilde{\Gamma}^\top \tilde{\Gamma}\right)^{-1} \tilde{\Gamma}^\top \ddot{R}^d=H_F F^d+\left(\tilde{\Gamma}^\top \tilde{\Gamma}\right)^{-1} \tilde{\Gamma}^\top\ddot{E}^d,
$$
where
$$
H_F=\left(\tilde{\Gamma}^\top \tilde{\Gamma}\right)^{-1} \tilde{\Gamma}^\top\Gamma.
$$
Similarly,
$$
\tilde{\Gamma}=\argmin_A\|\ddot{R}^d-A\tilde{F}^d\|_{\rm F}^2=\ddot{R}^d\tilde{F}^{d \top} (\tilde{F}^d\tilde{F}^{d \top})^{-1}=\Gamma H_\Gamma+\ddot{E}^d\tilde{F}^{d \top} (\tilde{F}^d\tilde{F}^{d \top})^{-1},
$$
where
$$
H_\Gamma=F^d\tilde{F}^{d \top} (\tilde{F}^d\tilde{F}^{d \top})^{-1}.
$$
The estimation error $\tilde{\Gamma}-\Gamma H_\Gamma$ can then be expressed as
\begin{gather}\label{eq:gamma_decomposition_naive}
	\ddot{E}^d \left(H_F F^d+\left(\tilde{\Gamma}^\top \tilde{\Gamma}\right)^{-1} \tilde{\Gamma}^\top\ddot{E}^d\right)^\top(\tilde{F}^d\tilde{F}^{d \top})^{-1}\\
	\nonumber =\ddot{E}^d F^{d \top}H_F^\top(\tilde{F}^d\tilde{F}^{d \top})^{-1}+\ddot{E}^d\ddot{E}^{d \top}\tilde{\Gamma}\left(\tilde{\Gamma}^\top \tilde{\Gamma}\right)^{-1}(\tilde{F}^d\tilde{F}^{d \top})^{-1}.
\end{gather}
The sample covariance of residuals,
\[
\ddot E^d \ddot E^{d\top} = \sum_{t=1}^T (X_t^\top X_t)^{-1} X_t^\top E^d_{t+1} E_{t+1}^{d\top} X_t (X_t^\top X_t)^{-1},
\]
has nonzero expectation and when $T/N$ does not vanish, the second term on the right hand side introduces non-negligible bias. 

To correct for this, we approximate the expectation of the noise covariance by
\[
\sum_{t=1}^T \hat \sigma_{t+1}^2 (X_t^\top X_t)^{-1},
\quad \text{where } \hat \sigma_{t+1}^2 = \frac{1}{N} \sum_{i=1}^N \hat \varepsilon_{i,t+1}^2,
\]
and $\hat \varepsilon_{i,t+1}$ are residuals from the current fit:
\[
\hat \varepsilon_{i,t+1} = r_{i,t+1} - (\hat \alpha_{O,it} + \tilde{\alpha}_{I,it}  + x_{it}^\top \tilde \Gamma \tilde f_{t+1}).
\]
Subtracting this estimated bias yields the \emph{debiased estimator}:
\begin{equation*}
	\hat \Gamma = \tilde \Gamma -
	\left( \sum_{t=1}^T \hat \sigma_{t+1}^2 (X_t^\top X_t)^{-1} \right)
	\tilde \Gamma (\tilde \Gamma^\top \tilde \Gamma)^{-1} (\tilde F^d \tilde F^{d\top})^{-1}.
\end{equation*}
The corresponding debiased estimate of the latent factors is
\begin{equation*}
	\hat F^d = (\hat \Gamma^\top \hat \Gamma)^{-1} \hat \Gamma^\top \ddot R^d.
\end{equation*}

This procedure removes the leading-order bias term of $\tilde \Gamma$ that arises when $T/N$ is not small. In Section~\ref{sec:inferential_theory}, we show that the resulting estimator admits a valid asymptotic normal distribution under mild regularity conditions, allowing for inference on both $\Gamma$ and the characteristic loadings even when the number of characteristics $L$ grows with $N$.

\subsection{Estimation of Pricing Errors}

Having estimated $\hat \Gamma$ and $\hat F^d$, we next turn to the estimation of inside and outside alphas.

\subsubsection{Inside Alpha ($\alpha_{I,t}$)}

By definition,
\[
\alpha_{I,t} = (I_N - P_{B_t}) X_t \eta = (P_{X_t} - P_{B_t})X_t \eta, \quad (P_{X_t} - P_{B_t})(\alpha_{O,t} + B_t f_{t+1})=0.
\]
A direct estimator of this quantity is
\[
\hat \alpha_{I,t} = (P_{X_t} - P_{X_t \hat \Gamma}) R_{t+1}.
\]
However, the convergence rate of this estimator is $\sqrt{L}/ \sqrt{N}$, which can be slow when $L$ is large. To obtain a more efficient estimator, we exploit the transformed model
\[
\ddot R_{t+1} = \eta + \Gamma \breve f_{t+1} + \ddot E_{t+1},
\]
which implies
\[
(I_L - P_\Gamma) \ddot R_{t+1} = \eta + (I_L - P_\Gamma) \ddot E_{t+1}.
\]
Hence, we can estimate $\eta$ by
\[
\hat \eta = (I_L - P_{\hat \Gamma}) \bar{\ddot{R}}
\quad \text{where } 
\bar{\ddot{R}} = \frac{1}{T} \sum_{t=1}^T \ddot R_{t+1}.
\]
Finally, substituting back yields a compact expression for inside alpha:
\[
\hat \alpha_{I,t} = (I_N - P_{X_t \hat \Gamma}) X_t \hat \eta 
= (I_N - P_{X_t \hat \Gamma}) X_t \bar{\ddot{R}}.
\]
This estimator enforces the orthogonality between $\alpha_{I,t}$ and factor loadings by construction and is computationally straightforward, requiring only matrix multiplications.

\subsubsection{Outside Alpha ($\alpha_{O,t}$)}

For the outside alpha, the estimation procedure consists of two steps. Note that $X_t^\top B^o_t = 0$, so projecting $R_{t+1}$ onto the orthogonal basis yields:
\[
(B_t^{o\top} B_t^o)^{-1} B_t^{o\top} R_{t+1} = \delta_{o,t} + (B_t^{o\top} B_t^o)^{-1} B_t^{o\top} E_{t+1}.
\]
Thus, an initial estimator of $\delta_{o,t}$ is
\[
\tilde \delta_{o,t} = (B_t^{o\top} B_t^o)^{-1} B_t^{o\top} R_{t+1}.
\]
Because we allow for a time-varying but sparse component $\xi_t$ such that $\delta_{o,t} = \zeta + \xi_t$, we estimate the persistent part $\zeta$ by time averaging:
\[
\tilde \zeta = \frac{1}{T} \sum_{t=1}^T \tilde \delta_{o,t},
\]
and then obtain a sparsity-regularized estimate of the transitory part via hard thresholding:
\[
\tilde \xi_{t,q} =
\begin{cases}
	\tilde \delta_{o,t,q} - \tilde \zeta_q, & \text{if } |\tilde \delta_{o,t,q} - \tilde \zeta_q| \geq \rho_t,\\
	0, & \text{otherwise},
\end{cases}
\]
where the threshold $\rho_t$ is chosen proportional to $\sqrt{(\log NT)/N}$ according to the analysis from Section \ref{sec:inferential_theory}. Additionally, since $\tilde \zeta$ has a bias term $\bar{\xi}= \frac{1}{T} \sum_{t=1}^T \xi_t$ in it, we further refine the estimator using $\tilde \xi_{t}$:
$$
\hat{\zeta} = \tilde \zeta - \frac{1}{T} \sum_{t=1}^T \tilde \xi_{t} .
$$
Similarly, we refine the estimator $\tilde \xi_{t,q}$ when $\tilde \xi_{t,q} \neq 0$:
\[
\hat \xi_{t,q} =
\begin{cases}
	\tilde \delta_{o,t,q} - \hat \zeta_q, & \text{if } \tilde \xi_{t,q} \neq 0,\\
	0, & \text{if }  \tilde \xi_{t,q} = 0.
\end{cases}
\]
The final estimator of outside alpha is then
\[
\hat \alpha_{O,t} = B^o_t (\hat \zeta + \hat \xi_t) .
\]

\subsection{Estimation Procedure}

We summarize the complete estimation procedure for $\Gamma$, the latent factors, and the two pricing error components below.  
The procedure relies only on standard linear algebra operations (matrix multiplications, singular value decomposition, and projection), and scales well for large panels.

\begin{breakablealgorithm}
	\caption{Estimation and Debiasing of Conditional Factor Model}
	\label{alg:estimation}
	\begin{algorithmic}[1]
		\Require Excess returns $\{R_{t+1}\}_{t=1}^T$, firm characteristics $\{X_t\}_{t=1}^T$, number of factors $K$, threshold $\rho_t$.
		
		\State \textbf{Step 1: Transformation and Initial Estimation of $\Gamma$}
		\State Compute $\ddot R_{t+1} = (X_t^\top X_t)^{-1} X_t^\top R_{t+1}$ and demean across $t$ to form $\ddot R^d$.
		\State Obtain top $K$ left singular vectors of $\ddot R^d$: $\tilde \Gamma \leftarrow \text{SVD}(\ddot R^d)$.
		\State Compute $\tilde F^d = (\tilde \Gamma^\top \tilde \Gamma)^{-1} \tilde \Gamma^\top \ddot R^d$.

        \State \textbf{Step 2: Initial Estimation of $\alpha_{I,t}$ and $f_{t+1}$}
        \State Estimate $\tilde \eta = (I_L - P_{\tilde \Gamma}) \bar{\ddot{R}}$, $\bar{\ddot{R}} = T^{-1}\sum_t \ddot R_{t+1}$.
		\State Compute $\tilde \alpha_{I,t} = (I_N - P_{X_t \tilde \Gamma}) X_t \tilde \eta$.
        \State Compute $\tilde{f}_{t+1} = (\tilde \Gamma^\top \tilde \Gamma)^{-1} \tilde \Gamma^\top \ddot R_{t+1} + (\tilde \Gamma^\top X_t^\top X_t \tilde \Gamma)^{-1} \tilde \Gamma^\top X_t^\top X_t \tilde \eta$.

		\State \textbf{Step 3: Debiasing of $\Gamma$}
		\State Compute residuals $\hat \varepsilon_{i,t+1} = r_{i,t+1} - (\hat{\alpha}_{O,it} + \tilde{\alpha}_{I,it} + x_{it}^\top \tilde \Gamma \tilde f_{t+1} )$.
		\State Estimate $\hat \sigma_{t+1}^2 = N^{-1}\sum_i \hat \varepsilon_{i,t+1}^2$.
		\State Apply bias correction:
		\[
		\hat \Gamma = \tilde \Gamma - 
		\Big(\sum_t \hat \sigma_{t+1}^2 (X_t^\top X_t)^{-1}\Big)
		\tilde \Gamma (\tilde \Gamma^\top \tilde \Gamma)^{-1}(\tilde F^d \tilde F^{d\top})^{-1}.
		\]
        \State Compute $\hat F^d = (\hat \Gamma^\top \hat \Gamma)^{-1} \hat \Gamma^\top \ddot R^d$.
		
		\State \textbf{Step 4: Inside Alpha}
		\State Repeat Step 2 with $\hat{\Gamma}$ to derive $\hat{\alpha}_{I,t}$ and $\hat{f}_{t+1}$.
		
		\State \textbf{Step 5: Outside Alpha}
		\State Construct $B^o_t = X^o_t[(X^{o\top}_t X^o_t)/N]^{-1/2}$, $X^o_t = (I_N - P_{X_t})\Omega_{N \times (N-L)}$.
		\State Compute $\tilde \delta_{o,t} = (B_t^{o\top}B_t^o)^{-1}B_t^{o\top}R_{t+1}$.
		\State Estimate $\tilde \zeta = T^{-1}\sum_t \tilde \delta_{o,t}$.
		\State Apply hard thresholding:
		\[
		\tilde \xi_{t,i} = 
		\begin{cases}
			\tilde \delta_{o,t,q} - \tilde \zeta_i, & |\tilde \delta_{o,t,q} - \tilde \zeta_q| \ge \rho_t,\\
			0, & \text{otherwise.}
		\end{cases}
		\]
        \State Refinement: estimate $\hat{\zeta} = \tilde \zeta - \frac{1}{T} \sum_{t=1}^T \tilde \xi_{t}$ and 
        \[
        \hat \xi_{t,q} =
        \begin{cases}
	      \tilde \delta_{o,t,q} - \hat \zeta_q, & \text{if } \tilde \xi_{t,q} \neq 0,\\
	      0, & \text{if }  \tilde \xi_{t,q} = 0.
        \end{cases}
        \]
		\State Compute $\hat \alpha_{O,t} = B^o_t(\hat \zeta + \hat \xi_t)$.
		\Ensure Outputs: Debiased $\hat \Gamma$, latent factors $\hat f_{t+1}$, inside alpha $\hat \alpha_{I,t}$, outside alpha $\hat \alpha_{O,t}$.
	\end{algorithmic}
\end{breakablealgorithm}

By transforming returns into characteristic space and exploiting low-rank structure, we obtain closed-form estimators for both $\Gamma$ and the pricing errors. The bias-correction step ensures valid inference even when $T$ is not small relative to $N$.  
Conceptually, our approach differs from the algorithmic methods in \cite{zhang2024testing}, which iteratively solve first-order conditions without theoretical guarantees.  
Instead, our estimators admit clear analytical forms, are grounded in the recent theory of debiased low-rank estimation, and directly link to the inferential results in Section~\ref{sec:inferential_theory}.

\section{Inference and Asymptotic Theory}\label{sec:inferential_theory}

This section develops the inferential theory for our estimators of characteristic loadings, factors, and pricing errors. While the estimation procedure in Section~\ref{sec:estimation} yields closed-form solutions, valid inference requires understanding their asymptotic behavior as both the cross-sectional and time-series dimensions grow. We show that the estimators admit standard Gaussian limits under mild regularity conditions, allowing conventional hypothesis testing even when the number of firm characteristics increases with the sample size.

\subsection{Setup and Regularity Conditions}

We first present a sequence of assumptions that ensure well-behaved moments, identification, and dependence properties of the data-generating process. For clarity, we group these conditions by theme.

\begin{assumption}[Characteristics and Identification]
	\label{asp:characteristics}
	Each firm $i$ at time $t$ is associated with an $L$-dimensional vector of characteristics $x_{it}$.
	\begin{itemize}
		\item[(i)] The second moments are uniformly bounded: $E[x_{it,l}^2] \le C$ for some constant $C>0$.
		\item[(ii)] The cross-sectional covariance matrix $Q_t = N^{-1}\sum_{i=1}^N x_{it}x_{it}^\top$ has eigenvalues bounded away from zero and infinity:
		\[
		c_1 < \psi_{\min}(Q_t) \le \psi_{\max}(Q_t) < c_2,
		\]
		for some positive constants $c_1$ and $c_2$, with probability approaching one. Here $\psi_{\min}(\cdot)$ and $\psi_{\max}(\cdot)$ are the smallest and largest nonzero eigenvalues, respectively.
	\end{itemize}
\end{assumption}

Assumption~\ref{asp:characteristics} ensures that characteristics are sufficiently informative and non-collinear. It parallels the ``pervasive'' condition in classical factor models \citep[see, e.g.,][]{fan2016projected, chen2023semiparametric} and is relatively mild since $L \ll N$ in most applications. 

\begin{assumption}[Factors and Loadings]
	\label{asp:factorloadings}
	Let $\Gamma$ denote the $L\times K$ matrix of characteristic loadings and $f_t$ the $K$-dimensional latent factor.
	\begin{itemize}
		\item[(i)] $\Gamma^\top \Gamma$ is well-conditioned: $c_1 < \psi_{\min}(\Gamma^\top\Gamma) \le \psi_{\max}(\Gamma^\top\Gamma) < c_2$ for some positive constants $c_1$ and $c_2$.
		\item[(ii)] $E[\|f_t\|^4] < C_1$ for some positive constant $C_1$.
		\item[(iii)] The de-meaned factor covariance satisfies $T^{-1}F^d (F^d)^\top \xrightarrow{p} \Sigma_f$, where $\Sigma_f$ is positive definite.
		\item[(iv)] The eigenvalues of $(\Gamma^\top\Gamma)\Sigma_f$ are distinct.
		\item[(v)] There exists a constant $C_2>0$ such that $E[\|B_{it}\|^2] \le C_2$ for all $i,t$.
		\item[(vi)] Identification: $\eta^\top\Gamma = 0$ and $\|\eta\|\le C_3$ for some constant $C_3>0$.
	\end{itemize}
\end{assumption}

These conditions guarantee identification of the factors and their characteristic-based loadings. Condition (i) is similar to the ``pervasive'' condition on factor loadings and common in the factor model literature. See, e.g., \cite{chen2023semiparametric}. Conditions (ii) - (iv) ensure factor uniqueness up to rotation and are also typical in the factor model literature. See, e.g., \cite{bai2003inferential,fan2016projected,chen2023semiparametric}. Condition (vi) enforces the orthogonality of inside alphas to factor loadings, which is essential for identifying pricing errors. See, also, \cite{kelly2019characteristics,kim2021arbitrage,chen2023semiparametric}.

\begin{assumption}[Idiosyncratic Noise]
	\label{asp:noise}
	Conditional on $(x_{it},f_{t+1})$, the idiosyncratic component $\epsilon_{it+1}$ satisfies:
	\begin{itemize}
		\item[(i)] $E[\epsilon_{it+1}] = 0$ and $E[\epsilon_{it+1}^2] = \sigma_{t+1}^2$;
		\item[(ii)] Sub-Gaussianity: $E[\exp(s\epsilon_{it+1})] \le \exp(C_1 s^2\sigma_{t+1}^2)$ for all $s\in\mathbb{R}$;
		\item[(iii)] Independence across $i$ and weak dependence across $t$:
		$\max_{i,t}\sum_{s}|\mathrm{Cov}(\epsilon_{it},\epsilon_{is})| \le C_2.$
	\end{itemize}
\end{assumption}

Assumption~\ref{asp:noise} allows for heteroskedasticity and mild serial dependence, both prevalent in asset-return data. 
Sub-Gaussianity simplifies the derivations without excluding heavy-tailed behavior under weak dependence.

\begin{assumption}[Sparsity of Outside Alphas]
	\label{asp:sparsity_heterodelta}
	Let $\delta_{o,t} = \zeta + \xi_t$ denote the outside-alpha component. 
	Then, for each coordinate $q$,
	\[
	\frac{1}{T}\sum_{s=1}^T |\xi_{s,q}| \ll \sigma_{t+1} \frac{\sqrt{\log(NT)}}{\sqrt{N}}, 
	\quad
	\frac{\sigma_{t+1} \sqrt{\log(NT)}}{|\xi_{t,q}|\sqrt{N}} \to 0
	\text{ for } q \in D_t,
	\]
	and $(\log N / N)|D_t|\to 0$ where $D_t = \{ 1 \leq q \leq N-L : \xi_{t,q} \neq 0 \}$.
\end{assumption}

This assumption imposes sparsity on transitory mispricing shocks, consistent with the view that only a small subset of firms experience idiosyncratic pricing deviations at any given time.

\begin{assumption}[Central Limit Conditions]
	\label{asp:clt}
	Define
	\[
	Q_f = T^{-1}\sum_t f_t^df_t^{d\top},\quad 
	Q_t^B = N^{-1}\sum_i B_{it}B_{it}^\top,\quad
	Q_t^{a,B} = N^{-1}\sum_i a_{it}B_{it}^\top,
	\]
	where $a_{it} = \eta^\top x_{it}$. Conditioning on $(x_{it},f_{t+1})_{i\leq N, t \leq T}$,
	\begin{align*}
		& (i) \ \ \frac{1}{\sqrt{NT}} \sum_{i=1}^N \sum_{t=1}^T \left(e_l^\top Q_t^{-1}x_{it}\right)  f_{t+1}^d  \epsilon_{i,t+1} \conD \calN\left( 0, \Sigma_{xf,l} \right), \\ 
		&(ii) \ \  \frac{1}{\sqrt{NTL}} \sum_{j=1}^N \sum_{s=1}^T g_{it,js} \epsilon_{j,s+1} \conD \calN\left( 0, \sigma_{I,it}^{2} \right), \\
		&(iii) \ \ \frac{1}{\sqrt{N}}\sum_{j=1}^N B^o_{t,jq} \epsilon_{j,t+1} \conD \calN\left( 0,  \sigma_{\delta,qt}^{2} \right),\\
		&(iv) \ \ \sigma_{o,it}^{-1} \left( \frac{1}{NT} \sum_{j=1}^N \sum_{s=1}^T B_{t,i}^{o \top} B_{s,j}^{o} \epsilon_{j,s+1}  +  \frac{1}{N} \sum_{j=1}^N  \left( \sum_{q \in D_t} B_{t,iq}^o B_{t,jq}^o \right) \epsilon_{j,t+1} \right) \conD \calN\left( 0, 1 \right),\\
		&\text{where} \\
		& g_{it,js} =  \left[1 - \left( Q^{a,B}_t (Q_t^B)^{-1} + \bar{\breve{f}}^\top \right) \left( Q^{f}\right)^{-1} f_{s+1}^d \right] \left(x_{it}^\top Q_t^{-1} x_{js} - B_{it}^\top (Q_t^{B})^{-1}  B_{js} \right)   \\  
		& \qquad \ \  - \left( B_{it}^\top (Q^{B}_t)^{-1} (Q^{f})^{-1} f_{s+1}^d \right) \left( a_{js} - Q^{a,B}_t  (Q^{B}_t)^{-1} B_{js} \right) ,
	\end{align*}
	for some positive values $\sigma_{I,it}$, $\sigma_{\delta,qt}$, $\sigma_{o,it}$, and a positive definite matrix $\Sigma_{xf,l}$.
%	
%	Under weak dependence, appropriately scaled sums of the residuals $\epsilon_{it+1}$ obey a central limit theorem, leading to asymptotic normality of the estimators for $\Gamma$, $\alpha_I$, and $\alpha_O$.
\end{assumption}

Assumption~\ref{asp:clt} provides a high-dimensional Lindeberg-type CLT that accommodates growing $L$ and heteroskedastic, weakly dependent errors, forming the statistical backbone of our inference. Because $(g_{it,js},B^{o}_{js})_{j\leq N, s \leq T}$ are functions of $(x_{js},f_{s+1})_{j \leq N, s \leq T}$, this assumption requires a weak dependence in the noise term, $(\epsilon_{js})_{j\leq N, s \leq T}$. For example, if $\epsilon_{it}$ are independent across $i$ and $t$ with $\bbE[\epsilon_{it}^2] = \sigma_{t}^2$, the condition will be satisfied by the Lindeberg theorem with the variances:
\begin{align}\label{eq:simplevariance}
	\nonumber &\Sigma_{xf,l} =  \lim_{N,T \rightarrow \infty} \frac{1}{T} \sum_{t=1}^T \sigma_{t+1}^2 \left[ Q_t^{-1} \right]_{ll} f_{t+1}^d f_{t+1}^{d \top}, \quad  
	\sigma_{I,it}^2 = \lim_{N,T \rightarrow \infty} \frac{1}{NTL} \sum_{j=1}^N \sum_{s=1}^T  \sigma^2_{s+1} g_{it,js}^2, \\
	&\sigma_{\delta,qt}^2 =  \sigma_{t+1}^2, \ \ \sigma_{o,it}^2 = \frac{\bar{\sigma}^2}{T} \left( \frac{1}{N} \norm{B_{t,i}^o}^2 \right) + \sigma_{t+1}^2 \frac{|D_t|}{N} \left( \frac{1}{|D_t|} \sum_{q \in D_t} B_{t,iq}^{o2}  \right) ,
\end{align}
where $\bar{\sigma}^2 = \frac{1}{T} \sum_{s=1}^T \sigma_{s+1}^2$. Because the size of $\norm{B_{t,i}^o}^2$ is close to $N-L$ and $B_{t,iq}^{o}$ is generally bounded, we can say roughly $\sigma_{o,it}^2 \asymp \frac{1}{T} + \frac{|D_t|}{N}$. In Assumption (ii), we adjusted the scale by including $\sqrt{L}$ in the denominator to avoid divergence. Without difficulty, we can show that the variances $\sigma_{I,it}^2$, $\sigma_{\delta,qt}^2$, and $\Sigma_{xf,l}$ are bounded under our weak dependence assumption. 

We are now in position to state the distributional properties of various parameters.

\subsection{Asymptotic Distributions}

We first derive the asymptotic distribution of the characteristic-loading matrix $\Gamma$. 
The spectral estimator is consistent but biased when $T$ is not small relative to $N$. 
The debiased estimator corrects this bias and enables valid inference.

\begin{theorem}[Asymptotic Normality of $\Gamma$]\label{thm:gamma_1}
	Suppose that Assumptions \ref{asp:characteristics} -- \ref{asp:sparsity_heterodelta}, \ref{asp:clt} (i) are satisfied.\\ 
	(a) If $L/N \rightarrow 0$, $T/N \rightarrow 0$, and $T/\left( \frac{N}{L} \right)^{20} \rightarrow 0$, For each $1 \leq l \leq L$,
	$$
	\sqrt{NT} \left( \tilde{\gamma}_l - H^\top_{\Gamma} \gamma_l \right) \conD \calN \left(0,  \boldsymbol{H}^\top \Sigma_f^{-1} \Sigma_{xf,l}   \Sigma_f^{-1}\boldsymbol{H} \right),
	$$
	where $\boldsymbol{H}$ is the limit of $H_{\Gamma}$ and $H^{-1}_{F}$.\\ 
	(b) If $L/N \rightarrow 0$, $T/N^3 \rightarrow 0$, and $\left(\frac{T}{N}\right) / \left( \frac{N}{L} \right)^{20} \rightarrow 0$, we have for each $1 \leq l \leq L$,
	$$
	\sqrt{NT} \left( \hat{\gamma}_{l} - H^\top_{\Gamma} \gamma_l \right) \conD \calN \left(0,  \boldsymbol{H}^\top \Sigma_f^{-1} \Sigma_{xf,l}   \Sigma_f^{-1}\boldsymbol{H} \right).
	$$
\end{theorem}

Here, we present the asymptotic normality of each $\gamma_l$ rather than that of $\Gamma$ because the dimension of $\Gamma$ diverges when $L \rightarrow \infty$. The conditions for (b), $T/N^3 \rightarrow 0$ and $\left(\frac{T}{N}\right) / \left( \frac{N}{L} \right)^{20} \rightarrow 0$, are milder than the conditions for (a), $T/N \rightarrow 0$ and $T/\left( \frac{N}{L} \right)^{20} \rightarrow 0$. Hence, when $N$ is not much larger than $T$ (or smaller than $T$), the debiased estimator $\hat{\gamma}_{l}$ can be useful. Theorem~\ref{thm:gamma_1} shows that the debiased estimator is asymptotically normal even when the time dimension is moderately large relative to $N$. This permits standard inference on the relationship between firm characteristics and factor exposures in typical empirical panels.

We next consider the component of mispricing explained by firm characteristics but orthogonal to factors.

\begin{theorem}[Asymptotic Normality of $\alpha_{I,it}$]\label{thm:alpha_I}
	Suppose that Assumptions \ref{asp:characteristics} -- \ref{asp:sparsity_heterodelta}, \ref{asp:clt} (ii) are satisfied.\\
	(a) If $L/N \rightarrow 0$, $T/N \rightarrow 0$ and $T/\left( \frac{N}{L} \right)^{20} \rightarrow 0$, we have
	$$
	V_{I,it}^{-1/2} \left( \tilde{\alpha}_{I,it} -  \alpha_{I,it} \right) \conD \calN(0,1),
	$$
	where $V_{I,it} = \sigma_{I,it}^2L/NT$.\\
	(b)
	If $L/N \rightarrow 0$, $T/N^3 \rightarrow 0$, $\left(\frac{T}{N}\right) / \left( \frac{N}{L} \right)^{20} \rightarrow 0$, then we have
	$$
	V_{I,it}^{-1/2} \left( \hat{\alpha}_{I,it} -  \alpha_{I,it} \right) \conD \calN(0,1).
	$$
\end{theorem}

Note that, because the convergence rate of $\hat{\alpha}_{I,it}$ is $\sqrt{L}/\sqrt{NT}$, the test using this estimator can have a higher power than that using $(P_{X_t} - P_{X_t \hat \Gamma}) R_{t+1}$ as an estimator. Similarly to Theorem \ref{thm:gamma_1}, the inferential theory based on $\hat{\alpha}_{I,it}$ requires milder conditions for $N$ and $T$ compared to that of $\tilde{\alpha}_{I,it}$. The convergence rate of $\hat{\alpha}_{I,it}$ is $\sqrt{L/(NT)}$, yielding high efficiency even in high-dimensional settings. This enables powerful tests for systematic pricing errors linked to observable fundamentals.

We now analyze the residual component $\alpha_O$, orthogonal to both factors and firm characteristics. 
The key intermediate parameter is the coefficient vector $\delta_{o,t}$.

\begin{theorem}[Asymptotic Normality of $\delta_{o,t}$]\label{thm:delta_heterodelta}
	Suppose that Assumptions \ref{asp:characteristics}, \ref{asp:noise}, and \ref{asp:clt} (iii) are satisfied. Then, we have
	$$
	V_{\delta,tq}^{-1/2}  \left( \tilde{\delta}_{o,t,q} -  \delta_{o,t,q} \right) \conD \calN(0,1), \qquad \text{where  } V_{\delta,tq} = \sigma_{\delta,qt}^2/N.
	$$
\end{theorem}

Importantly, this result is still valid without the assumptions regarding sub-Gaussianity and cross-sectionally independent noise as long as noise is weakly dependent across $i$. Moreover, it does not require the sparsity condition. This result can be utilized to conduct an outside alpha test whose null hypothesis is $H_o : \delta_{o,t} = 0$ for all $t$, because based on the asymptotic normality above, we can have 
$$
\bbP \left( \max_{t \leq T,q \leq N - L } \abs{ \hat{V}^{-1/2}_{\delta,tq} \left(\tilde{\delta}_{o,t,q} - \delta_{o,t,q} \right)}  >  \Phi^{-1} (1 - a /(2T(N-L)) ) \right) \leq a + o(1),
$$
e.g., \cite{belloni2018high}. Theorem~\ref{thm:delta_heterodelta} allows testing for the existence of outside alphas via the null $H_0: \delta_{o,t} = 0$ for all $t$. The test can be implemented using extreme-value approximations as in \citet{belloni2018high}, providing a way to detect residual anomalies beyond characteristic-based mispricing.

To extend inference from $\delta_{o,t}$ to $\alpha_{O,it}$, we impose mild regularity conditions controlling approximation bias.

\begin{assumption}[Regularity for Outside-Alpha Bias Control]
	\label{asp:regularity_heterodelta}
	Conditional on $(x_{it})$, the following hold:
	\begin{itemize}
		\item[(i)] $\frac{|D_t|}{NT}\frac{1}{|D_t|}\sum_{q\in D_t}B_{o,t,iq}^2 \ll \sigma_{o,it}^2;$
		\item[(ii)] $\frac{1}{NT}\sum_{s=1}^T |D_s|\frac{1}{|D_s|}\sum_{q\in D_s\setminus D_t} B_{o,t,iq}^2 \ll \sigma_{o,it}^2;$
		\item[(iii)] $\frac{1}{T}\sum_{s=1}^T |D_s| \left(\frac{1}{|D_s|}\sum_{q\in D_s\setminus D_t} B_{o,t,iq}\bar{\xi}_q\right) \ll \sigma_{o,it}.$
	\end{itemize}
\end{assumption}

Assumption~\ref{asp:regularity_heterodelta} is mild and automatically satisfied when the number of firms with nonzero transitory shocks is small relative to $N$ and $T$. It ensures that cross-sectional spillovers from temporary idiosyncratic shocks are asymptotically negligible.

In the case of the first relation, the order of the left side is roughly $\frac{|D_t|}{NT}$ while that of $\sigma_{o,it}^2$ is roughly $\frac{1}{T} + \frac{|D_t|}{N}$ as we noted in \eqref{eq:simplevariance}. Hence, when $N,T \rightarrow \infty$, it would be satisfied. Similarly, because the order of the left side of the second relation is roughly $\frac{|\bar{D}_\star|}{NT}$ where $|\bar{D}_\star| = \frac{1}{T} \sum_{s=1}^T |D_s|$, the second condition would be satisfied. Lastly, the third relation would be satisfied by the sparsity of $\xi_{t}$. For instance, if $\{\xi_{t,q}\}$ is nonzero at a small number of time periods by the sparsity, the order of $\bar{\xi}_{q}$ would be roughly $\frac{1}{T}$. Hence, the order of the left side is roughly $\frac{|\bar{D}_\star|}{T}$ and less than $\frac{1}{\sqrt{T}} + \frac{\sqrt{|D_t|}}{\sqrt{N}}$, when $|\bar{D}_\star|$ is small due to the sparsity of $\xi$. Then, under the above conditions, we have the following asymptotic normality.

\begin{theorem}[Asymptotic Normality of $\alpha_{O,it}$]\label{thm:alpha_O_heterodelta}
	Suppose that Assumptions \ref{asp:characteristics}, \ref{asp:noise}, \ref{asp:sparsity_heterodelta}, \ref{asp:clt} (iv), \ref{asp:regularity_heterodelta} are satisfied. Additionally, if $(\epsilon_{it})_{i\leq N , t \leq T}$ is dependent across $t$, assume that
	$$
	\bbE \left[ \abs{ \frac{1}{\sqrt{NT}} \sum_{s=1}^T \sum_{j=1}^N B_{s,jq}^{o} \epsilon_{j,s+1} }^\alpha \right] \quad \text{is bounded}    
	$$
	for some integer $\alpha \geq 1$ where $N = O(T^{\alpha/2})$. Then, we have
	$$
	V_{o,it}^{-1/2} \left( \hat{\alpha}_{O,it} -  \alpha_{O,it} \right) \conD \calN(0,1),
	$$
	where $V_{o,it} = \sigma_{o,it}^2 $ is in Assumption \ref{asp:clt}.
\end{theorem}

Theorem~\ref{thm:alpha_O_heterodelta} completes the inferential theory by establishing Gaussian limits for the outside-alpha estimator. Together with Theorems~\ref{thm:gamma_1}--\ref{thm:delta_heterodelta}, it provides a comprehensive inferential framework for both systematic and idiosyncratic components of mispricing.

\medskip
\medskip
Our inferential results provide the following empirical tools:
\begin{itemize}
	\item \textbf{Testing characteristic relevance:} Wald-type tests on each $\gamma_l$ identify which firm attributes significantly explain factor exposures.
	\item \textbf{Evaluating systematic mispricing:} Tests on $\alpha_I$ detect whether pricing errors align with observable fundamentals.
	\item \textbf{Detecting residual anomalies:} Tests on $\alpha_O$ assess whether idiosyncratic mispricing remains after accounting for all systematic sources.
\end{itemize}

These tools yield a unified econometric framework that is both theoretically grounded and empirically tractable, enabling rigorous inference in large-scale panels of asset returns with rich firm characteristics.

\section{Application to U.S. Stock Data}\label{sec:empirical}

We now illustrate the empirical relevance of our framework by applying it to U.S. equity returns. This section evaluates the magnitude, dynamics, and economic interpretation of both inside and outside alphas estimated using our methodology. 
The goal is to demonstrate how the inferential theory developed in Section~\ref{sec:inferential_theory} translates into concrete insights about mispricing and factor structure in the cross-section of stock returns.

\subsection{Data and Methods}

\paragraph{Data.}  
We examine monthly excess returns on U.S.\ stocks from January~2000 through December~2019, yielding $T = 240$ time periods. Our data are drawn from the same sources as \citet{zhang2024testing}, covering $N = 973$ continuously observed firms. We use the $36$ firm characteristics from \citet{kelly2019characteristics} and \citet{chen2023semiparametric}, augmented by a constant, as potential explanatory variables. 
These characteristics span size, value, profitability, investment, momentum, liquidity, and trading frictions, and are detailed in Appendix \ref{appendix:characteristics}.

Following standard practice, each characteristic $x_{i,t,l}$ is transformed into a rank-normalized variable across firms at time $t$:
\[
x_{i,t,l} = -0.5 + \frac{z_{i,t,l}}{N},
\]
where $z_{i,t,l}$ denotes the cross-sectional rank of firm~$i$. This transformation mitigates the influence of outliers and ensures scale invariance.

\paragraph{Estimation.}  
We implement the debiased estimation procedure from Section~\ref{sec:estimation}. Given that $N$ is of the same order of magnitude as $T$, we employ the debiased estimators $\widehat{\Gamma}$ and $\widehat{\alpha}_{I,it}$ to obtain valid inference under finite-sample bias. The rank of $\Gamma$ (the number of latent factors $K$) is selected using the eigenvalue-ratio criterion proposed by \citet{chen2023semiparametric}. For the orthogonal complement $X_t^o$ in constructing $B_t^o$, we adopt the specification in Section~\ref{sec:model}. The threshold parameter $\rho_t$ in the sparse outside-alpha estimation is set to
\[
\rho_t = \widehat{\sigma}_{t+1} \frac{(\log NT)^{0.6}}{\sqrt{N}},
\]
where $\widehat{\sigma}_t^2$ is the cross-sectional variance of residuals at time $t$. All variances used in inference are estimated under the assumption of independence and heteroskedasticity across time.

\subsection{Empirical Findings}

We now examine the estimated pricing errors and factor structure implied by the model.  
Throughout, we report results for $K = 1$ to $10$, highlighting $K = 5$ as the benchmark case selected by the data.

\subsubsection{Testing for Outside Alphas}

We first test whether the model admits a nontrivial outside-alpha component ($\alpha_O$) and whether these effects vary over time. The corresponding hypotheses are
\[
H_0^{(1)}: \delta_{o,t} = 0 \quad \text{for all } t,
\qquad 
H_0^{(2)}: \delta_{o,t} = \delta_o \quad \text{for all } t.
\]
The test statistics follow from Theorem~\ref{thm:delta_heterodelta}:
\[
T\text{-stat}_1 = \max_{t \le T, q \le N-L} |\widehat{\tau}_{1,tq}|, 
\quad
\widehat{\tau}_{1,tq} = \widehat{V}_{\delta,tq}^{-1/2}\,\tilde{\delta}_{o,t,q},
\]
\[
T\text{-stat}_2 = \max_{t \le T, q \le N-L} |\widehat{\tau}_{2,tq}|, 
\quad
\widehat{\tau}_{2,tq} = \widehat{V}_{\delta,tq}^{-1/2}\!\left(
\tilde{\delta}_{o,t,q} - \frac{1}{T}\sum_{s=1}^T \tilde{\delta}_{o,s,q}
\right).
\]
Table~\ref{tab:sigtest} reports these statistics for $K = 1,\dots,10$.  
Under the null, the extreme-value bound from \citet{belloni2018high} provides asymptotically valid $p$-values:
\[
P\!\left(
\max_{t,q} |\widehat{\tau}_{tq}| > \Phi^{-1}(1 - a/(2T(N-L)))
\right) \le a + o(1).
\]

\begin{table}[h]
  \centering
  \footnotesize
  \caption{$\alpha_O$ test and heterogeneous $\delta_{o,t}$ test}
 \begin{tabular}{c|cccccccccc}
    \hline
    \hline
    K     & 1     & 2     & 3     & 4     & 5*   & 6     & 7     & 8     & 9     & 10 \\[0.3em]
    \hline
    $T-stat_{1}$ & 15.45  & 16.05  & 16.95  & 17.16  & 17.30  & 16.88  & 17.49  & 17.69  & 17.81  & 17.81 \\[0.3em]
    p-value ($T-stat_{1}$) & \multicolumn{10}{c}{$<10^{-10}$} \\[0.4em]
    $T-stat_{2}$  & 16.06  & 16.34  & 16.96  & 17.17  & 17.30  & 16.89  & 17.50  & 17.70  & 17.81  & 17.82 \\[0.3em]
    p-value ($T-stat_{2}$) & \multicolumn{10}{c}{$<10^{-10}$} \\[0.4em]
    \hline
    \end{tabular}%
    	\\[0.3em]  \noindent {\footnotesize Footnote: The critical values for significance levels $5\%$ and $1\%$ are 5.18 and 5.47, respectively.}
  \label{tab:sigtest}%
\end{table}%

%\paragraph{Results.}
As shown in Table~\ref{tab:sigtest}, both $T\text{-stat}_1$ and $T\text{-stat}_2$ exceed the 1\% critical value (5.47) by a wide margin across all $K$. The associated $p$-values are below $10^{-10}$, decisively rejecting both null hypotheses.  
Hence, the data exhibit statistically and economically significant outside alphas, and these effects are time-varying.  
This finding underscores that idiosyncratic mispricing persists beyond the span of firm characteristics and evolves dynamically over time.

\subsubsection{Testing for Inside and Outside Pricing Errors}

Next, we test for the joint existence of both inside and outside alphas at the firm-month level using
\[
T\text{-stat}_O = \max_{i,t} |\widehat{\tau}_{O,it}|, \quad
\widehat{\tau}_{O,it} = \widehat{V}_{O,it}^{-1/2}\widehat{\alpha}_{O,it},
\]
\[
T\text{-stat}_I = \max_{i,t} |\widehat{\tau}_{I,it}|, \quad
\widehat{\tau}_{I,it} = \widehat{V}_{I,it}^{-1/2}\widehat{\alpha}_{I,it}.
\]
The null hypothesis is $H_0: \alpha_{\iota,it}=0$ for all $(i,t)$ and $\iota \in \{O,I\}$.  
Critical values are again obtained using the extreme-value approximation in \citet{belloni2018high}.  
Table~\ref{tab:alphatest_het} reports the resulting statistics and model $R^2$ values.

\begin{table}[h]
	\centering
	\footnotesize
	\caption{Alpha test and $R^2$}
	\begin{tabular}{c|cccccccc}
		\hline
		\hline
		K     & $T-stat_o$ & p-value ($T_o$) & $T-stat_I$ & p-value ($T_I$) & 10\%  & 5\%   & 1\%   &  $R^2$ \\[0.3em]
		\hline
		1     &  36.618 & $< 10^{-10}$ & 22.401 & $< 10^{-10}$ &  &  &  & 6.34\% \\[0.3em]
		2     &  38.026  & $< 10^{-10}$ & 22.395 & $< 10^{-10}$ &       &       &       & 7.80\% \\[0.3em]
		3     & 40.153  & $< 10^{-10}$  & 16.230 & $< 10^{-10}$ &       &       &       & 11.50\% \\[0.3em]
		4     &  40.647 & $< 10^{-10}$  & 14.327 & $< 10^{-10}$ &       &       &       & 12.47\% \\[0.3em]
		5*    & 40.971 & $< 10^{-10}$  & 14.350 & $< 10^{-10}$ &  5.056     &  5.186      &  5.478      & 14.36\% \\[0.3em]
		6     & 40.001  & $< 10^{-10}$  & 15.583 & $< 10^{-10}$ &       &       &       & 21.06\% \\[0.3em]
		7     & 41.435 & $< 10^{-10}$  & 14.362 & $< 10^{-10}$ &      &       &       & 22.41\% \\[0.3em]
		8     & 41.916 & $< 10^{-10}$  & 14.859 & $< 10^{-10}$ &       &       &       & 26.14\% \\[0.3em]
		9     & 42.192 & $< 10^{-10}$  & 13.459 & $< 10^{-10}$ &       &       &       & 26.83\% \\[0.3em]
		10    & 42.201 & $< 10^{-10}$  & 13.565 & $< 10^{-10}$ &       &       &       & 27.21\% \\
		\hline
	\end{tabular}%
	\\[0.3em] 
	\raggedright
	\noindent {\footnotesize Footnote: `5*' means that the estimated $K$ is $5$. $10\%$, $5\%$, and $1\%$ denote the critical values for each significance level. These critical values are the same over $K$ because $N$ and $T$ are the same.}
	\label{tab:alphatest_het}%
\end{table}%

%\paragraph{Results.}  
For all $K$, both $T\text{-stat}_O$ and $T\text{-stat}_I$ reject the null hypothesis at significance levels below $10^{-10}$.  
Hence, both inside and outside alphas are pervasive in the cross-section of returns.  
The explanatory power of the model increases with the number of factors, with $R^2$ rising from $6.3\%$ for $K=1$ to $27.2\%$ for $K=10$.  
At the empirically selected $K=5$, the model explains $14.4\%$ of total variation in returns, suggesting a balance between parsimony and explanatory strength.  
These results affirm the empirical relevance of decomposing mispricing into characteristic-driven and residual components.

\subsubsection{Dynamics and Economic Interpretation of Inside Alphas}

We now explore the temporal and cross-sectional behavior of the inside-alpha component $\widehat{\alpha}_I$, which captures systematic mispricing linked to firm characteristics but orthogonal to factor betas.

\begin{figure}[h]
	\centering
	\includegraphics[width=0.85\textwidth]{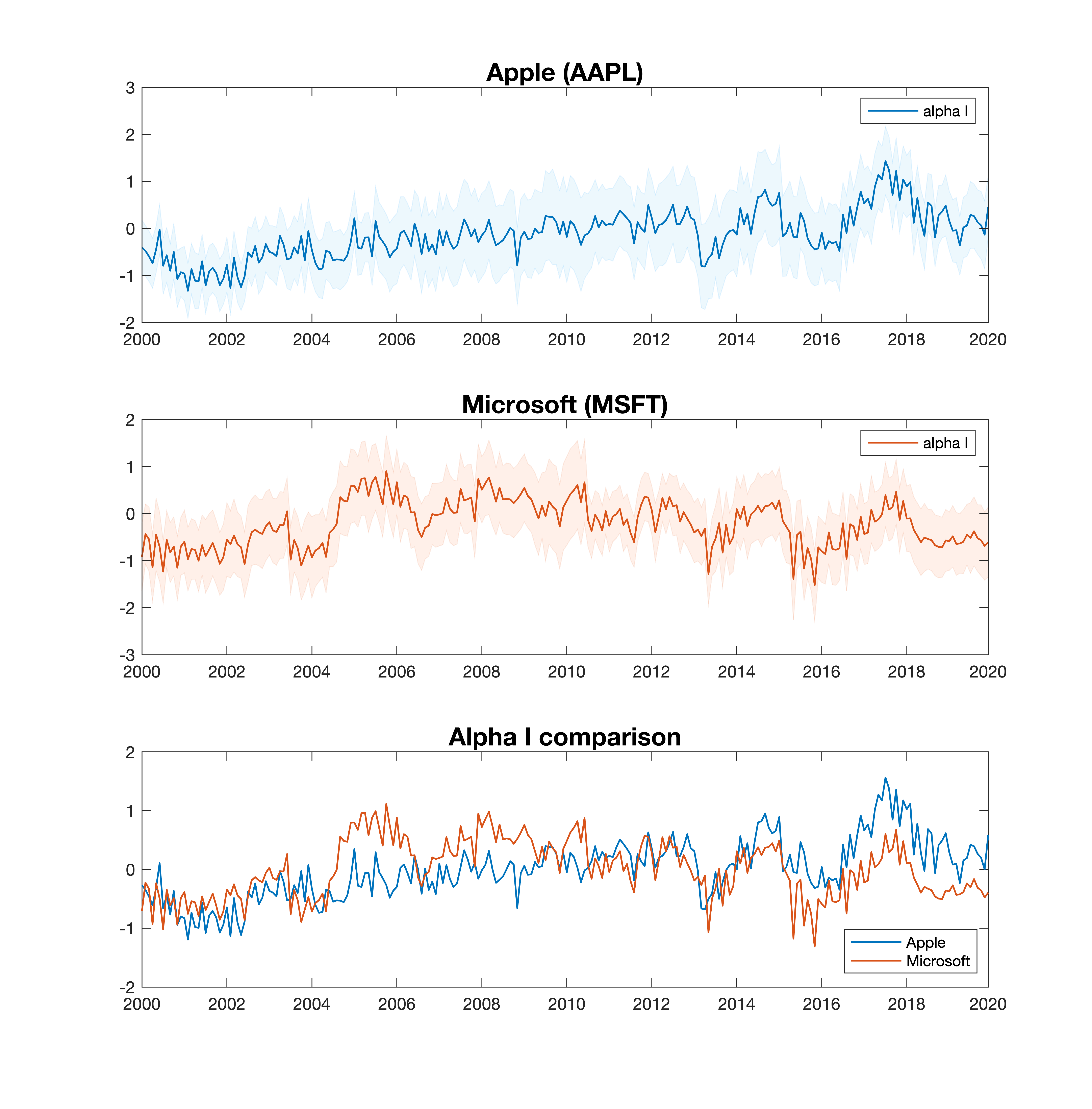}
	\caption{The dynamics of $\alpha_I$ of typical tech firms: \small In the comparison graph, we plot the centered $\alpha_I$. For the confidence band, we adjust the critical values using a FDR control method (Benjamini–Yekutieli procedure).}
	\label{fig:Tech_comp}
\end{figure}

\begin{figure}[h]
	\centering
	\includegraphics[width=0.85\textwidth]{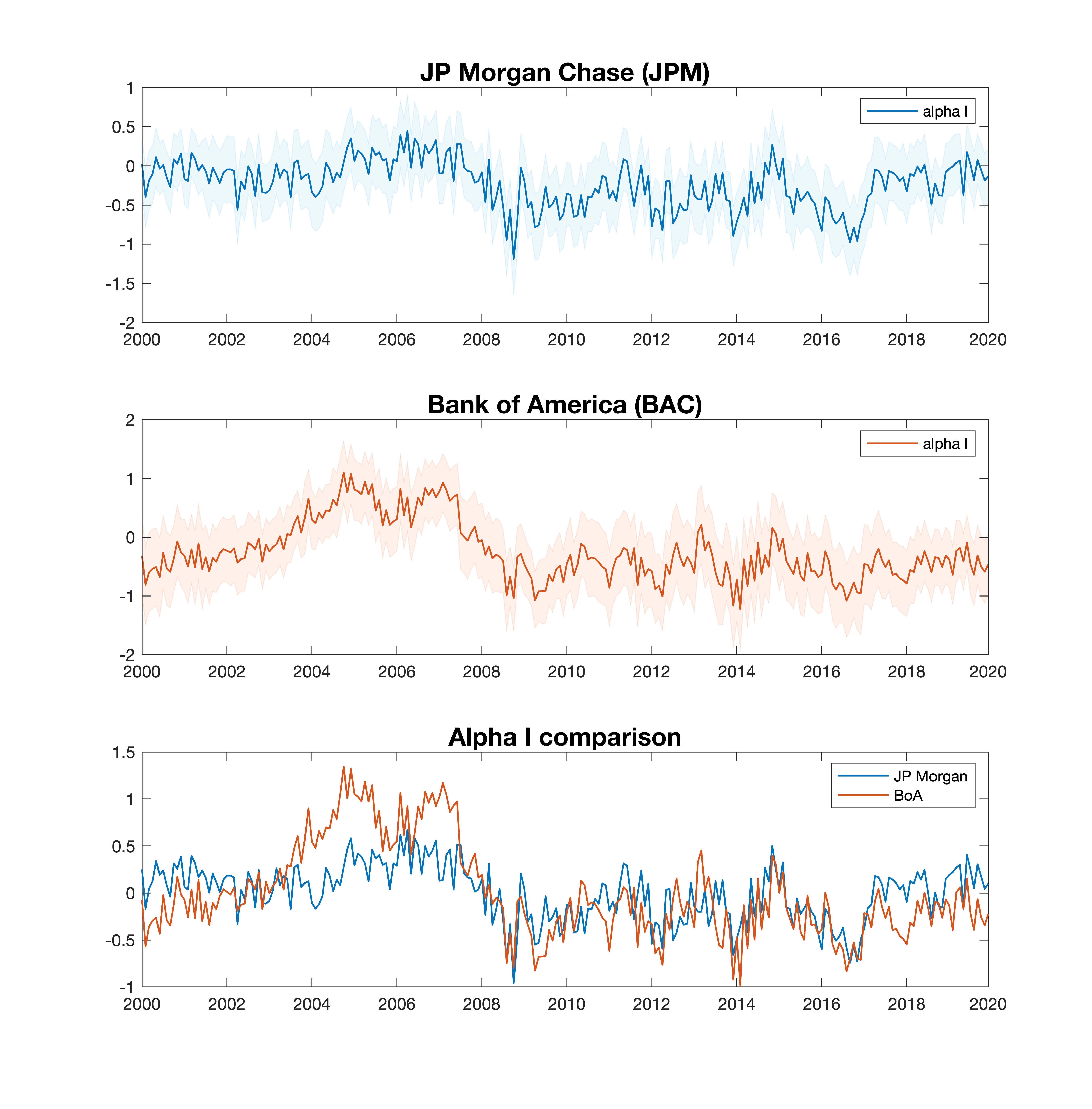}
	\caption{The dynamics of the monthly $\alpha_I$ of typical finance firms.}
	\label{fig:Finance_comp}
\end{figure}

\begin{figure}[h]
	\centering
	\includegraphics[width=\textwidth]{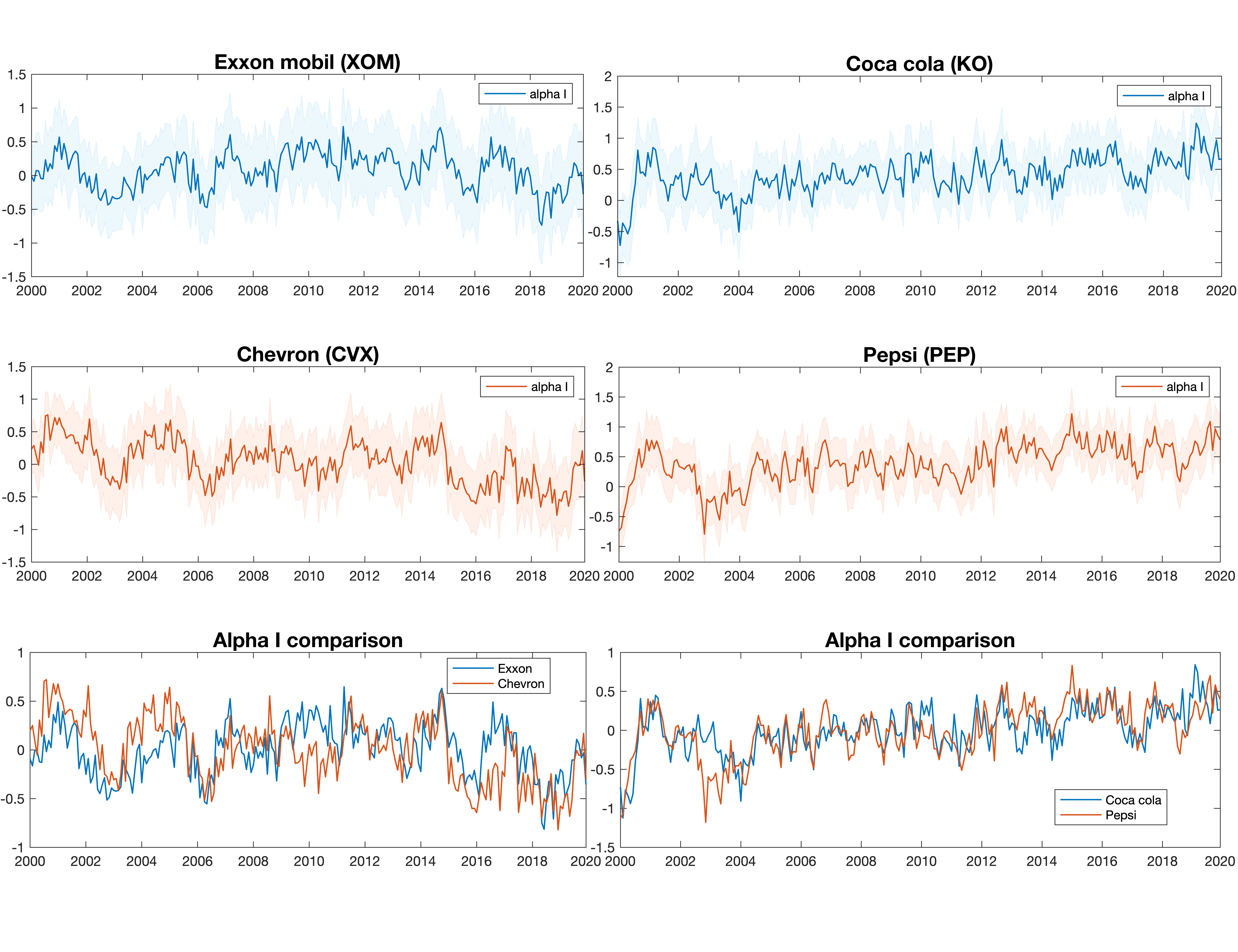}
	\caption{The dynamics of the monthly $\alpha_I$ of typical oil and beverage companies.}
	\label{fig:together_comp}
\end{figure}

\begin{figure}[htp]
	\centering
	\includegraphics[width=\textwidth]{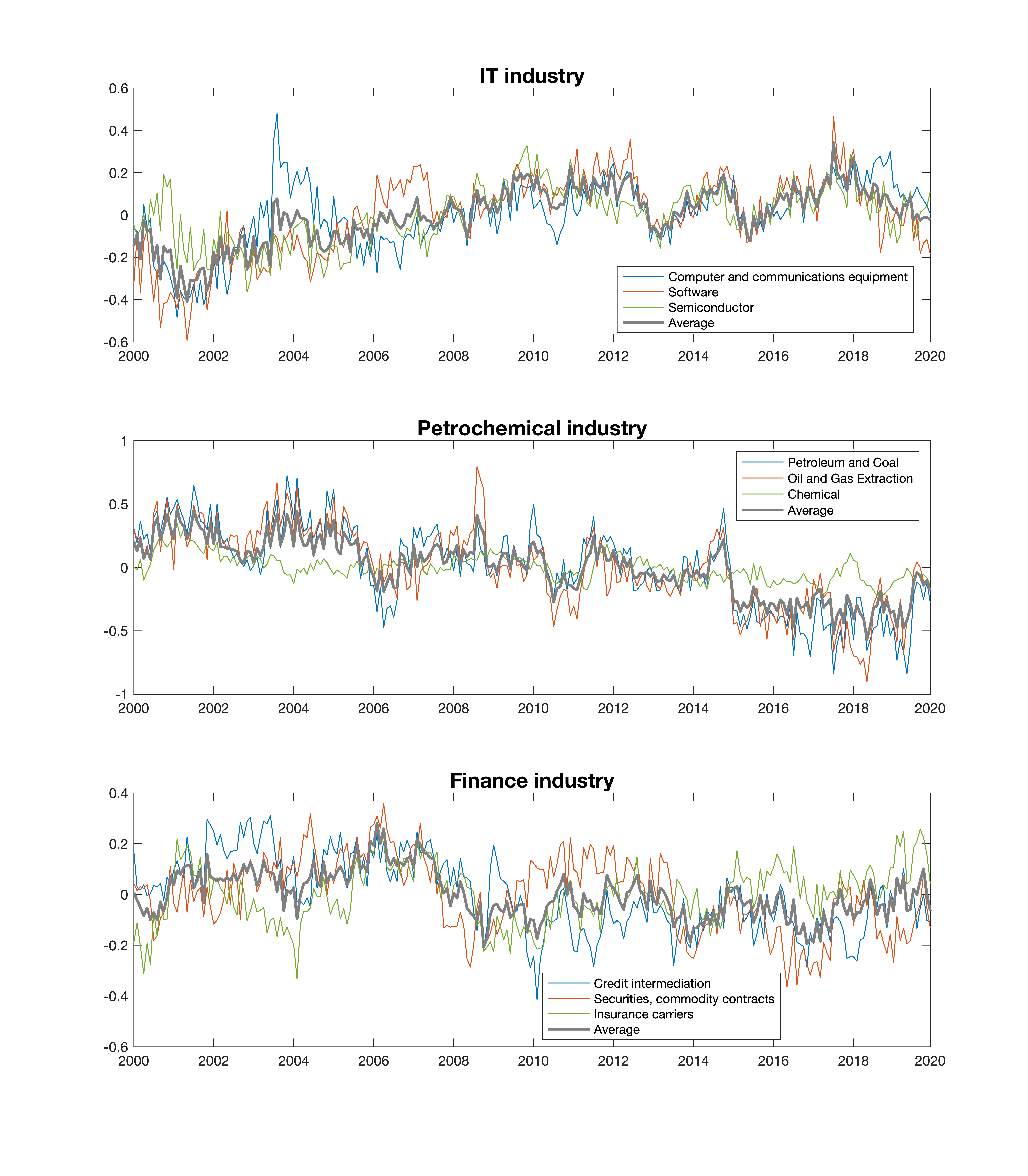}
	\caption{The dynamics of the monthly sector-averaged $\alpha_I$: \small `computer and communications equipment' denotes NAICS 3341\&3342, `software' denotes denotes NAICS 5112, and `semiconductor' denotes NAICS 3344. `Petroleum and Coal' denotes NAICS 324, `oil and gas extraction' denotes denotes NAICS 211, and `chemical' denotes NAICS 325. `credit intermediation' denotes NAICS 522, `securities, commodity contracts' denotes denotes NAICS 523, and `insurance carriers' denotes NAICS 524. Here, we use the centered $\alpha_I$ and the line `Average' denotes the average of $\alpha_I$ of the sectors.}
	\label{fig:Industry_1}
\end{figure}

\begin{figure}[htp]
	\centering
	\includegraphics[width=\textwidth]{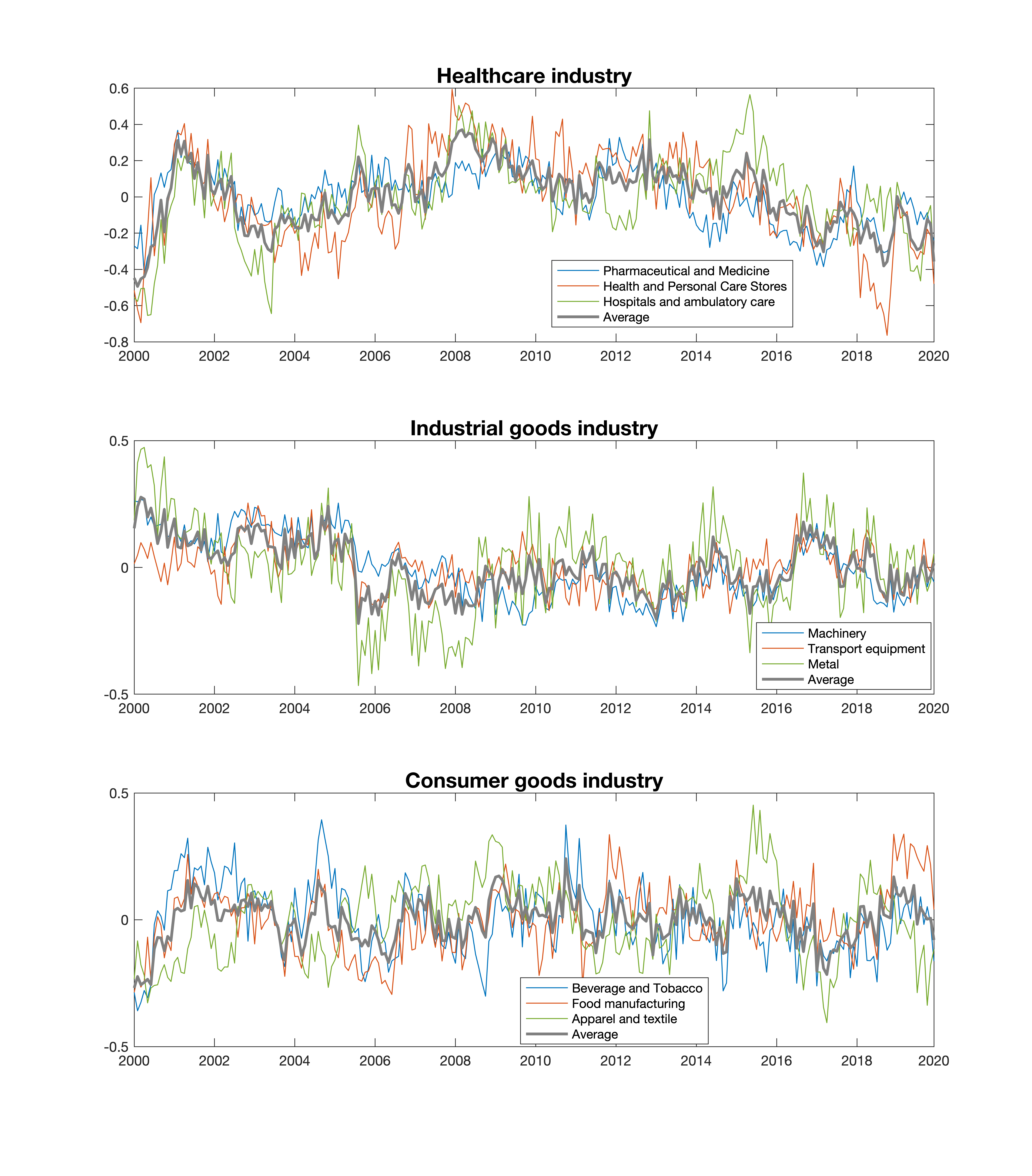}
	\caption{The dynamics of the monthly sector-averaged $\alpha_I$: \small `pharmaceutical and medicine' denotes NAICS 3254, `health and personal care stores' denotes denotes NAICS 446, and `hospitals and ambulatory care' denotes NAICS 62. `machinery' denotes NAICS 333, `transportation equipment' denotes denotes NAICS 336, and `metal' denotes NAICS 331. `beverage and tobacco' denotes NAICS 312, `food manufacturing' denotes denotes NAICS 511, and `apparel and textile' denotes NAICS 313--316. Here, we use the centered $\alpha_I$ and the line `Average' denotes the average of $\alpha_I$ of the sectors.}
	\label{fig:Industry_2}
\end{figure}

Figures~\ref{fig:Tech_comp}-\ref{fig:Industry_2} plot the estimated monthly inside alphas for representative firms and sector averages, together with 95\% confidence intervals adjusted via the false discovery rate (FDR) control of \citet{benjamini2001control}.  
In what follows, we discuss several representative patterns.

\paragraph{Technology Sector.}  
Figure~\ref{fig:Tech_comp} depicts $\widehat{\alpha}_I$ for Apple and Microsoft.  
Both exhibit pronounced co-movement: alphas were low during the early 2000s following the dot-com crash, remained resilient through the 2008 financial crisis, and trended upward post-2010.  
The alignment of $\alpha_I$ across these firms suggests that inside alphas capture persistent industry-level fundamentals rather than firm-specific anomalies.

\paragraph{Financial Sector.}  
Figure~\ref{fig:Finance_comp} plots $\widehat{\alpha}_I$ for J.P.\ Morgan Chase and Bank of America.  
Both series decline sharply during the 2007–2008 crisis, indicating that beyond the market-wide factor exposure, financial firms suffered deterioration in fundamentals not captured by standard betas.  
Post-crisis, their inside alphas recover gradually and move in tandem, again pointing to a strong sectoral component.

\paragraph{Energy and Consumer Sectors.}  
Figures~\ref{fig:together_comp} display $\widehat{\alpha}_I$ for representative oil and consumer goods firms.  
Within-industry alphas exhibit substantial co-movement, most notably for ExxonMobil and Chevron, consistent with shared exposure to oil prices and global supply conditions. 

\paragraph{Industry-Level Evidence.}  
Figure~\ref{fig:Industry_1} and Figure~\ref{fig:Industry_2} summarize sector-level average inside alphas based on NAICS classifications.  
Inside alphas display clear industry patterns: 
the IT sector shows sharp declines during the dot-com crash but little response to the financial crisis; 
the petrochemical and finance sectors experience simultaneous declines during 2008–2009; 
and the healthcare and consumer goods sectors maintain positive alphas during downturns, consistent with their resilience and inelastic demand. Overall, inside alphas track industry fundamentals and sectoral shocks rather than aggregate macroeconomic fluctuations, reinforcing their interpretation as characteristic-linked systematic mispricing.

\subsubsection{Dynamics of Outside Alphas}

\begin{figure}[h]
	\centering
	\includegraphics[width=\textwidth]{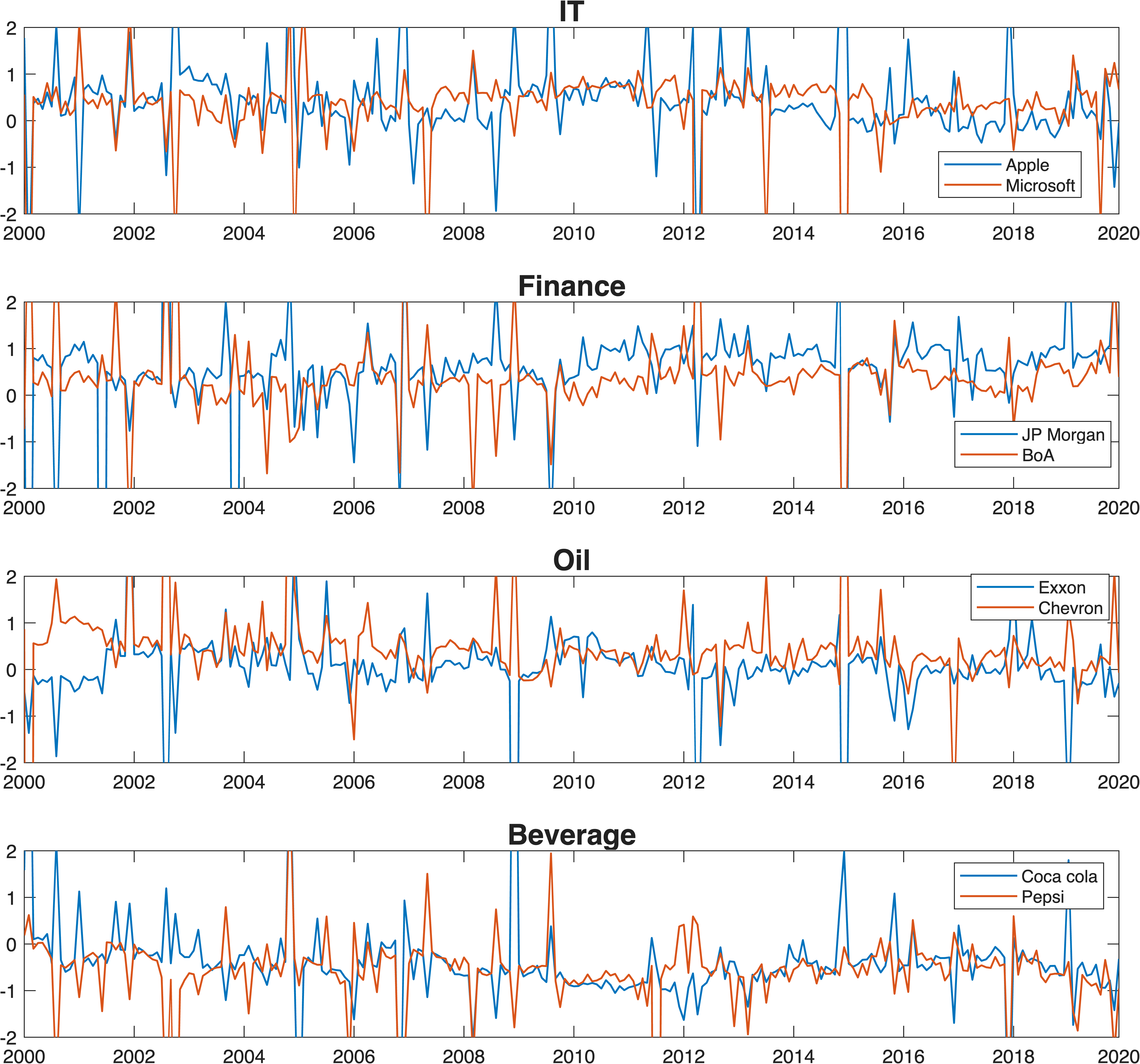}
	\caption{The dynamics of the monthly $\alpha_O$ of typical companies.}
	\label{fig:alpha_o_ind}
\end{figure}

\begin{figure}[h]
	\centering
	\includegraphics[width=0.9\textwidth]{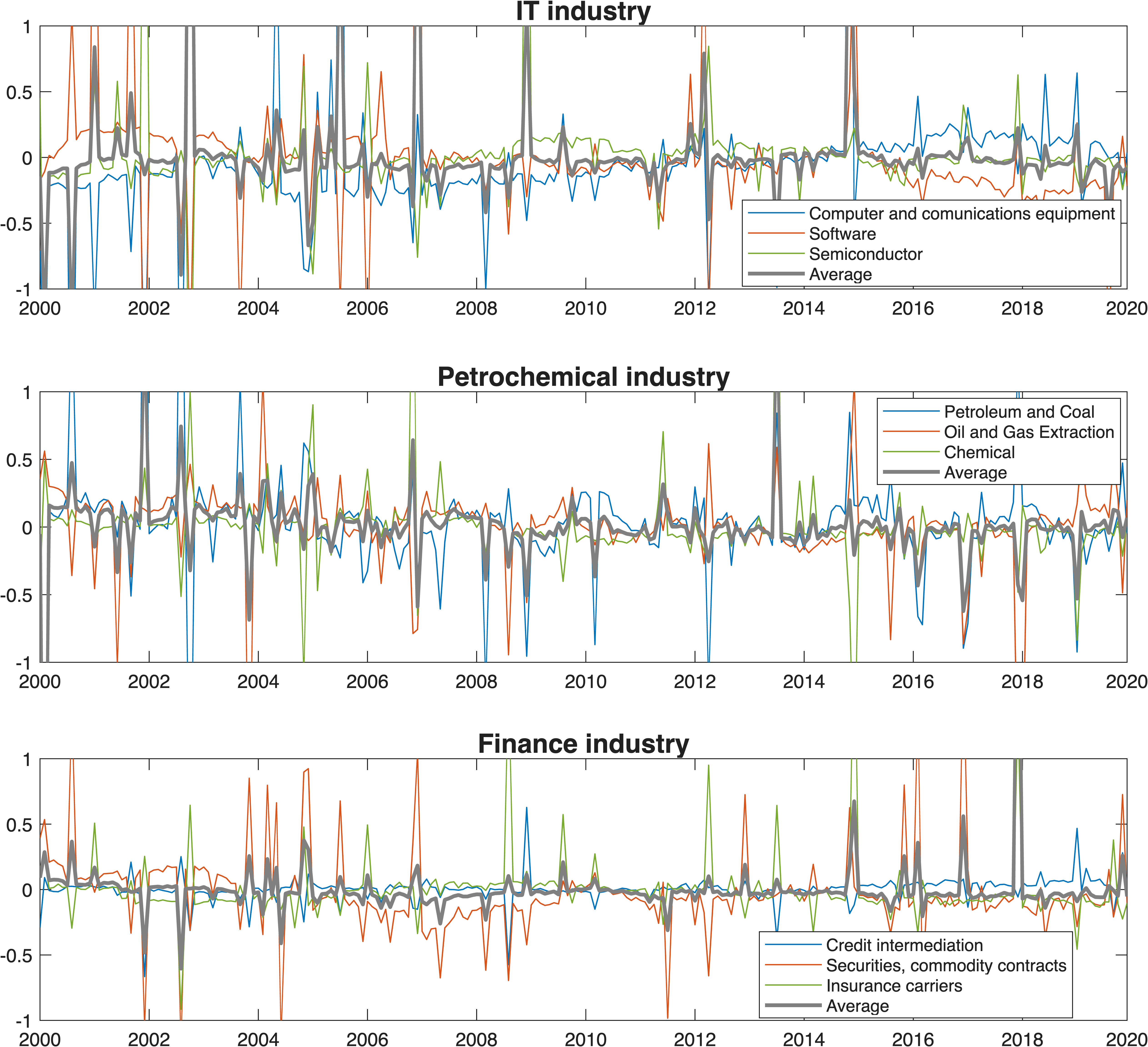}
	\caption{The dynamics of the monthly sector-averaged $\alpha_O$.}
	\label{fig:alpha_o_sec}
\end{figure}

We next examine the residual component $\widehat{\alpha}_O$, orthogonal to both characteristics and factors.  
Figures~\ref{fig:alpha_o_ind} and~\ref{fig:alpha_o_sec} plot representative firm-level and sector-averaged series.  
Unlike $\alpha_I$, the outside alphas exhibit no clear co-movement across firms or industries, suggesting that they primarily reflect idiosyncratic, transient deviations from fundamental value.  
This distinction between structured and residual mispricing provides new evidence on how inefficiencies manifest in the cross-section of returns.

\subsubsection{Factor Loadings and Characteristic Relevance}

Finally, we investigate the estimated $\widehat{\Gamma}$ matrix to assess which characteristics drive variation in factor exposures.  
We compute the Wald statistic
\[
W_l = \widehat{\gamma}_l^\top \widehat{V}_{\gamma_l}^{-1} \widehat{\gamma}_l,
\]
which follows a $\chi^2(K)$ distribution under $H_0: \gamma_l = 0$.  
Table~\ref{tab:gammatest} reports the results for $K = 1$–$10$, with Bonferroni-adjusted critical values.

\begin{table}[htb]
\scriptsize
  \centering
  \caption{Gamma test}
    \begin{tabular}{c|cccccccccc}
    \hline
    \hline
    Rank  & 1     & 2     & 3     & 4     & 5*    & 6     & 7     & 8     & 9     & 10 \\
    \hline
    inv   & 0.4   & 0.4   & 2.4   & 2.7   & 3.2   & 4.4   & 5.4   & 8.3   & 9.3   & 11.0 \\
    dpi2a & 0.7   & 0.7   & 0.9   & 10.3   & 10.7  & 16.2  & 17.5  & 18.6  & 22.4  & 23.0 \\
    noa   & 2.8   & 2.9   & 6.5   & 8.5   & 16.5  & 40.5** & 43.1** & 60.3** & 61.6** & 69.5** \\
    lbm   & 8.5   & 25.5** & 39.4** & 48.8** & 53.7** & 66.4** & 71.8** & 87.5** & 90.5** & 92.7** \\
    s2p   & 0.0   & 6.9   & 10.9   & 11.6   & 12.3  & 20.8  & 29.2**  & 42.5** & 50.7** & 52.9** \\
    strev & 3.3   & 4.0   & 19.4** & 21.4** & 32.1** & 40.9** & 51.1** & 55.3** & 56.9** & 80.3** \\
    q     & 22.1** & 30.0** & 39.7** & 53.8** & 60.3** & 70.6** & 76.8** & 95.1** & 99.3** & 101.8** \\
    imom  & 0.9   & 1.0   & 1.3   & 3.2   & 7.5   & 8.9   & 9.2   & 37.0** & 48.3** & 59.7** \\
    prof  & 1.2   & 6.0   & 7.5   & 12.0  & 12.6  & 22.2*  & 26.1*  & 32.2** & 55.4** & 62.8** \\
    mom   & 2.8   & 6.2   & 19.8** & 21.0* & 38.1** & 45.7** & 51.0** & 80.6** & 99.1** & 145.0** \\
    ol    & 64.1** & 68.9** & 82.1** & 87.3** & 99.1** & 120.2** & 126.6** & 157.4** & 163.9** & 166.0** \\
    d2a   & 0.7   & 1.2   & 4.3   & 10.4   & 11.3   & 16.1  & 18.0  & 31.7** & 32.2** & 32.9* \\
    lme   & 20.4** & 26.1** & 131.1** & 158.8** & 141.7** & 188.2** & 195.6** & 229.2** & 252.4** & 255.3** \\
    bidask & 0.0   & 2.4   & 3.4   & 12.8  & 25.0** & 29.3** & 41.3** & 52.7** & 77.6** & 86.5** \\
    ltrev & 7.3   & 9.2   & 21.7** & 34.5** & 36.1** & 39.3** & 40.4** & 42.8** & 46.9** & 63.2** \\
    lev   & 0.1   & 0.2   & 0.3   & 7.9   & 7.5   & 11.9  & 12.8  & 15.9  & 16.3  & 16.7 \\
    cto   & 54.7** & 57.1** & 77.7** & 81.3** & 92.6** & 101.5** & 106.6** & 132.7** & 139.9** & 142.0** \\
    ca    & 1.0   & 10.0   & 11.9  & 12.9  & 13.3  & 18.8  & 19.1  & 47.0** & 47.7** & 50.1** \\
    sga2s & 0.1   & 105.8** & 128.6** & 132.7** & 144.4** & 170.3** & 176.5** & 202.9** & 217.8** & 221.2** \\
    at    & 24.1** & 37.1** & 106.7** & 110.6** & 121.9** & 144.4** & 149.7** & 176.0** & 184.5** & 186.6** \\
    ato   & 1.9   & 19.9** & 28.0** & 33.4** & 56.4** & 64.9** & 76.0** & 85.4** & 88.0** & 88.9** \\
    fc2y  & 3.8   & 190.7** & 226.0** & 251.9** & 260.1** & 286.5** & 298.6** & 325.6** & 331.7** & 336.5** \\
    e2p   & 17.7** & 20.5** & 24.9** & 28.9** & 34.0** & 39.4** & 50.5** & 58.1** & 60.1** & 62.6** \\
    fcf   & 1.6   & 1.8   & 2.3   & 3.9   & 4.1   & 4.6   & 9.4   & 10.0   & 13.4  & 14.6 \\
    pm    & 1.1   & 1.8   & 1.9   & 33.6** & 73.0** & 96.6** & 102.5** & 128.6** & 133.3** & 137.2** \\
    lturn & 17.1** & 17.4** & 41.4** & 42.3** & 47.9** & 107.4** & 108.2** & 222.6** & 224.6** & 247.3** \\
    a2me  & 20.5** & 21.1** & 30.3** & 43.1** & 47.6** & 76.6** & 101.3** & 116.9** & 146.6** & 151.2** \\
    roe   & 13.7** & 18.3** & 22.8** & 25.3** & 27.9** & 39.1** & 41.3** & 53.7** & 56.4** & 63.7** \\
    beta  & 0.0   & 26.6** & 257.9** & 322.4** & 349.8** & 447.3** & 457.1** & 685.3** & 985.5** & 1097.4** \\
    suv\_m & 0.5   & 0.9   & 1.5   & 1.8   & 2.4   & 5.6   & 6.8   & 13.8  & 15.4  & 16.7 \\
    oa    & 0.0   & 4.4   & 6.5   & 6.9   & 9.4   & 13.8  & 14.9  & 16.9  & 17.0  & 19.8 \\
    roa   & 2.9   & 5.9   & 10.6   & 11.4   & 13.5  & 31.0** & 35.2** & 50.8** & 52.9** & 54.8** \\
    pcm   & 7.8   & 17.2** & 19.4** & 33.9** & 118.5** & 136.8** & 152.3** & 176.9** & 185.0** & 189.1** \\
    rna   & 5.1   & 36.6** & 49.8** & 56.2** & 77.1** & 83.5** & 95.2** & 103.9** & 105.3** & 107.2** \\
    w52h  & 3.9   & 4.6   & 13.4  & 13.6  & 16.4  & 26.8** & 36.1** & 160.2** & 168.3** & 172.3** \\
    ivol & 0.0   & 2.8   & 4.3   & 12.5  & 13.3  & 16.1  & 17.5  & 18.4  & 43.0** & 49.6** \\
   \hline
    5\%   & 10.3  & 13.2  & 15.6  & 17.8  & 19.8  & 21.7  & 23.6  & 25.4  & 27.1  & 28.8 \\
    1\%   & 13.3  & 16.4  & 19.0  & 21.3  & 23.5  & 25.5  & 27.5  & 29.4  & 31.2  & 33.0 \\
    \hline
    \end{tabular}%
   \label{tab:gammatest}%
   \footnotesize
    	\\[0.3em]  \noindent {\small Footnote: ** and * denote that a variable significantly affects beta at 1\% and 5\% levels, respectively. `5\%' and `1\%' denote the critical values adjusted with Bonferoni correction.}
\end{table}%

%\paragraph{Results and Interpretation.}  
The number of statistically significant characteristics increases with $K$, as additional latent factors capture more structure in the cross-section.  
When $K = 10$, $30$ of $36$ characteristics significantly affect factor loadings.  
Variables such as book-to-market (LBM), Tobin’s $Q$, operating leverage (OL), market equity (LME), and capital turnover (CTO) consistently exhibit large test statistics, indicating that firm size, value, and operating efficiency are fundamental determinants of risk exposures.  
By contrast, investment (INV), leverage (LEV), and free cash flow (FCF) are generally insignificant.

\begin{figure}[h]
	\centering
	\includegraphics[width=0.9\textwidth]{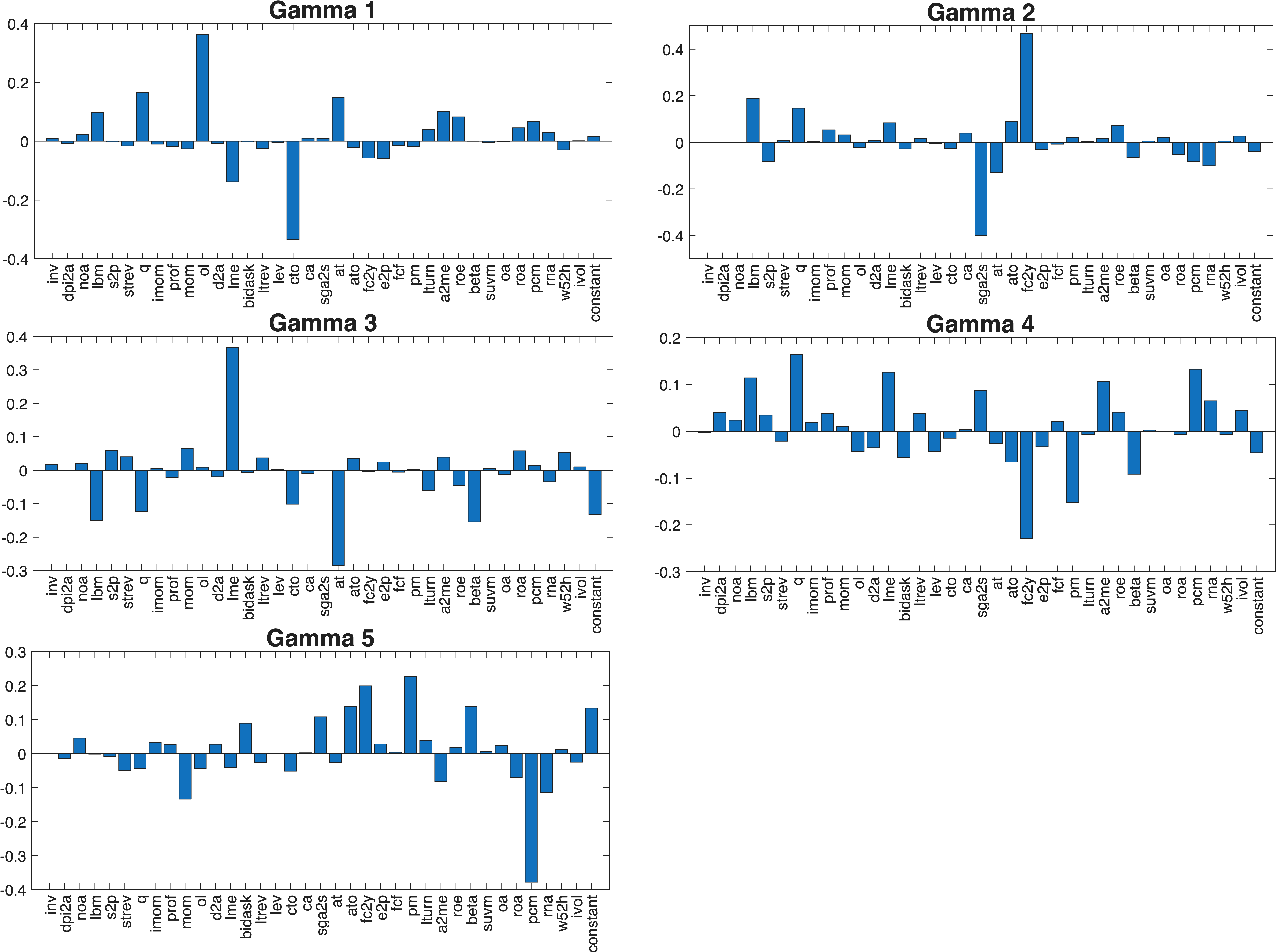}
	\caption{$\Gamma$ estimates when $K = 5$.}
	\label{fig:Gamma_est}
\end{figure}

Figure~\ref{fig:Gamma_est} visualizes the estimated $\widehat{\Gamma}$ when $K = 5$.  
The first factor loads primarily on operating leverage and capital turnover, while the second is driven by cost ratios (SG\&A-to-sales and fixed costs-to-sales), which together form a “cost” factor.  
The third factor contrasts market capitalization and book assets, resembling a value-like factor similar to the HML component in \citet{fama1993common} and \citet{kelly2019characteristics}.  
Later factors are less interpretable, reflecting more diffuse combinations of firm attributes.

\subsection{Summary and Discussion}
Taken together, our empirical findings confirm three key messages.  
First, both inside and outside alphas are statistically significant, highlighting that mispricing has distinct structured and idiosyncratic components.  
Second, inside alphas exhibit clear industry-level co-movement tied to fundamentals, while outside alphas capture transitory, firm-specific deviations.  
Third, characteristic-based factor loadings reveal economically interpretable dimensions of risk, including value, cost, and size components.  

These results validate the inferential theory developed in Section~\ref{sec:inferential_theory} and underscore the usefulness of our decomposition for understanding how firm fundamentals, latent factors, and residual mispricing jointly shape the cross-section of asset returns.

\section{Concluding Remarks}\label{sec:conclusion}

This paper develops a unified econometric framework for modeling and inferring pricing errors in factor models that combine latent factors with firm characteristics. 
Our approach decomposes mispricing into two orthogonal components---\emph{inside alpha}, which is systematically related to firm fundamentals but orthogonal to factor loadings, and \emph{outside alpha}, which is orthogonal to both factors and characteristics. 
This decomposition reconciles the statistical efficiency of latent-factor approaches with the economic interpretability of characteristic-based models, thereby providing a coherent foundation for studying both systematic and idiosyncratic sources of mispricing.

Methodologically, we contribute a new class of low-rank estimators equipped with explicit debiasing and valid inferential theory. 
The resulting estimators admit closed-form expressions and Gaussian asymptotics even when the number of characteristics grows with the sample size, relaxing the restrictive conditions typically imposed in earlier work such as \citet{kelly2019characteristics} and \citet{zhang2024testing}. 
Our theoretical results establish the asymptotic normality of characteristic loadings, inside alphas, and outside alphas, allowing standard hypothesis tests on both factor structure and pricing errors. 
These inferential tools make it possible to distinguish between characteristic-driven and residual components of mispricing in a statistically rigorous way.

Empirically, applying the framework to U.S.\ equities from 2000--2019 reveals several new insights. 
Both inside and outside alphas are statistically significant, but they exhibit distinct economic patterns. 
Inside alphas display pronounced industry-level co-movement that aligns with persistent fundamentals such as technological change and sectoral shocks, while outside alphas behave as transient, firm-specific deviations that likely reflect liquidity frictions, behavioral biases, or short-term constraints. 
In addition, characteristic-based factor loadings highlight the importance of value, cost, and size dimensions in shaping cross-sectional risk exposures. 
Taken together, these results demonstrate that pricing errors in equity markets are structured, multi-layered phenomena rather than purely idiosyncratic residuals.

More broadly, our analysis bridges the gap between statistical and economic perspectives on asset pricing. 
By explicitly connecting latent factors to firm characteristics and by distinguishing between systematic and residual mispricing, the framework opens new avenues for understanding the sources and persistence of return anomalies. 
Future research could extend this setting to dynamic environments with time-varying characteristics, international markets, or alternative asset classes, as well as explore the interaction between inside and outside alphas in explaining cross-sectional risk premia. 
We hope that the theoretical tools and empirical evidence developed here will serve as a foundation for future studies at the intersection of econometrics, machine learning, and financial economics.

\bibliographystyle{apalike}
\bibliography{Instrumented_PCA}

@article{bryzgalova2019forest,
	title={Forest through the trees: Building cross-sections of stock returns},
	author={Bryzgalova, Svetlana and Pelger, Markus and Zhu, Jason},
	journal={The Journal of Finance},
	year={2019},
	publisher={Wiley Online Library}
}

@article{feng2024deep,
	title={Deep learning in characteristics-sorted factor models},
	author={Feng, Guanhao and He, Jingyu and Polson, Nicholas G and Xu, Jianeng},
	journal={Journal of Financial and Quantitative Analysis},
	volume={59},
	number={7},
	pages={3001--3036},
	year={2024},
	publisher={Cambridge University Press}
}

@article{hou2020replicating,
	title={Replicating Anomalies},
	author={Hou, Kewei and Xue, Chen and Zhang, Lu},
	journal={The Review of Financial Studies},
	volume={33},
	number={5},
	pages={2019--2133},
	year={2020},
	publisher={JSTOR}
}

@article{harvey2016and,
	title={… and the cross-section of expected returns},
	author={Harvey, Campbell R and Liu, Yan and Zhu, Heqing},
	journal={The Review of Financial Studies},
	volume={29},
	number={1},
	pages={5--68},
	year={2016},
	publisher={Oxford University Press}
}

@article{choi2024high,
	title={High dimensional factor analysis with weak factors},
	author={Choi, Jungjun and Yuan, Ming},
	journal={arXiv preprint arXiv:2402.05789},
	year={2024}
}

@article{benjamini2001control,
	title={The control of the false discovery rate in multiple testing under dependency},
	author={Benjamini, Yoav and Yekutieli, Daniel},
	journal={Annals of statistics},
	pages={1165--1188},
	year={2001},
	publisher={JSTOR}
}

@article{fama1993common,
	title={Common risk factors in the returns on stocks and bonds},
	author={Fama, Eugene F and French, Kenneth R},
	journal={Journal of financial economics},
	volume={33},
	number={1},
	pages={3--56},
	year={1993},
	publisher={Elsevier}
}

@article{bai2003inferential,
  title={Inferential theory for factor models of large dimensions},
  author={Bai, Jushan},
  journal={Econometrica},
  volume={71},
  number={1},
  pages={135--171},
  year={2003},
  publisher={Wiley Online Library}
}

@article{fan2016projected,
  title={Projected principal component analysis in factor models},
  author={Fan, Jianqing and Liao, Yuan and Wang, Weichen},
  journal={Annals of statistics},
  volume={44},
  number={1},
  pages={219},
  year={2016},
  publisher={NIH Public Access}
}

@article{zhang2024testing,
	title={Testing Pricing Errors of Models with Latent Factors and Firm Characteristics as Covariances},
	author={Zhang, Chu},
	journal={Management Science},
	volume={70},
	number={3},
	pages={1706--1728},
	year={2024},
	publisher={INFORMS}
}

@article{kelly2019characteristics,
  title={Characteristics are covariances: A unified model of risk and return},
  author={Kelly, Bryan T and Pruitt, Seth and Su, Yinan},
  journal={Journal of Financial Economics},
  volume={134},
  number={3},
  pages={501--524},
  year={2019},
  publisher={Elsevier}
}

@techreport{chen2023semiparametric,
  title={Semiparametric conditional factor models: Estimation and inference},
  author={Chen, Qihui and Roussanov, Nikolai and Wang, Xiaoliang},
  year={2023},
  institution={National Bureau of Economic Research}
}

@article{chernozhukov2023inference,
  title={Inference for low-rank models},
  author={Chernozhukov, Victor and Hansen, Christian and Liao, Yuan and Zhu, Yinchu},
  journal={The Annals of statistics},
  volume={51},
  number={3},
  pages={1309--1330},
  year={2023},
  publisher={Institute of Mathematical Statistics}
}

@book{vershynin2018high,
  title={High-dimensional probability: An introduction with applications in data science},
  author={Vershynin, Roman},
  volume={47},
  year={2018},
  publisher={Cambridge university press}
}

@article{kim2021arbitrage,
  title={Arbitrage portfolios},
  author={Kim, Soohun and Korajczyk, Robert A and Neuhierl, Andreas},
  journal={The Review of Financial Studies},
  volume={34},
  number={6},
  pages={2813--2856},
  year={2021},
  publisher={Oxford University Press}
}

@article{belloni2018high,
  title={High-dimensional econometrics and regularized GMM},
  author={Belloni, Alexandre and Chernozhukov, Victor and Chetverikov, Denis and Hansen, Christian and Kato, Kengo},
  journal={arXiv preprint arXiv:1806.01888},
  year={2018}
}

@article{sharpe1964capital,
  title={Capital asset prices: A theory of market equilibrium under conditions of risk},
  author={Sharpe, William F},
  journal={The journal of finance},
  volume={19},
  number={3},
  pages={425--442},
  year={1964},
  publisher={Wiley Online Library}
}

@article{ross1976arbitrage,
  title={The arbitrage theory of capital asset pricing},
  author={Ross, Stephen A},
  journal={Journal of Economic Theory},
  volume={13},
  number={3},
  pages={341--360},
  year={1976},
  publisher={Elsevier BV}
}

@article{fama1973risk,
  title={Risk, return, and equilibrium: Empirical tests},
  author={Fama, Eugene F and MacBeth, James D},
  journal={Journal of political economy},
  volume={81},
  number={3},
  pages={607--636},
  year={1973},
  publisher={The University of Chicago Press}
}

@article{connor1986performance,
  title={Performance measurement with the arbitrage pricing theory: A new framework for analysis},
  author={Connor, Gregory and Korajczyk, Robert A},
  journal={Journal of financial economics},
  volume={15},
  number={3},
  pages={373--394},
  year={1986},
  publisher={Elsevier}
}

@article{connor1988risk,
	title={Risk and return in an equilibrium APT: Application of a new test methodology},
	author={Connor, Gregory and Korajczyk, Robert A},
	journal={Journal of financial economics},
	volume={21},
	number={2},
	pages={255--289},
	year={1988},
	publisher={Elsevier}
}

@misc{chamberlain1982arbitrage,
  title={Arbitrage, factor structure, and mean-variance analysis on large asset markets},
  author={Chamberlain, Gary and Rothschild, Michael},
  year={1982},
  publisher={National Bureau of Economic Research Cambridge, Mass., USA}
}

@article{fan2022structural,
  title={Structural deep learning in conditional asset pricing},
  author={Fan, Jianqing and Ke, Zheng Tracy and Liao, Yuan and Neuhierl, Andreas},
  journal={Available at SSRN 4117882},
  year={2022}
}

@article{gu2021autoencoder,
  title={Autoencoder asset pricing models},
  author={Gu, Shihao and Kelly, Bryan and Xiu, Dacheng},
  journal={Journal of Econometrics},
  volume={222},
  number={1},
  pages={429--450},
  year={2021},
  publisher={Elsevier}
}

@article{chernozhuokov2022improved,
  title={Improved central limit theorem and bootstrap approximations in high dimensions},
  author={Chernozhuokov, Victor and Chetverikov, Denis and Kato, Kengo and Koike, Yuta},
  journal={The Annals of Statistics},
  volume={50},
  number={5},
  pages={2562--2586},
  year={2022},
  publisher={Institute of Mathematical Statistics}
}

@article{newey1987simple,
  title={A Simple, Positive Semi-Definite, Heteroskedasticity and Autocorrelation Consistent Covariance Matrix},
  author={Newey, Whitney K and West, Kenneth D},
  journal={Econometrica},
  pages={703--708},
  year={1987}
}

@article{hansen2007asymptotic,
  title={Asymptotic properties of a robust variance matrix estimator for panel data when T is large},
  author={Hansen, Christian B},
  journal={Journal of Econometrics},
  volume={141},
  number={2},
  pages={597--620},
  year={2007},
  publisher={Elsevier}
}

@article{bai2020standard,
  title={Standard errors for panel data models with unknown clusters},
  author={Bai, Jushan and Choi, Sung Hoon and Liao, Yuan},
  journal={Journal of Econometrics},
  pages={105004},
  year={2020},
  publisher={Elsevier}
}

%%%%%%%%%%%%%%%%%%%%%%%%%%%%%%%%%%%%%%%%%%%%%%%%%%%%

\newpage

%\renewcommand{\theHpart}{A\arabic{part}}
%\part{} % Start the document part

\appendix

{\LARGE 
\begin{center}
    APPENDIX
\end{center}
}

%\setcounter{table}{0}
%\renewcommand{\thetable}{A\arabic{table}}
%\renewcommand{\thefigure}{A\arabic{figure}}    
%\setcounter{figure}{0}  
%\setcounter{equation}{0} 
%\renewcommand{\theequation}{A\arabic{equation}}

%\addcontentsline{toc}{section}{Appendix} % Add the appendix text to the document TOC
%\part{Appendix} % Start the appendix part
%\parttoc % Insert the appendix TOC
%\newpage 

\section{Table for characteristics} \label{appendix:characteristics}

\begin{table}[h]
\scriptsize
  \centering
  \caption{Firm Characteristics}
    \begin{tabular}{c|c}
    \hline
    \hline
    Symbol & Description \\
    \hline
    INV   & Investment, percentage year-on-year growth rate of total assets. \\
    DPI2A & Changes in PPE and inventory scaled by lagged AT. \\
    NOA   & Net operating assets: operating assets minus operating liabilities, scaled by lagged AT. \\
    LBM   & Log book-to-market ratio of equity. \\
    S2P   & Sales-to-price ratio: net sales, scaled by market equity. \\
    STREV & Short-term reversal: one-month return. \\
    Q     & Tobin’s Q: market equity plus book debt, divided by AT. \\
    IMOM  & Intermediate momentum: cumulative return from month -12 to month -7. \\
    PROF  & Profitability: gross profitability scaled by book equity. \\
    MOM   & Momentum: cumulative return from month -12 to month -2. \\
    OL    & Operating leverage: sum of cost of goods sold and SG\&A expenses scaled by AT. \\
    D2A   & Depreciation and amortization scaled by AT. \\
    LME   & Log market capitalization: log of market equity. \\
    BIDASK & Bid-ask spread: average daily bid-ask spread in the month. \\
    LTREV & Long-term reversal: cumulative return from month -36 to month -13. \\
    LEV   & Leverage: (LTD+DCL)/(LTD+CL+BE) with LTD: long-term debt, DCL: debt in current liabilities,\\
    &and BE: book equity. \\
    CTO   & Capital turnover: ratio of net sales to lagged total assets. \\
    CA    & Cash and short-term investment, scaled by total assets. \\
    SGA2S & SG\&A-to-sales: selling, general and administrative expenses scaled by net sales. \\
    AT    & Total assets. \\
    ATO   & Net sales scaled by lagged, unscaled net operating assets. \\
    FC2Y  & Fixed costs-to-sales: SG\&A plus advertising expenses and R\&D expenses, divided by net sales. \\
    E2P   & Earnings-to-price ratio: income before extraordinary items divided by lagged market equity. \\
    FCF   & Net income, depreciation, and amortization less change in working capital and capital expenditure,\\
    &scaled by book equity. \\
    PM    & Profit margin: operating income after depreciation scaled by sales. \\
    LTURN & Turnover: dollar trading volume over market equity. \\
    A2ME  & Assets-to-market cap: total assets over lagged market equity. \\
    ROE   & Income before extraordinary items scaled by lagged book equity. \\
    BETA  & Market beta: market beta estimated with past one year’s daily data. \\
    SUV\_m & Standardized unexplained volume: Standardized residual from regressing trading volume on\\
    & absolute values of positive and negative returns. \\
    OA    & Operating accruals: changes in noncash working capital minus depreciation, scaled by lagged AT. \\
    ROA   & Return on assets: income before extraordinary items divided by lagged AT. \\
    PCM   & Price-to-cost margin: net sales minus costs of goods sold, scaled by net sales. \\
    RNA   & Return on net operating assets: operating income after depreciation scaled by lagged, unscaled net\\
    &operating assets. \\
    W52H  & Stock price relative to its 52-week high price. \\
    IVOL  & Idiosyncratic volatility with respect to the Fama–French three-factor model. \\
    \hline
    \end{tabular}%
  \label{tab:character}%
\end{table}%

\section{Simulated experiment}

To demonstrate the finite sample performance of our methodology and the validity of our inferential theory, we conducted a simulation experiment.

\subsection{Finite sample performance of inferential theory}

We calibrate the simulated data to our model estimated from US monthly stock returns in the empirical study and set $N=973$, $T=240$, $L=37$, and $K=5$. For the parameters $\Gamma$ and $\eta$, we use the estimated values from the model. For the characteristics, we first estimate the $(L-1) \times (L-1)$ covariance matrix $\Sigma_x$ from the characteristics data excluding the constant term, generate $x_{it}$ from $\calN (0,\Sigma_x)$, and include the constant. To generate the sparse $\xi_t$, we first randomly choose 71 periods and for each $t$ in the chosen periods, we draw 3 values from $uniform[\xi_{center}-0.5,\xi_{center}+0.5] $ where $\xi_{center}$ is the average of maximum and minimum of absolute value of nonzero estimated $\xi_{t,q}$s from the stock returns data. Then, we assign these values to 3 randomly chosen elements in $\xi_t$ and set other elements to be zero. For the remaining (240 - 71) periods, we set $\xi_t$ to be zero. Lastly, we set a sign of element randomly. Here, $71$ is the number of periods where $\hat{\xi}_t \neq \mathbf{0}$ in the empirical study and $3$ is the average of the number of nonzero $\hat{\xi}_{t,q}$ over the periods where $\hat{\xi}_t \neq \mathbf{0}$. In addition, for $\zeta$, we use the estimated $\zeta$ from the empirical study. In addition, for each $1\leq k \leq 5$, we generate $\breve{f}_{t+1,k}$ from a normal distribution whose mean and variance are the estimated values from the data. Lastly, we generate $\epsilon_{i,t+1}$ from $\calN(0,\sigma^2)$ where $\sigma$ is estimated from the data. The number of simulations is set to 1,000.

First, to study the advantage of debiased estimators for $\Gamma$ and $\alpha_I$, we compare the histograms and kernel density estimates of the t-statistics of the plain $\Gamma$ estimator and the debiased $\Gamma$ estimator. For the inference of $\Gamma$, we report the results of $\gamma_{1,1}$. For the inference of $\alpha_{I,it}$, we report the results of a randomly chosen $i$ at the last period $T$.

\begin{figure}[htb]
\centering
\includegraphics[width = \textwidth]{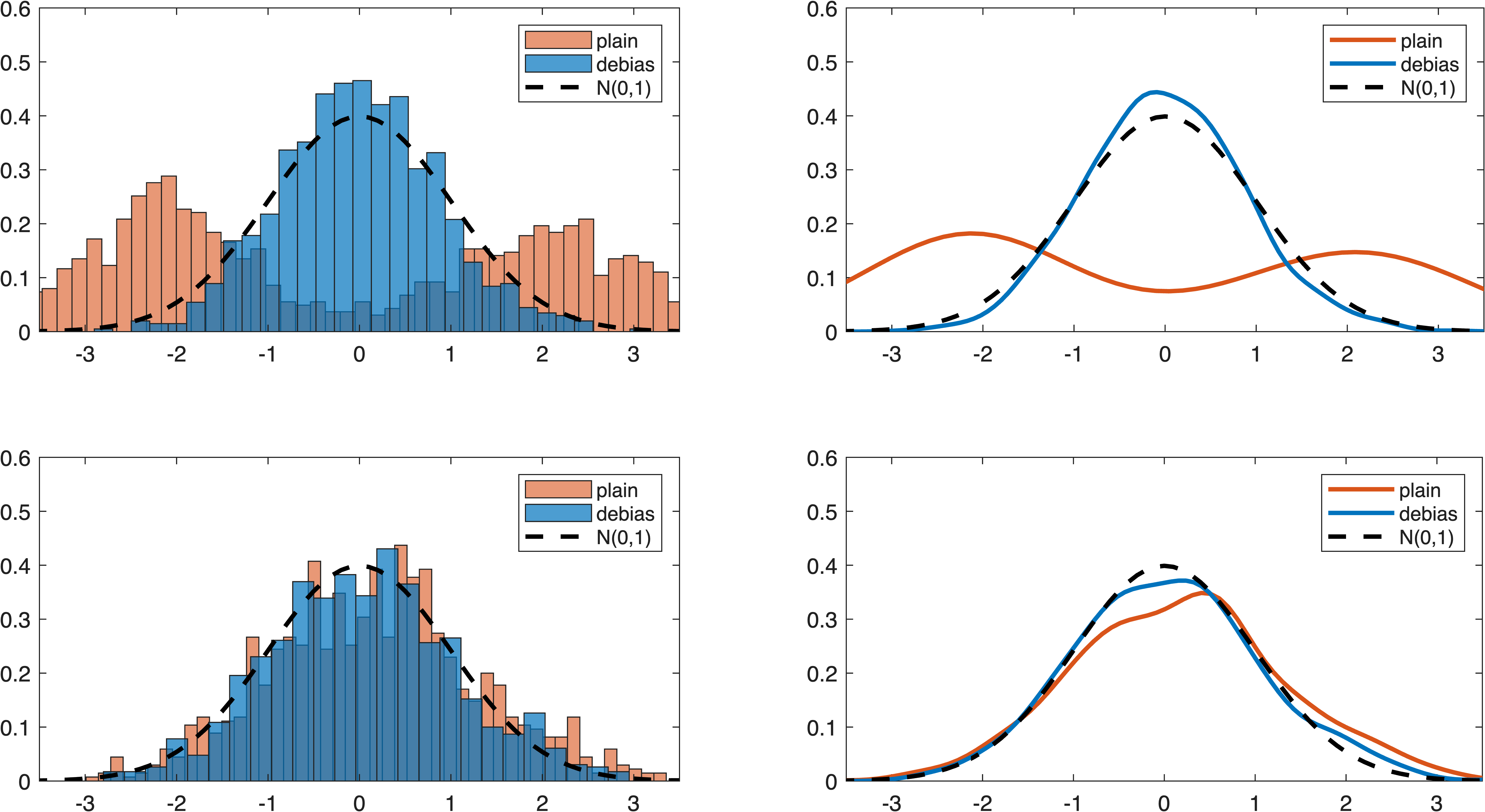}
\caption{Histograms for standardized estimates}\label{fig:histrogram_het_gamma_alphaI}
\end{figure}

Figure \ref{fig:histrogram_het_gamma_alphaI} shows the histograms and kernel density estimates of the t-statistics. The upper panel shows the results for the $\Gamma$ estimates and the the lower panel shows those for the $\alpha_I$ estimates. In the case of $\Gamma$ estimates, the plain estimator has a fairly large bias. Here, the sign of bias depends on that of rotation in each iteration. In addition, we can also check that the plain estimator of $\alpha_I$ is more biased compared to the debiased estimator. Besides, Table \ref{tab:coverage_prob_bias} shows the coverage probabilities of the (asymptotic) confidence intervals. From the table, we can confirm that the coverage probabilities of the debiased estimators are closer to the target probabilities.

\begin{table}[htb]
\caption{Coverage probability of the confidence interval.}
	\centering
	 \begin{tabular}{c|ccc}
    \hline
    \hline
    Target prob. & 90\%  & 95\%  & 99\% \\[0.2em]
    \hline
    $\tilde{\alpha}_{I,it}$ & 83.3\% & 89.9\% & 96.9\% \\[0.2em]
    $\hat{\alpha}_{I,it}$  & 88.0\% & 94.1\% & 98.7\% \\[0.2em]
    $\tilde{\gamma}_{1,1}$ & 30.7\% & 42.7\% & 71.2\%  \\[0.2em]
    $\hat{\gamma}_{1,1}$ & 94.2\% & 97.3\% & 99.8\%  \\[0.2em]
    \hline
    \end{tabular}%
	\label{tab:coverage_prob_bias}%
\end{table}%

Next, we present the coverage probabilities of the confidence interval and the histograms for the standardized estimates (t-statistics) for $\alpha_O$. For the inference of $\alpha_{O,it}$, we report the results of a randomly chosen $i$ at the last period $T$. In the simulations, we always make $T$ to be in the 71 chosen periods so that $\xi_T \neq \mathbf{0}$. Here, for the truncation level, we set $\rho = 1.5 \times \sigma \frac{\sqrt{\log NT}}{\sqrt{N}}$. For the inference of $\delta_{o,t,q}$, we report the results of a randomly chosen $q$ at the last period $T$.

\begin{figure}[htb]
\centering
\includegraphics[width = 0.9\textwidth]{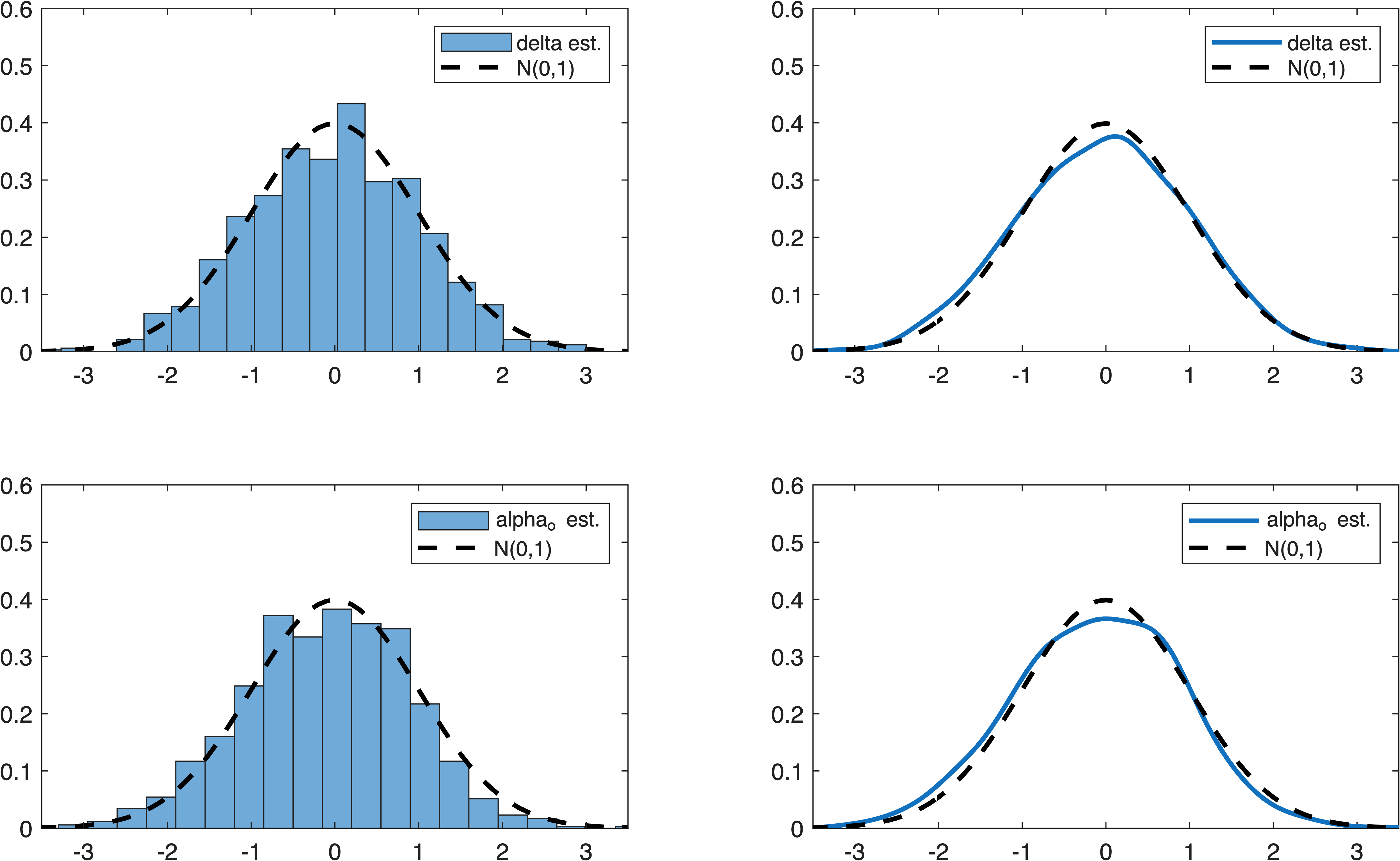}
\caption{Histograms for standardized estimates}\label{fig:histogram_het_alpha_o_final}
\end{figure}

Figure \ref{fig:histogram_het_alpha_o_final} shows the histograms of the standardized estimates (t-statistics) along with the standard normal distribution and Table \ref{tab:coverage_prob_het} shows the coverage probabilities of the (asymptotic) confidence intervals. From the figure and table, we can confirm the asymptotic normality of our estimates.

\begin{table}[htb]
\caption{Coverage probability of the confidence interval}
	\centering
	 \begin{tabular}{c|ccc|ccc}
    \hline
    \hline
    Parameter & \multicolumn{3}{c|}{$\delta_{o,t,q}$} & \multicolumn{3}{c}{$\alpha_{O,it}$} \\[0.2em]
    \hline
    Target prob. & 90\%  & 95\%  & 99\%  & 90\%  & 95\%  & 99\% \\[0.2em]
    \hline
    Coverage prob. & 89.3\% & 95.2\% & 99.3\% & 89.8\% & 95.1\% & 99.1\% \\[0.2em]
    \hline
    \end{tabular}%
	\label{tab:coverage_prob_het}%
\end{table}%

\subsection{Power comparison with other methods}

In this section, to show the relative advantage of our inference method, we compare the power of our alpha test with that of other methods in the case where the true model is close to the null hypothesis. Specifically, we consider the following model:
\begin{gather*}
R_{t+1} = \alpha_{O,t}  +   B_t f_{t+1}  + E_{t+1}, \qquad t= 1, \dots, T,
\end{gather*}
where $B_t = X_t \Gamma$, $\alpha_{O,t} = B^{o}_{t} \delta_o$ and test the null hypothesis that there is no outside pricing error, $\alpha_{O,t}$. Here, the inside pricing error $\alpha_{I,t}$ is set to zero and $\delta_o$ is time invariant, so that the model can belong to both the model of \cite{zhang2024testing} and that of this paper. In addition, as in \cite{zhang2024testing}, we define the basis $B^{o}_{t}$ such that
\begin{gather*}
 B^{o}_{t} =  X^{o}_{t} (X^{o \top}_{t} X^{o}_{t}/N)^{-1/2}, \quad X^{o}_{t} = \left[I_N - P_{X,t} \right] 
\begin{pmatrix}
I_{N-L} \\
\boldsymbol{0}_{L \times (N-L)}
\end{pmatrix}	
.
\end{gather*}
For the characteristics, we set $L = 10$ and generate $x_{it}$ from the standard normal distribution and include the constant. For the factors, we set $K = 2$ and for each $t$, generate factors from $\calN ( 0, \diag(2,1)^2)$. In addition, we generate each element of $\Gamma$ from $\calN ( 0, 1/L)$ and fix it for all iterations. Noises are generated from a standard normal distribution. On top of that, to generate the case where the true model is close to the null hypothesis, but, different from it, we set $\delta_1 = 0.01 \sim 0.06$ and $\delta_q = 0$ for all $2 \leq q \leq N-L$. 

\begin{figure}[h]
\centering
\includegraphics[width = \textwidth]{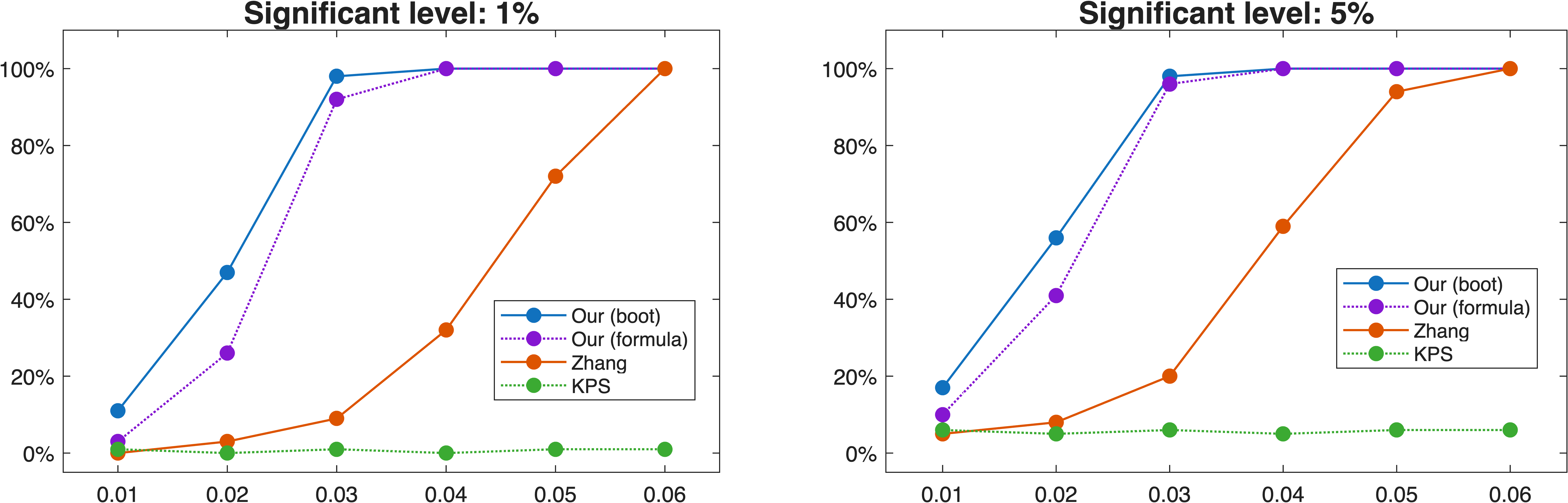}
\caption{Rejection probabilities with diverse $\delta_1$}\label{fig:power_graph}
\end{figure}

Figure \ref{fig:power_graph} shows the rejection probabilities of several tests when $N=T=200$. Here, `Our (formula)' uses the test statistics $\max_{i \leq N ,t \leq T} \abs{\hat{V}^{-1/2}_{o,it} \hat{\alpha}_{O,it}}$ and for the critical values, utilizes the relation
$$
\bbP \left( \max_{i \leq N ,t \leq T} \abs{\hat{V}^{-1/2}_{o,it} \hat{\alpha}_{O,it} }  >  \Phi^{-1} (1 - a /(2NT) ) \right) \leq a + o(1).
$$
under the null. On the other hand, `Ours (boot)' utilizes the bootstrap method in \cite{belloni2018high} and \cite{chernozhuokov2022improved} to derive the distribution of the test statistics and the corresponding critical values, while it uses the same test statistics, $\max_{i \leq N ,t \leq T} \abs{\hat{V}^{-1/2}_{o,it} \hat{\alpha}_{O,it}}$. In addition, `Zhang' denotes the method in \cite{zhang2024testing} using the test statistics $||\hat{\delta}_o||$ with the bootstrap method where $\hat{\delta}_o$ is the estimator from the iterative numerical algorithm. Lastly, `KPS' denotes the alpha test method introduced in \cite{kelly2019characteristics}. Here, the number of simulations is 100 and the number of bootstraps is set to 500.

We can check that our methods have more power than other methods. In the case of the method in \cite{kelly2019characteristics}, it cannot detect the existence of alpha because their model only considers the inside pricing error. In addition, when it comes to the inference method in \cite{zhang2024testing}, it seems to be weak in the tough case where the true model is close to the null. Among our methods, the bootstrap method is slightly better than the method based on the formula using the inverse cumulative distribution function.

Moreover, Tables \ref{tab:power99} and \ref{tab:power95} show the rejection probabilities with other sample sizes when the level is 99\% and 95\%, respectively. We can find similar patterns to Figure \ref{fig:power_graph}. Our methods show the highest rejection probabilities and have better power than others. On the other hand, the method in \cite{kelly2019characteristics} show the lowest rejection probabilities.

\begin{table}[H]
\small
  \centering
  \caption{Rejection probabilities with level 99\%}
    \begin{tabular}{ccccccc}
    \hline
    \hline
    \multirow{2}[2]{*}{Sample Size (N,T)} & \multirow{2}[2]{*}{Inference method} & \multicolumn{5}{c}{$\delta_1$} \\
          &       & 0.01  & 0.02  & 0.03  & 0.04  & 0.05 \\
    \hline
    \multirow{4}[2]{*}{(500,200)} & Ours (boot) &   18\%    & 96\%  & 100\% & 100\% & 100\% \\[0.2em]
          & Ous (formula) &   4\%    & 82\%  & 100\% & 100\% & 100\% \\[0.2em]
          & Zhang &    0\%   & 0\%   & 12\%  & 56\%  & 97\% \\[0.2em]
          & KPS   &    0\%   & 0\%   & 0\%   & 0\%   & 0\% \\[0.2em]
    \hline
    \multirow{4}[2]{*}{(300,300)} & Ours (boot) & 7\%  & 96\%  & 100\% & 100\% & 100\% \\[0.2em]
          & Ous (formula) & 0\%   & 86\%  & 100\% & 100\% & 100\% \\[0.2em]
          & Zhang & 0\%   & 0\%   & 26\%  & 90\%  & 100\% \\[0.2em]
          & KPS   & 0\%   & 0\%   & 0\%   & 0\%   & 1\% \\[0.2em]
    \hline
    \multirow{4}[2]{*}{(200,500)} & Ours (boot) & 32\%  & 98\% & 100\% & 100\% & 100\% \\[0.2em]
          & Ous (formula) & 15\%   & 94\%  & 100\% & 100\% & 100\% \\[0.2em]
          & Zhang & 0\%   & 10\%   & 73\%  & 100\% & 100\% \\[0.2em]
          & KPS   & 2\%   & 1\%   & 0\%   & 0\%   & 1\% \\[0.2em]
    \hline
    \end{tabular}%
  \label{tab:power99}%
\end{table}%

% Table generated by Excel2LaTeX from sheet 'Table (95%)'
\begin{table}[H]
\small
  \centering
  \caption{Rejection probabilities with level 95\%}
    \begin{tabular}{ccccccc}
    \hline
    \hline
    \multirow{2}[2]{*}{Sample Size (N,T)} & \multirow{2}[2]{*}{Inference method} & \multicolumn{5}{c}{$\delta_1$} \\
          &       & 0.01  & 0.02  & 0.03  & 0.04  & 0.05 \\
    \hline
    \multirow{4}[2]{*}{(500,200)} & Ours (boot) &   22\%    & 100\% & 100\% & 100\% & 100\% \\[0.2em]
          & Ours (formula) &   8\%    & 93\%  & 100\% & 100\% & 100\% \\[0.2em]
          & Zhang &   0\%    & 8\%   & 37\%  & 86\%  & 100\% \\[0.2em]
          & KPS   &  6\%     & 8\%   & 8\%   & 10\%   & 10\% \\[0.2em]
    \hline
    \multirow{4}[2]{*}{(300,300)} & Ours (boot) & 15\%  & 100\%  & 100\% & 100\% & 100\% \\[0.2em]
          & Ours (formula) & 3\%   & 96\%  & 100\% & 100\% & 100\% \\[0.2em]
          & Zhang & 0\%   & 12\%  & 59\%  & 100\% & 100\% \\[0.2em]
          & KPS   & 3\%   & 4\%   & 3\%   & 4\%  & 2\% \\[0.2em]
    \hline
    \multirow{4}[2]{*}{(200,500)} & Ours (boot) & 41\%  & 99\% & 100\% & 100\% & 100\% \\[0.2em]
          & Ours (formula) & 19\%  & 94\%  & 100\% & 100\% & 100\% \\[0.2em]
          & Zhang & 0\%   & 27\%  & 97\%  & 100\% & 100\% \\[0.2em]
          & KPS   & 6\%   & 3\%   & 5\%   & 6\%   & 4\% \\[0.2em]
    \hline
    \end{tabular}%
  \label{tab:power95}%
\end{table}%

\section{Variance estimators}\label{sec:variance_est}

Here, we present the variance estimators we used in the empirical study and simulated experiment. We estimate the variances in our inferential theory under the assumption that the noises are independent across $i$ and $t$, and there is heteroskedasticity across $t$. We may also be able to consider more robust estimators as in \cite{newey1987simple,hansen2007asymptotic,bai2020standard}.

Let $\hat{\sigma}^2_{t+1} = \frac{1}{N} \sum_{i=1}^N \hat{\epsilon}_{i,t+1}^2$, where $\hat{m}_{i,t+1}= \hat{\alpha}_{O,it} + \hat{\alpha}_{I,it} + x_{it}^\top \hat{\Gamma} \hat{f}_{t+1}$, and $\hat{\epsilon}_{i,t+1} = r_{i,t+1} - \hat{m}_{i,t+1}$. Then, we define
\begin{align*}
&\widehat{(\boldsymbol{H}^\top \Sigma_f^{-1} \Sigma_{xf,l}   \Sigma_f^{-1}\boldsymbol{H})} = \left( \frac{1}{T}\sum_t \hat{f}_{t+1}^d \hat{f}_{t+1}^{d \top}\right)^{-1} \frac{1}{T} \sum_{t=1}^T \hat{\sigma}_{t+1}^2 [Q_t^{-1}]_{ll} \hat{f}_{t+1}^{d} \hat{f}_{t+1}^{d \top} \left( \frac{1}{T}\sum_t \hat{f}_{t+1}^d \hat{f}_{t+1}^{d \top}\right)^{-1},\\
& \hat{\sigma}_{I,it}^2 = \frac{1}{TL} \sum_{s=1}^T \hat{\sigma}_{s+1}^2 (\hat{a}_s \hat{A} - \hat{b}_s \hat{B}) Q_s (\hat{A}^\top \hat{a}_s - \hat{B}^\top \hat{b}_s),\\
&\hat{a}_s = 1 - (\hat{\eta}^\top Q_t \hat{\Gamma} (\hat{\Gamma}^\top Q_t \hat{\Gamma} )^{-1} + (\hat{\bar{\breve{f}}})^\top ) \left( \frac{1}{T} \sum_{u=1}^T \hat{f}_{u+1}^d \hat{f}_{u+1}^{d \top} \right)^{-1} \hat{f}_{s+1}^d, \ \       \hat{\bar{\breve{f}}} = (\hat{\Gamma}^\top   \hat{\Gamma})^{-1} \hat{\Gamma}^\top \bar{\ddot{R}}, \\
&\hat{b}_s = x_{it}^\top \hat{\Gamma} (\hat{\Gamma}^\top Q_t \hat{\Gamma})^{-1} \left( \frac{1}{T} \sum_{u=1}^T \hat{f}_{u+1}^d \hat{f}_{u+1}^{d \top} \right)^{-1} \hat{f}_{s+1}^d, \\
&\hat{A} = x_{it}^\top Q_t^{-1} - x_{it}^\top \hat{\Gamma} (\hat{\Gamma}^\top Q_t \hat{\Gamma} )^{-1} \hat{\Gamma}^\top , \quad \hat{B} = \hat{\eta}^\top - \hat{\eta}^\top Q_t \hat{\Gamma} (\hat{\Gamma}^\top Q_t \hat{\Gamma} )^{-1} \hat{\Gamma}^\top,\\
& \hat{\bar{\sigma}}^2 = \frac{1}{T} \sum_{s=1}^T \hat{\sigma}^2_{s+1}, \quad \hat{\sigma}_{\delta,qt}^2 = \hat{\sigma}^2_{t+1}.
\end{align*}

Then, the following theorem shows the consistency of the variance estimators.

\begin{theorem}\label{thm:variance_estimation}
Assume that the noises are independent across $i$ and $t$, and there is heteroskedasticity across $t$. Then, we have\\ (i) $\norm{\widehat{(\boldsymbol{H}^\top \Sigma_f^{-1} \Sigma_{xf,l}   \Sigma_f^{-1}\boldsymbol{H})} - \boldsymbol{H}^\top \Sigma_f^{-1} \Sigma_{xf,l}   \Sigma_f^{-1}\boldsymbol{H}} = o_p(1)$; (ii) $\abs{\hat{\sigma}_{I,it}^2 - \sigma_{I,it}^2} = o_p(1)$; (iii) $\abs{\hat{\bar{\sigma}}^2 - \bar{\sigma}^2} = o_p(1)$; 
(iv) $\abs{\hat{\sigma}_{\delta,qt}^2 - \sigma_{\delta,qt}^2} = o_p(1)$ and $\abs{\hat{\sigma}_{t+1}^2 - \sigma_{t+1}^2} = o_p(1)$.
\end{theorem}

Here, we can use the variance estimators using `tilde ( $\widetilde{}$ )' estimators instead of the current estimators using `hat ( $\widehat{}$ )' estimators and will get the same result.

\section{Dependence on units of characteristics in \cite{zhang2024testing}}\label{sec:units}

\cite{zhang2024testing} uses the following transformed characteristics instead of $X_t$:
$$
X_t^\star =  X_t + \mathbf{1}_N \psi^\top  ,
$$
where $\psi$ is some $L \times 1 $ vector and $\mathbf{1}_N$ is the $N \times 1 $ vector of ones. In addition, the systematic risk has the following form:
$$
B_t f_{t+1} = X_t^\star \Gamma f_{t+1} = (X_t + \mathbf{1}_N \psi^\top) \Gamma f_{t+1} 
$$
and the inside pricing error has the following form:
\begin{gather*}
\alpha_{I,t} = B_t^I \delta_I ,\qquad \text{where } B_t^I = S^{\star} (S^{\star \top} S^{\star})^{-1/2},\\
   S^\star = \left[ X_t^\star \left(X_t^{\star \top} X_t^\star \right)^{-1} X_t^{\star \top}  - X_t^\star \Gamma \left(\Gamma^{\top} X_t^{\star \top} X_t^\star \Gamma \right)^{-1} \Gamma^\top X_t^{\star \top} \right] \begin{pmatrix}
I_{L-K} \\
O_{(N-L+K) \times (L-K)} 
\end{pmatrix}
.
\end{gather*}

Let $W_d$ be the $L \times L$ diagonal matrix representing the change in units of characteristics. Consider the case where we use characteristics with different units $\dot{X}_t = X_t W_d$ instead of $X_t$, and put `dot' ( $\dot{}$ ) on the parameters of this case. Then, to preserve the systematic risk and the inside pricing error, we need to have
\begin{gather}\label{eq:preserve}
\dot{\Gamma} = W_d^{-1} \Gamma  ,  \quad  \dot{\psi} = W_d \psi \quad  \text{  and  } \quad \dot{\theta} = \theta, 
\end{gather}
where $\theta = \Gamma^\top \psi$. On the other hand, \cite{zhang2024testing} derives $\psi$ such that
$$
\psi = \Gamma (\Gamma^\top \Gamma)^{-1} \theta .
$$
So, if we use $\dot{X}_t$, then we would have $\dot{\psi} = \dot{\Gamma} (\dot{\Gamma}^\top \dot{\Gamma})^{-1} \dot{\theta}$. However, it doesn't satisfy \eqref{eq:preserve} because
$$
\dot{\psi} =  \dot{\Gamma} (\dot{\Gamma}^\top \dot{\Gamma})^{-1} \dot{\theta} = 
W_d^{-1} \Gamma \left(  \Gamma^\top  W_d^{-2} \Gamma \right)^{-1} \theta \ \ \neq \ \  W_d \Gamma (\Gamma^\top \Gamma)^{-1} \theta = W_d \psi ,
$$
unless $W_d = a I_L$ for some $a \neq 0$. Hence, the sizes of the systematic risk and the inside pricing error are changed depending on the units we use for the characteristics, which is undesirable.

On the other hand, in the case of our paper, because
$$
P_{X,t} = P_{\dot{X},t} \quad \text{and} \quad B_t = X_t \Gamma = \dot{X}_t \dot{\Gamma} = \dot{B}_t
$$
where $\dot{X}_t = X_t W_d$ and $\dot{\Gamma} = W_d^{-1} \Gamma$, we have
$$
(I_N - P_{\dot{B},t}) X_t \eta = (I_N - P_{\dot{B},t}) \dot{X}_t \dot{\eta}
$$
where $\dot{\eta} =  W_d^{-1} \eta $. Hence, our model is robust to the change of units of the characteristics.

\section{$\Omega$ matrix}\label{sec:omega}

This section presents the $\Omega$ matrix we used in our empirical analysis. Here, we consider a different $\Omega$ matrix for $\alpha_{O,t}$ from that in \cite{zhang2024testing} because $\Omega$ in \cite{zhang2024testing} with our data makes the sizes of some $B_{t,jq}^o$ quite large. To remedy this issue, we introduce a new $\Omega$ such that:
\begin{gather*}
B^{o}_{t} =  X^{o}_{t} (X^{o \top}_{t} X^{o}_{t}/N)^{-1/2}, \quad X^{o}_{t} = \left[I_N - P_{X,t} \right] \Omega, \quad \Omega_{N \times (N-L)} = \begin{pmatrix}
\Psi_{(N-L) \times (N-L)} \\
\Theta_{L \times (N-L)}
\end{pmatrix}
\end{gather*}
where
\begin{align*}
&\Psi =  {\scriptsize
\begin{bmatrix}
1 & 1.01 & 1 & 1.01 & 1 & 1.01 & 0 & 0 & 0 \\ 
0 & 1 & 1.01 & 1 & 1.01 & 1 & 1.01 & 0 & 0 \\
0 & 0 & 1 & 1.01 & 1 & 1.01 & 1 & 1.01 & 0 \\
0 & 0 & 0 & 1 & 1.01 & 1 & 1.01 & 1 & 1.01 \\
1 & 0 & 0 & 0 & 1 & 1.01 & 1 & 1.01 & 1 \\
1.01 & 1 & 0 & 0 & 0 & 1 & 1.01 & 1 & 1.01 \\
1 & 1.01 & 1 & 0 & 0 & 0 & 1 & 1.01 & 1 \\
1.01 & 1 & 1.01 & 1 & 0 & 0 & 0 & 1 & 1.01 \\
1 & 1.01 & 1 & 1.01 & 1 & 0 & 0 & 0 & 1 
\end{bmatrix} 
} \otimes I_{(N-L)/9} , \\
\\
&\Theta = \left[ \textbf{1}_{\lfloor (N-L)/L \rfloor}^\top \otimes I_{L} \quad \textbf{0}_{L \times (N-L - \lfloor (N-L)/L \rfloor \cdot L)} \right],
\end{align*}
which is designed to make $\Omega = [\Psi^\top \ \ \Theta^\top ]^\top$ to have the same number of nonzero elements in each column and it makes $B_{t,jq}^o$s not too large in our data. Here, we put $1.01$ in some elements instead of $1$ just to make $\Omega$ have a full column rank. Importantly, the representable set of $\alpha_{O,t}$ is not changed even if we use this basis, because $ \alpha_{O,t} = B^{o}_{t} \delta_{o,t} $ can be represented with any other basis $\dot{B}^{o}_{t}$ such that $ \alpha_{O,t} = \dot{B}^{o}_{t} \dot{\delta}_{o,t} $
with $\dot{\delta}_{o,t} = ( \dot{B}^{o \top}_{t} \dot{B}^{o}_{t} )^{-1} \dot{B}^{o \top}_{t} B^{o}_{t} \delta_{o,t} $.

\section{Proofs}

\subsection{Proof of main results}

\subsubsection{Proof of Theorem \ref{thm:gamma_1} (a)}

First of all, by Lemma \ref{lem:gamma_f_final_bound}, we have 
\begin{gather*}
\norm{ \tilde{f}_{t+1}^{d} - H_{F} f^d_{t+1}} , \ \ \frac{1}{T}\sum_{t=1}^T \norm{ \tilde{f}_{t+1}^{d} - H_{F} f^d_{t+1}} = O_p\left( \frac{L^{10}}{N^{10}} + \frac{L}{N\sqrt{T}} + \frac{1}{\sqrt{N}} \right),\\
\frac{1}{T}\sum_{t=1}^T \norm{ \tilde{f}_{t+1}^{d} - H_{F} f^d_{t+1}}^2 = O_p\left( \left( \frac{L^{10}}{N^{10}} + \frac{L}{N\sqrt{T}} + \frac{1}{\sqrt{N}} \right)^2 \right).
\end{gather*}
In addition, by Lemma \ref{lem:rotation_matrix_tech}, $\norm{H_{F}}, \norm{H_{F}^{-1}} = O_p(1)$. Note that
\begin{align}\label{eq:gamma_clt_expansion}
&\tilde{\gamma}_l -  H_{\Gamma}^{\top} \gamma_l  =  \left(e_l^\top \otimes  \left(\frac{1}{T}\sum_t \tilde{f}_{t+1}^{d} \tilde{f}_{t+1}^{d \top} \right)^{-1}  \right)
\frac{1}{T} \sum_{t=1}^T \ddot{E}_{t+1}^d \otimes  \left(\tilde{f}_{t+1}^{d} -  H_{F} f_{t+1}^{d} \right)\\
\nonumber &  +\left( e_l \otimes \left[ \left(\frac{1}{T}\sum_t \tilde{f}_{t+1}^{d} \tilde{f}_{t+1}^{d \top} \right)^{-1} - \left(\frac{1}{T}\sum_t H_{F}f_{t+1}^{d} f_{t+1}^{d\top} H_{F}^{\top} \right)^{-1} \right] \right)
\frac{1}{T} \sum_{t=1}^T \ddot{E}_{t+1}^d \otimes \left( H_{F} f_{t+1}^{d} \right) \\
\nonumber &  + H_{F}^{- \top}\left( e_l \otimes   \left(\frac{1}{T}\sum_t f_{t+1}^{d} f_{t+1}^{d\top}  \right)^{-1} \right)
\frac{1}{T} \sum_{t=1}^T \ddot{E}_{t+1}^d \otimes \ f_{t+1}^{d},
\end{align}
where $H_{\Gamma}^{\top} 
=  \left(\frac{1}{T}\sum_t \tilde{f}_{t+1}^{d} \tilde{f}_{t+1}^{d \top} \right)^{-1} \left( \frac{1}{T}\sum_t \tilde{f}_{t+1}^{d} f_{t+1}^{d \top}  \right) $. For the first term, because 
\begin{align*}
\norm{\frac{1}{T} \sum_{t=1}^T e_l^\top \ddot{E}_{t+1} \otimes  \left(\tilde{f}_{t+1}^{d} -  H_{F} f_{t+1}^{d} \right)}
&\leq \left( \frac{1}{T} \sum_{t=1}^T (e_l^\top \ddot{E}_{t+1})^2 \right)^{1/2} \left( \frac{1}{T} \sum_{t=1}^T \norm{\tilde{f}_{t+1}^{d} -  H_{F} f_{t+1}^{d}}^2 \right)^{1/2}\\
&= O_p\left( \frac{1}{\sqrt{N}} \left( \frac{L^{10}}{N^{10}} + \frac{L}{N\sqrt{T}} + \frac{1}{\sqrt{N}} \right) \right),
\end{align*}
by Lemma \ref{lem:error_tech}, we know the order of the first term is $O_p\left( \frac{1}{\sqrt{N}} \left( \frac{L^{10}}{N^{10}} + \frac{L}{N\sqrt{T}} + \frac{1}{\sqrt{N}} \right) \right)$. For the second term, note that
\begin{align*}
&\norm{\left(\frac{1}{T}\sum_t \tilde{f}_{t+1}^{d} \tilde{f}_{t+1}^{d \top} \right)^{-1} - \left(\frac{1}{T}\sum_t H_{F}f_{t+1}^{d} f_{t+1}^{d\top} H_{F}^{\top} \right)^{-1}} \\
&\lesssim \norm{\left(\frac{1}{T}\sum_t \tilde{f}_{t+1}^{d} \tilde{f}_{t+1}^{d \top} \right)^{-1}} \norm{\left(\frac{1}{T}\sum_t H_{F}f_{t+1}^{d} f_{t+1}^{d\top} H_{F}^{\top} \right)^{-1}}\\
& \ \ \times \norm{\frac{1}{T}\sum_t \tilde{f}_{t+1}^{d} \tilde{f}_{t+1}^{d \top}  - \frac{1}{T}\sum_t H_{F}f_{t+1}^{d} f_{t+1}^{d\top} H_{F}^{\top} } \\
& = O_p \left( \frac{L^{10}}{N^{10}} + \frac{L}{N\sqrt{T}} + \frac{1}{\sqrt{N}} \right) 
\end{align*}
as mentioned in the proof of Lemma \ref{lem:Gamma_bound}. Moreover, 
$$
\frac{1}{T} \sum_{t=1}^T e_l^\top \ddot{E}_{t+1}^d \otimes \left( H_{F} f_{t+1}^{d} \right) = H_{F}\frac{1}{T}\sum_{t=1}^T e_l^\top \ddot{E}_{t+1} \otimes f_{t+1}^{d} 
$$ 
and $\norm{\frac{1}{T}\sum_{t=1}^T e_l^\top \ddot{E}_{t+1} \otimes  f_{t+1}^{d} }= O_p\left( \frac{1}{\sqrt{NT}} \right)$ because 
$$
\frac{1}{T}\sum_{t=1}^T e_l^\top \ddot{E}_{t+1} \otimes  f_{t+1}^{d} = \frac{1}{NT} \sum_{it} (e_l^\top Q_t^{-1} x_{it})f_{t+1}^{d} \epsilon_{i,t+1}  = \frac{1}{NT} A^\top \vect(E)
$$ 
where
$A$ is the $NT \times K$ matrix of $(e_l^\top Q_t^{-1} x_{it})f_{t+1}^{d}$ and
$$
\bbE\norm{\frac{1}{NT}A^\top \vect(E)}_F^2 \lesssim \frac{1}{N^2T^2} \norm{\bbE[\vect(E)\vect(E)^\top]} \norm{A}_F^2 = O_p\left( \frac{1}{NT} \right)
$$
since $\norm{A}_F^2 \leq \max_t |e_l^\top Q_t^{-1} e_l| \sum_t ||f_{t+1}^{d}||^2  = O_p(NT)$. Hence, the order of the second term is $O_p\left( \frac{1}{\sqrt{NT}} \left( \frac{L^{10}}{N^{10}} + \frac{L}{N\sqrt{T}} + \frac{1}{\sqrt{N}} \right) \right)$. Lastly, we show that the third term converges to a normal distribution. Note that
\begin{align*}
\sqrt{NT} \frac{1}{T} \sum_{t=1}^T e_l^\top \ddot{E}_{t+1}^d \otimes f_{t+1}^{d}  = \frac{1}{\sqrt{NT}} \sum_{it} (e_l^\top Q_t^{-1} x_{it})f_{t+1}^{d} \epsilon_{i,t+1} \conD \calN(0, \Sigma_{xf,l}).
\end{align*}
In addition, because $H_{F}^{- \top} \conP \boldsymbol{H}^\top$ by Lemma \ref{lem:rotation_matrix_tech} and $   \frac{1}{T}\sum_{t=1}^T f_t^d f_t^{d \top} \conP \Sigma_f$ by Assumption \ref{asp:factorloadings}, we have 
$$
\sqrt{NT} H_{F}^{- \top}\left( e_l \otimes   \left(\frac{1}{T}\sum_t f_{t+1}^{d} f_{t+1}^{d\top}  \right)^{-1} \right)
\frac{1}{T} \sum_{t=1}^T \ddot{E}_{t+1}^d \otimes \ f_{t+1}^{d} \conD \calN \left(0, \boldsymbol{H}^\top \Sigma_f^{-1} \Sigma_{xf,l} \Sigma_f^{-1}\boldsymbol{H} \right) .
$$
Since the first and second terms are $o_p(1/\sqrt{NT})$ under our assumption, we have
$$
\sqrt{NT} \left( \hat{\gamma}_l -  H_{\Gamma}^{\top} \gamma_l \right) \conD \calN \left(0, \boldsymbol{H}^\top \Sigma_f^{-1} \Sigma_{xf,l} \Sigma_f^{-1}\boldsymbol{H} \right) . \ \ \square
$$

\subsubsection{Proof of Theorem \ref{thm:gamma_1} (b)}

The first term of \eqref{eq:gamma_clt_expansion} can be represented as:
\begin{align*}
\frac{1}{T} \sum_{t=1}^T e_l^\top \ddot{E}_{t+1}^d \otimes  \left(\tilde{f}_{t+1}^{d} -  H_{F} f_{t+1}^{d} \right) 
 = \tilde{\Gamma}^{\top} \frac{1}{T} \sum_{t=1}^T (e_l^\top \ddot{E}_{t+1}) \times \ddot{E}_{t+1}  + \tilde{\Gamma}^{\top} \bar{ \ddot{E}} \times \frac{1}{T} \sum_{t=1}^T (e_l^\top \ddot{E}_{t+1}).
\end{align*}
By Lemma \ref{lem:error_tech}, we have $||\bar{ \ddot{E}}||| = O_p\left(\frac{\sqrt{L}}{\sqrt{NT}}\right)$, $||\frac{1}{T} \sum_{t=1}^T (e_l^\top \ddot{E}_{t+1})|| = O_p\left(\frac{1}{\sqrt{NT}}\right)$. Hence, the second term of the above equation is $O_p\left( \frac{\sqrt{L}}{NT}\right) = o_p\left( \frac{1}{\sqrt{NT}}\right)$. For the first term, we use the debasing method. Note that
\begin{align*}
&\tilde{\Gamma}^{\top} \frac{1}{T} \sum_{t=1}^T (e_l^\top \ddot{E}_{t+1}) \times \ddot{E}_{t+1}  - \tilde{\Gamma}^{\top} \frac{1}{T} \sum_{t=1}^T  \hat{\sigma}_{t+1}^2  (X_t^\top X_t)^{-1}e_l \\
&= \tilde{\Gamma}^{\top} \frac{1}{N^2 T} \sum_{i,j,t}(e_l^\top Q_t^{-1} x_{it}) Q_t^{-1} x_{jt} \epsilon_{i,t+1} \epsilon_{j,t+1} 
- \tilde{\Gamma}^{\top} \frac{1}{N^2 T} \sum_{i,t}(e_l^\top Q_t^{-1} x_{it}) Q_t^{-1} x_{it}  \hat{\sigma}_{t+1}^2 \\
&=\tilde{\Gamma}^{\top} \frac{1}{N^2 T} \sum_{i,j,t}a_{ijt} (u_{ij,t+1} - \bbE[u_{ij,t+1}])  +
\tilde{\Gamma}^{\top} \frac{1}{N^2 T} \sum_{i,t}a_{iit} (\bbE[\epsilon_{i,t+1}^2] - \hat{\sigma}_{t+1}^2),
\end{align*}
where $u_{ij,t+1} = \epsilon_{i,t+1} \epsilon_{j,t+1}$ and $a_{ijt} = (e_l^\top Q_t^{-1} x_{it}) Q_t^{-1} x_{jt}$. Then, by Lemma \ref{lem:debiasing_tech}, the first term of the last equation is $O_p\left( \frac{\sqrt{L}}{N \sqrt{T} } \right) = o_p\left( \frac{1}{\sqrt{NT}} \right)$. In addition, the second term of the last equation is also $o_p\left( \frac{1}{\sqrt{NT}} \right)$ by Lemma \ref{lem:debiasing_tech}. Hence, we have
\begin{align*}
\sqrt{NT} \left( \hat{\gamma}_{l} -  H_{\Gamma}^{\top} \gamma_l \right) &= \sqrt{NT} H_{F}^{- \top}\left( e_l \otimes   \left(\frac{1}{T}\sum_t f_{t+1}^{d} f_{t+1}^{d\top}  \right)^{-1} \right)
\frac{1}{T} \sum_{t=1}^T \ddot{E}_{t+1}^d \otimes \ f_{t+1}^{d} + o_p(1)\\
&  \conD \calN \left(0, \boldsymbol{H}^\top \Sigma_f^{-1} \Sigma_{xf,l} \Sigma_f^{-1}\boldsymbol{H} \right). \ \ \square
\end{align*}

\subsubsection{Proof of Theorem \ref{thm:alpha_I}}
(a) CLT for $\tilde{\alpha}_{I,it}$: Note that
\begin{align}\label{eq:decomp_alpha_i_2}
\nonumber \tilde{\alpha}_{I,it}  - \alpha_{I,it} 
&= e_i^\top \left( P_{B,t} - P_{\tilde{B},t} \right) X_t \bar{\ddot{R}} + \left( X_{it}^\top - e_i^\top P_{B,t} X_t  \right) \bar{\ddot{E}}\\  
&= e_i^\top \left( P_{B,t} - P_{\tilde{B},t} \right) X_t (\eta + \Gamma \bar{\breve{f}}) + \left( x_{it}^\top - e_i^\top P_{B,t} X_t  \right) \bar{\ddot{E}} + e_i^\top \left( P_{B,t} - P_{\tilde{B},t} \right) X_t \bar{\ddot{E}}.
\end{align}
By Lemma \ref{lem:projection_tech}, the first term can be represented like
\begin{align*}
e_i^\top \left( P_{B,t} - P_{\tilde{B},t} \right) X_t (\eta + \Gamma \bar{\breve{f}}) &= -B_{it}^\top \left( B_t^\top B_t \right)^{-1} H_{\Gamma}^{- \top } 
 \left(\tilde{B}_t -B_t H_{\Gamma}  \right)^\top (I_{N} - P_{B,t}) X_t (\eta + \Gamma \bar{\breve{f}}) \\
 &- e_i^\top (I_{N} - P_{B,t})  \left(\tilde{B}_t -B_t H_{\Gamma}\right) H_{\Gamma}^{-1}  \left( B_t^\top B_t \right)^{-1} B_t^\top X_t (\eta + \Gamma \bar{\breve{f}}) \\
 &+ O_p\left(  \frac{L}{NT} + \frac{L}{N^2} + \left( \frac{L}{N} \right)^{20 +1 }\right).
\end{align*}
Here, we use Lemma \ref{lem:alpha_I_1_tech} to show the higher order terms are $O_p\left(  \frac{L}{NT} + \frac{L}{N^2} + \left( \frac{L}{N} \right)^{20 +1 }\right)$. Then, the first part of the dominating term can be represented as
\begin{align*}
&B_{it}^\top \left( B_t^\top B_t \right)^{-1} H_{\Gamma}^{- \top} 
 \left(\tilde{B}_t -B_t H_{\Gamma}  \right)^\top (I_{N} - P_{B,t}) X_t (\eta + \Gamma \bar{\breve{f}})\\
& = B_{it}^\top \left( B_t^\top B_t \right)^{-1} H_{\Gamma}^{- \top} 
 \left(\tilde{\Gamma}_t -\Gamma_t H_{\Gamma}  \right)^\top X_t^\top (I_{N} - P_{B,t}) X_t (\eta + \Gamma \bar{\breve{f}})) \\
& = \left(  (\eta + \Gamma \bar{\breve{f}}))^\top X_t^\top (I_{N} - P_{B,t}) X_t \otimes B_{it}^\top \left( B_t^\top B_t \right)^{-1} H_{\Gamma}^{-\top} \right) \vect \left[ \left(\tilde{\Gamma} -\Gamma H_{\Gamma}  \right)^\top \right] \\
& = \left( \eta^\top X_t^\top (I_{N} - P_{B,t}) X_t \otimes B_{it}^\top \left( B_t^\top B_t \right)^{-1} H_{\Gamma}^{-\top } \right) \vect \left[ \left(\tilde{\Gamma} -\Gamma H_{\Gamma}  \right)^\top \right].
\end{align*}
Here, we use the relation that $\Gamma^\top X_t^\top M_{B,t} = 0$ where $M_{B,t} = I_{N} - P_{B,t} $. In addition, we have
\begin{align*}
\vect \left[ \left(\tilde{\Gamma} -\Gamma H_{\Gamma}  \right)^\top \right]
& = \left( I_L \otimes  \left(\frac{1}{T}\sum_t \tilde{f}_{t+1}^{d} \tilde{f}_{t+1}^{d \top} \right)^{-1}  \right)
\frac{1}{T} \sum_{t=1}^T \ddot{E}_{t+1}^d \otimes  \left(\tilde{f}_{t+1}^{d} -  H_{F} f_{t+1}^{d} \right)   \\
& \ \ +\left( I_L \otimes  \left(\frac{1}{T}\sum_t \tilde{f}_{t+1}^d \tilde{f}_{t+1}^{d \top} \right)^{-1}  \right)
\frac{1}{T} \sum_{t=1}^T \ddot{E}_{t+1}^d \otimes \left( H_{F} f_{t+1}^{d} \right)\\
& = \left( I_L \otimes H_{F}^{-\top} \left(\frac{1}{T}\sum_t f_{t+1}^{d} f_{t+1}^{d \top} \right)^{-1}  \right)
\frac{1}{T} \sum_{t=1}^T \ddot{E}_{t+1}^d \otimes  f_{t+1}^{d} \\
& \ \ + O_p\left( \frac{\sqrt{L}}{\sqrt{N}} \left( \frac{L}{N\sqrt{T}} + \left( \frac{L}{N} \right)^{10} + \frac{1}{\sqrt{N}} \right) \right).
\end{align*}
Here, we derive the order of the residual terms using the proof of Lemma \ref{lem:Gamma_bound} with $a_{NT} = \frac{L}{N\sqrt{T}} + \left( \frac{L}{N} \right)^{10} + \frac{1}{\sqrt{N}} $. Then, since $\norm{\left( \eta^\top X_t^\top M_{B,t} X_t \otimes B_{it}^\top \left( B_t^\top B_t \right)^{-1} H_{\Gamma}^{\top -1} \right)} = O_p(1)$, we have
\begin{align*}
&B_{it}^\top \left( B_t^\top B_t \right)^{-1} H_{\Gamma}^{- \top} 
 \left(\tilde{B}_t -B_t H_{\Gamma}  \right)^\top (I_{N} - P_{B,t}) X_t (\eta + \Gamma \bar{f})\\
&= \left( \eta^\top( X_t^\top M_{B,t} X_t/N) \otimes B_{it}^\top \left( B_t^\top B_t/N \right)^{-1}\left( \frac{1}{T} \sum_{s = 1}^T f_{s+1}^{d} f_{s+1}^{d^\top} \right)^{-1} \right) \frac{1}{NT} \sum_{j=1}^N \sum_{s=1}^T \left( Q_t^{-1} x_{js} \otimes f_{s+1}^{d} \right) \epsilon_{j,s+1}\\
&\ \ + O_p\left( \frac{\sqrt{L}}{\sqrt{N}} \left( \frac{L}{N\sqrt{T}} + \left( \frac{L}{N} \right)^{10} + \frac{1}{\sqrt{N}} \right) \right)\\
&= \frac{1}{NT} \sum_{j=1}^N \sum_{s=1}^T \left(\eta^\top - \eta^\top Q_t \Gamma  (Q^{B}_t)^{-1} \Gamma^\top \right) x_{js}  \left( B_{it}^\top (Q^{B}_t)^{-1} (Q^{f})^{-1} f_{s+1}^d \right)  \epsilon_{j,s+1} \\
&\ \ + O_p\left( \frac{\sqrt{L}}{\sqrt{N}} \left( \frac{L}{N\sqrt{T}} + \left( \frac{L}{N} \right)^{10} + \frac{1}{\sqrt{N}} \right) \right),
\end{align*}
where $Q^B_t = B_t^\top B_t/N$ and $Q^f = F^{d\top} F^{d}/T$ since $ \eta^\top( X_t^\top M_{B,t} X_t/N) = \eta^\top Q_t - \eta^\top Q_t \Gamma  (Q^{B}_t)^{-1} \Gamma^\top Q_t$. Similarly, the second part of the dominating term can be represented as
\begin{align*}
&e_i^\top (I_{N} - P_{B,t})  \left(\tilde{B}_t -B_t H_{\Gamma}\right) H_{\Gamma}^{-1}  \left( B_t^\top B_t \right)^{-1} B_t^\top X_t (\eta + \Gamma \bar{\breve{f}})\\
& = \left(e_i^\top M_{B,t} X_t \otimes \left( \eta^\top Q_t \Gamma (Q_t^B)^{-1} + \bar{\breve{f}}^\top \right)  H_{\Gamma}^{- \top } \right) \vect \left[ \left(\hat{\Gamma} -\Gamma H_{\Gamma}  \right)^\top \right]  \\
& =\left(e_i^\top M_{B,t} X_t \otimes \left( \eta^\top Q_t \Gamma (Q_t^B)^{-1} + \bar{\breve{f}}^\top \right)  \left(Q^f \right)^{-1} \right) \frac{1}{NT} \sum_{j=1}^N \sum_{s=1}^T \left( Q_t^{-1} x_{js} \otimes f_{s+1}^d \right) \epsilon_{j,s+1} \\
&\ \ + O_p\left( \frac{\sqrt{L}}{\sqrt{N}} \left( \frac{L}{N\sqrt{T}} + \left( \frac{L}{N} \right)^{10} + \frac{1}{\sqrt{N}} \right) \right)\\
& = \frac{1}{NT} \sum_{j=1}^N \sum_{s=1}^T \left(x_{it}^\top Q_t^{-1} - B_{it}^\top (Q_t^B)^{-1} \Gamma^\top \right) x_{js} \left( \eta^\top Q_t \Gamma (Q_t^B)^{-1} + \bar{\breve{f}}^\top \right) \left( Q^{f}\right)^{-1} f_{s+1}^d \epsilon_{j,s+1} \\
&\ \ + O_p\left( \frac{\sqrt{L}}{\sqrt{N}} \left( \frac{L}{N\sqrt{T}} + \left( \frac{L}{N} \right)^{10} + \frac{1}{\sqrt{N}} \right) \right),
\end{align*}
because $e_i^\top M_{B,t} X_t = x_{it}^\top - B_{it}^\top (Q_t^{B})^{-1} \Gamma^\top Q_t$. In addition, the second term in \eqref{eq:decomp_alpha_i_2} can be represented as 
$$
\left( x_{it}^\top - e_i^\top P_{B,t} X_t  \right) \bar{\ddot{E}} = \frac{1}{NT} \sum_{j=1}^N \sum_{s=1}^T \left(x_{it}^\top Q_t^{-1}- B_{it}^\top (Q_t^{B})^{-1} \Gamma^\top \right)  x_{js} \epsilon_{j,s+1} .
$$
Moreover, the third term in \eqref{eq:decomp_alpha_i_2} can be bounded like
\begin{gather*}
\norm{e_i^\top \left( P_{B,t} - P_{\tilde{B},t} \right) X_t \bar{\ddot{E}}} \leq \norm{e_i^\top \left( P_{B,t} - P_{\tilde{B},t} \right) X_t} \norm{\bar{\ddot{E}}} = O_p\left( \frac{\sqrt{L}}{\sqrt{NT}} \left(   \frac{\sqrt{L}}{\sqrt{NT}} + \frac{\sqrt{L}}{N} + \left( \frac{L}{N} \right)^{10 +\frac{1}{2} }    \right)  \right), 
\end{gather*}
by Lemma \ref{lem:error_tech} and the bound that $\norm{e_i^\top \left( P_{B,t} - P_{\tilde{B},t} \right) X_t} = O_p\left(   \frac{\sqrt{L}}{\sqrt{NT}} + \frac{\sqrt{L}}{N} + \left( \frac{L}{N} \right)^{10 + \frac{1}{2} } \right)$ since
\begin{align*}
e_i^\top (P_{B,t} - P_{\tilde{B},t}) X_t
&= x_{it}^\top \Gamma \left( \Gamma^\top Q_t \Gamma  \right)^{-1} \Gamma^\top Q_t  - x_{it}^\top \tilde{\Gamma} \left( \tilde{\Gamma}^\top Q_t \tilde{\Gamma}  \right)^{-1} \tilde{\Gamma}^\top Q_t  \\
& \lesssim \norm{x_{it}^\top \Gamma H_{\Gamma} - x_{it}^\top \tilde{\Gamma}}
\norm{ \left( H_{\Gamma}^\top \Gamma^\top Q_t \Gamma  H_{\Gamma} \right)^{-1} } \norm{H_{\Gamma}^\top \Gamma^\top Q_t } \\
& +  \norm{B_{it}^\top H_{\Gamma} } \norm{ \left( H_{\Gamma}^\top \Gamma^\top Q_t \Gamma  H_{\Gamma} \right)^{-1} - \left( \tilde{\Gamma}^\top Q_t \tilde{\Gamma} \right)^{-1} } \norm{H_{\Gamma}^\top \Gamma^\top Q_t } \\
& +  \norm{B_{it}^\top  H_{\Gamma} } \norm{\left( \tilde{\Gamma}^\top Q_t \tilde{\Gamma} \right)^{-1} } \norm{\tilde{\Gamma}^\top - H_{\Gamma}^\top \Gamma^\top} \norm{ Q_t }  \\
& = O_p\left( \frac{\sqrt{L}}{\sqrt{NT}} + \frac{\sqrt{L}}{N} + \left( \frac{L}{N} \right)^{10 + \frac{1}{2}} \right)
\end{align*}
by Lemma \ref{lem:alpha_I_1_tech}. In summary, we have
\begin{align*}
 \frac{\sqrt{NT}}{\sqrt{L}} \sigma_{I,it}^{-1} \left(\tilde{\alpha}_{I,it}  - \alpha_{I,it}\right) = \sigma_{I,it}^{-1} \frac{1}{\sqrt{NTL}} \sum_{j=1}^N \sum_{s=1}^T  g_{it,js} \epsilon_{j,s+1} + o_p\left( 1 \right).
\end{align*}
where
\begin{align*}
g_{it,js} =& \left[1 - \left( \eta^\top Q_t \Gamma (Q_t^B)^{-1} + \bar{\breve{f}}^\top \right) \left( Q^{f}\right)^{-1} f_{s+1}^d \right] \left(x_{it}^\top Q_t^{-1}- B_{it}^\top (Q_t^{B})^{-1} \Gamma^\top \right) x_{js}   \\  
& - \left( B_{it}^\top (Q^{B}_t)^{-1} (Q^{f})^{-1} f_{s+1}^d \right) \left(\eta^\top - \eta^\top Q_t \Gamma  (Q^{B}_t)^{-1} \Gamma^\top \right) x_{js} .
\end{align*}
By Assumption \ref{asp:clt}, the first term converges to a standard normal distribution. It completes the proof.\\
(b) CLT for $\hat{\alpha}_{I,it}$: The proof is basically the same as that of (a). The only difference is that  we use the bounds from Lemma \ref{lem:alpha_I_1_tech} like $\norm{\hat{B}_t -B_t H_{\Gamma}}  = O_p\left( \frac{\sqrt{L}}{\sqrt{T}} \right)$, $\norm{\hat{B}_{it} -H_{\Gamma}^\top B_{it} } = O_p\left( \frac{\sqrt{L}}{\sqrt{NT}} \right)$, $\norm{\left(\hat{B}_t^{\top} \hat{B}_t / N \right)^{-1} - \left(H_{\Gamma}^\top B_t^\top B_t H_{\Gamma} / N \right)^{-1}} =  O_p\left( \frac{\sqrt{L}}{\sqrt{NT}} \right)$ to show the higher order terms are $O_p\left( \frac{L}{N^{3/2}\sqrt{T}}\right)$. In addition, the order of the residuals in the dominating parts becomes $o_p \left( \frac{\sqrt{L}}{\sqrt{NT}} \right)$. Hence, we have 
\begin{align*}
 \frac{\sqrt{NT}}{\sqrt{L}} \sigma_{I,it}^{-1} \left(\hat{\alpha}_{I,it}  - \alpha_{I,it}\right) = \sigma_{I,it}^{-1} \frac{1}{\sqrt{NTL}} \sum_{j=1}^N \sum_{s=1}^T  g_{it,js} \epsilon_{j,s+1} + o_p\left( 1 \right).
\end{align*}
By Assumption \ref{asp:clt}, the first term converges to a standard normal distribution, and the second term converges to $0$. It completes the proof. $\square$

\begin{comment}
\subsubsection{Proof of Theorem \ref{thm:alpha_O}}
(i) Because 
$$
\sqrt{NT} \left( \hat{\delta}_q - \delta_q \right) =  \frac{1}{\sqrt{NT}} \sum_{s=1}^T \sum_{j=1}^N B_{s,jq}^{o} \epsilon_{j,s+1}
$$  
where $B_{s,jq}^{o} = e_j^\top B_s^o e_q$, we have by Assumption \ref{asp:clt} that
$$
\sqrt{NT} \sigma_{\delta,q}^{-1} \left( \hat{\delta}_q - \delta_q \right) \conD \calN (0,1).
$$
(ii) Because 
$$
\hat{\alpha}_{O,it} - \alpha_{O,it} = B^{o \top}_{t,i} (\hat{\delta} - \delta) = \frac{1}{NT} \sum_{j=1}^N \sum_{s=1}^T B^{o \top}_{t,i} B^{o}_{s,j} \epsilon_{j,s+1},
$$
we have
$$
\sqrt{NT} \norm{B^{o \top}_{t,i}}^{-1} \left( \hat{\alpha}_{O,it} - \alpha_{O,it} \right) =  \frac{1}{\sqrt{NT}}\sum_{j=1}^N \sum_{s=1}^T h_{js} \epsilon_{j,s+1},
$$
where $h_{js} =  \norm{B^{o}_{t,i}}^{-1} B^{o \top}_{t,i} B^{o}_{s,j}$. Hence, by Assumption \ref{asp:clt}, 
$$
\sqrt{NT} \sigma_{o,it}^{-1} \norm{B^{o \top}_{t,i}}^{-1} \left( \hat{\alpha}_{O,it} - \alpha_{O,it} \right) \conD \calN(0,1). \ \ \square
$$

\end{comment}

\subsubsection{Proof of Theorem \ref{thm:delta_heterodelta}}

Because 
$$
\sqrt{N} \left( \tilde{\delta}_{o,tq} - \delta_{o,tq} \right) =  \frac{1}{\sqrt{N}} \sum_{j=1}^N B_{t,jq}^{o} \epsilon_{j,s+1}
$$  
where $B_{t,jq}^{o} = e_j^\top B_t^o e_q$, we have by Assumption \ref{asp:clt} that
$$
\sqrt{N} \sigma_{\delta,qt}^{-1} \left( \tilde{\delta}_{o,tq} - \delta_{o,tq} \right) \conD \calN (0,1). \ \ \square
$$

\subsubsection{Proof of Theorem \ref{thm:alpha_O_heterodelta}}
Let $\dot{\xi}_{t} = \tilde{\delta}_{o,t} - \tilde{\zeta}$. First, we denote by $\calW$ the event that for all $t$,
$$
\max_{1 \leq q \leq N - L}\abs{\frac{1}{N} \sum_{j=1}^{N} B_{t,jq}^{o} \epsilon_{j,t+1} - \frac{1}{NT} \sum_{s=1}^T \sum_{j=1}^{N} B_{s,jq}^{o} \epsilon_{j,s+1} } \leq  (C_u + 0.05) \sigma_{t+1} \frac{\sqrt{\log NT}}{\sqrt{N}} 
$$
where $C_u>0$ is the universal constant in Lemma \ref{lem:uniformbound_1}. Then, by Lemmas \ref{lem:uniformbound_1} and \ref{lem:uniformbound_2}, we know $\Pr(\calW) \rightarrow 1$. Set $\rho_t = C_\rho \sigma_{t+1} \frac{\sqrt{\log NT}}{\sqrt{N}} $ where $C_\rho = C_u + 0.1$. Then, because $\xi_{t,q} = 0$ if $q \notin D_t$, on the event $\calW$, we have for all $t$,
\begin{align*}
\max_{q \notin D_t} \abs{\dot{\xi}_{t,q}} 
&\leq \max_{q \notin D_t}\abs{\frac{1}{N} \sum_{j=1}^{N} B_{t,jq}^{o} \epsilon_{j,t+1} - \frac{1}{NT} \sum_{s=1}^T \sum_{j=1}^{N} B_{s,jq}^{o} \epsilon_{j,s+1} } + \max_{q \notin D_t} \abs{ \frac{1}{T} \sum_{s=1}^T \xi_{s,q} } \\
&\leq  C_\rho \sigma_{t+1} \frac{\sqrt{\log NT}}{\sqrt{N}} = \rho_t.
\end{align*}
Hence, on the event $\calW$, we have $\tilde{\xi}_{t,q} = 0$ for all $t$ and $q \notin D_t$. Here, we use the relation that
$$
\dot{\xi}_{t} = \xi_t + \frac{1}{N} B_t^{o \top} E_{t+1} - \frac{1}{NT} \sum_{s=1}^T B_s^{o \top} E_{s+1} - \frac{1}{T} \sum_{s=1}^T \xi_s .
$$

In addition, on the event $\calW$, we have for all $t$,
\begin{align*}
\min_{q \in D_t }\abs{\dot{\xi}_{t,q}}   &\geq \min_{q \in D_t} \abs{\xi_{t,q}} - \max_{q \in D_t}\abs{\frac{1}{N} \sum_{j=1}^{N} B_{t,jq}^{o} \epsilon_{j,t+1} - \frac{1}{NT} \sum_{s=1}^T \sum_{j=1}^{N} B_{s,jq}^{o} \epsilon_{j,s+1} } - \max_{q \in D_t} \abs{ \frac{1}{T} \sum_{s=1}^T \xi_{s,q} } \\
& \geq \min_{q \in D_t} \abs{\xi_{t,q}} - C_\rho \sigma_{t+1} \frac{\sqrt{\log NT}}{\sqrt{N}} \\
& \gg C_\rho  \sigma_{t+1} \frac{\sqrt{\log NT}}{\sqrt{N}} ,
\end{align*}
by Assumption \ref{asp:sparsity_heterodelta}. So, on the event $\calW$, we have $\tilde{\xi}_{t,q} = \dot{\xi}_{t,q}$ for all $t$ and $q \in D_t$. In addition, on the event $\calW$, $D_t$ is the same as $\calD_t$ for all $t$, where $\calD_t = \{ 1 \leq q \leq N-L: \tilde{\xi}_{t,q} \neq 0 \}$. Note that
$$
V_{o,it}^{-1/2} \left(\hat{\alpha}_{O,it} - \alpha_{O,it}\right) =  V_{o,it}^{-1/2}\sum_{q=1}^{N - L} B_{t,iq}^o  \left( \hat{\zeta}_{q} - \zeta_{q} \right) 
+ V_{o,it}^{-1/2}\sum_{q=1}^{N - L} B_{t,iq}^o  \left( \tilde{\xi}_{t,q} - \xi_{t,q} \right) 
+ V_{o,it}^{-1/2}\sum_{q \in \calD_t} B_{t,iq}^o \tilde{\bar{\xi}}_q,
$$
where 
$$
\hat{\zeta} - \zeta = \frac{1}{NT} B_s^{o \top} E_{s+1} + \frac{1}{T} \sum_{s=1}^T (\xi_{s} - \tilde{\xi}_s).
$$
Then, the first part can be decomposed into
$$
V_{o,it}^{-1/2}\sum_{q=1}^{N - L} B_{t,iq}^o  \left( \hat{\zeta}_{q} - \zeta_{q} \right) = V_{o,it}^{-1/2}\frac{1}{NT} \sum_{j=1}^N \sum_{s=1}^T B_{t,i}^{o \top} B_{s,j}^{o} \epsilon_{j,s+1}  - V_{o,it}^{-1/2}\sum_{q=1}^{N-L} B_{t,iq}^o \left( \tilde{\bar{\xi}}_q - \bar{\xi}_q \right) .
$$
In addition, the second part can be decomposed into
\begin{align*}
& V_{o,it}^{-1/2}\sum_{q=1}^{N - L} B_{t,iq}^o  \left( \tilde{\xi}_{t,q} - \xi_{t,q} \right) \\
&= V_{o,it}^{-1/2}\sum_{q \in D_t} B_{t,iq}^o  \left( \tilde{\xi}_{t,q} - \xi_{t,q} \right) + V_{o,it}^{-1/2} \sum_{q \notin D_t} B_{t,iq}^o  \left( \tilde{\xi}_{t,q} - \xi_{t,q} \right) \\
&= V_{o,it}^{-1/2}\sum_{q \in D_t} B_{t,iq}^o  \left(  \dot{\xi}_{t,q} - \xi_{t,q} \right) + V_{o,it}^{-1/2}\sum_{q \in D_t} B_{t,iq}^o  \left(  \tilde{\xi}_{t,q} - \dot{\xi}_{t,q} \right)
+ V_{o,it}^{-1/2} \sum_{q \notin D_t} B_{t,iq}^o  \left( \tilde{\xi}_{t,q} - \xi_{t,q} \right).
\end{align*}
Note that, on the event $\calW$, the second term is zero since $\tilde{\xi}_{t,q} = \dot{\xi}_{t,q}$ for all $q \in D_t$. In addition, $\Pr(\calW) \rightarrow 1$. Hence, w.p.c. to 1, the second term is zero, and for any $\varepsilon > 0$, we have
$$
\Pr\left(\abs{V_{o,it}^{-1/2}\sum_{q \in D_t} B_{t,iq}^o  \left(  \tilde{\xi}_{t,q} - \dot{\xi}_{t,q} \right)} \geq \varepsilon \right)
\leq \Pr\left(\abs{V_{o,it}^{-1/2}\sum_{q \in D_t} B_{t,iq}^o  \left(  \tilde{\xi}_{t,q} - \dot{\xi}_{t,q} \right)} \neq 0 \right) \rightarrow 0 .
$$
Hence, the second term is $o_p(1)$. Similarly, on the event $\calW$, the third term is zero, since $\tilde{\xi}_{t,q} = \xi_{t,q} = 0$ for all $q \notin D_t$. Hence, we know that the third term is $o_p(1)$. Hence, we have
\begin{align*}
&V_{o,it}^{-1/2}\sum_{q=1}^{N - L} B_{t,iq}^o  \left( \tilde{\xi}_{t,q} - \xi_{t,q} \right) = V_{o,it}^{-1/2}\sum_{q \in D_t} B_{t,iq}^o  \left(  \dot{\xi}_{t,q} - \xi_{t,q} \right) + o_p(1) \\
& = V_{o,it}^{-1/2}\frac{1}{N} \sum_{j=1}^{N} \sum_{q \in D_t} B_{t,iq}^o B_{t,jq}^{o} \epsilon_{j,t+1} - V_{o,it}^{-1/2}\frac{1}{NT} \sum_{j=1}^{N}\sum_{s=1}^{T} \sum_{q \in \calD_t} B_{t,iq}^o B_{s,jq}^{o} \epsilon_{j,s+1}
- V_{o,it}^{-1/2}\sum_{q \in D_t} B_{t,iq}^o \bar{\xi}_q + o_p(1)\\
& = V_{o,it}^{-1/2}\frac{1}{N} \sum_{j=1}^{N} \sum_{q \in D_t} B_{t,iq}^o B_{t,jq}^{o} \epsilon_{j,t+1} - V_{o,it}^{-1/2}\sum_{q \in D_t} B_{t,iq}^o \bar{\xi}_q + o_p(1).
\end{align*}
Here, the last equation comes from the fact that $\bbE \left[ \left(\frac{1}{NT} \sum_{j=1}^{N}\sum_{s=1}^{T} \sum_{q \in D_t} B_{t,iq}^o B_{s,jq}^{o} \epsilon_{j,s+1} \right)^2 \right] \lesssim \frac{|D_t|}{NT}  \left( \frac{1}{|D_t|} \sum_{q \in D_t} B_{t,iq}^{o 2} \right) $. Therefore, we have
\begin{align*}
 V_{o,it}^{-1/2} \left(\hat{\alpha}_{O,it} - \alpha_{O,it}\right) &=  V_{o,it}^{-1/2}\frac{1}{NT} \sum_{j=1}^N \sum_{s=1}^T B_{t,i}^{o \top} B_{s,j}^{o} \epsilon_{j,s+1} +  V_{o,it}^{-1/2}\frac{1}{N} \sum_{j=1}^{N} \sum_{q \in D_t} B_{t,iq}^o B_{t,jq}^{o} \epsilon_{j,t+1} \\
 & \ \   + V_{o,it}^{-1/2}\sum_{q \notin D_t} B_{t,iq}^o \left( \tilde{\bar{\xi}}_q - \bar{\xi}_q   \right) + o_p(1) .
\end{align*}
Here, we use the fact that $V_{o,it}^{-1/2}\sum_{q \in D_t} B_{t,iq}^o \tilde{\bar{\xi}}_q - V_{o,it}^{-1/2}\sum_{q \in \calD_t} B_{t,iq}^o \tilde{\bar{\xi}}_q = o_p(1)$ because, on the event $\calW$, $D_t = \calD_t$ and $V_{o,it}^{-1/2}\sum_{q \in D_t} B_{t,iq}^o \tilde{\bar{\xi}}_q = V_{o,it}^{-1/2}\sum_{q \in \calD_t} B_{t,iq}^o \tilde{\bar{\xi}}_q$. Lastly, we show that $V_{o,it}^{-1/2}\sum_{q \notin D_t} B_{t,iq}^o \left( \tilde{\bar{\xi}}_q - \bar{\xi}_q   \right) \conP 0$. By using the same argument as above, we have
\begin{align*}
&V_{o,it}^{-1/2}\sum_{q \notin D_t} B_{t,iq}^o \left( \tilde{\bar{\xi}}_q - \bar{\xi}_q   \right) \\
&=   V_{o,it}^{-1/2} \frac{1}{T} \sum_{s=1}^T \sum_{q \notin D_t} B_{t,iq}^o \left( \tilde{\xi}_{s,q} - \xi_{s,q}   \right) \\
& = V_{o,it}^{-1/2} \frac{1}{T} \sum_{s=1}^T \sum_{q \in D_s / D_t} B_{t,iq}^o \left( \dot{\xi}_{s,q} - \xi_{s,q} \right) + o_p(1) \\
& = V_{o,it}^{-1/2} \frac{1}{T} \sum_{s=1}^T \sum_{q \in D_s / D_t} B_{t,iq}^o \left( \frac{1}{N} \sum_{j=1}^N B_{s,jq}^o \epsilon_{j,s+1} - \frac{1}{NT} \sum_{s'=1}^T \sum_{j=1}^N B_{s',jq}^o \epsilon_{j,s'+1}  - \frac{1}{T} \sum_{s'=1}^T \xi_{s',q} \right) + o_p(1) .
\end{align*}
Some calculation shows that 
$$
\bbE\left[\left(  \frac{1}{T} \sum_{s=1}^T \sum_{q \in D_s / D_t} B_{t,iq}^o \frac{1}{N} \sum_{j=1}^N B_{s,jq}^o \epsilon_{j,s+1}  \right)^2 \right] \lesssim \frac{1}{NT} \frac{1}{T} \sum_{s=1}^T \sum_{q \in D_s / D_t} B_{t,iq}^{o2} 
$$
Hence, the first term is $o_p(1)$. In addition, we can also show that
$$
\bbE\left[\left(  \frac{1}{T} \sum_{s=1}^T \sum_{q \in D_s / D_t} B_{t,iq}^o \frac{1}{NT} \sum_{s'=1}^T \sum_{j=1}^N B_{s',jq}^o \epsilon_{j,s'+1}  \right)^2 \right] \lesssim \frac{1}{NT} \frac{1}{T} \sum_{s=1}^T \sum_{q \in D_s / D_t} B_{t,iq}^{o2} .
$$
So, the second term is $o_p(1)$. Moreover, the third term also converges to $0$ under our assumption. Hence, we have
\begin{align*}
 V_{o,it}^{-1/2} \left(\hat{\alpha}_{O,it} - \alpha_{O,it}\right) &=  V_{o,it}^{-1/2} \left( \frac{1}{NT} \sum_{j=1}^N \sum_{s=1}^T B_{t,i}^{o \top} B_{s,j}^{o} \epsilon_{j,s+1} +  \frac{1}{N} \sum_{j=1}^{N} \sum_{q \in D_t} B_{t,iq}^o B_{t,jq}^{o} \epsilon_{j,t+1} \right) + o_p(1) \\
 & \conD \calN (0,1)  \ \ \square
\end{align*}

\subsubsection{Proof of Theorem \ref{thm:variance_estimation}}

(i) Let $\varpi_t = [Q_t^{-1}]_{ll}$. Then, we know $\max_t |\varpi_t|$ is bounded. Note that
\begin{align*}
&\norm{\frac{1}{T} \sum_{t=1}^T \hat{\sigma}_{t+1}^2 \varpi_t \hat{f}_{t+1}^{d} \hat{f}_{t+1}^{d \top}  -   \frac{1}{T} \sum_{t=1}^T \sigma_{t+1}^2 \varpi_t H_{F} f^d_{t+1}f^{d \top}_{t+1} H_{F}^\top } \\
&\lesssim
\norm{\frac{1}{T} \sum_{t=1}^T (\hat{\sigma}_{t+1}^2 - \sigma_{t+1}^2) \varpi_t f^d_{t+1}f^{d \top}_{t+1}}
+ \norm{\frac{1}{T} \sum_{t=1}^T \sigma_{t+1}^2 \varpi_t \left(H_{F} f^d_{t+1}f^{d \top}_{t+1} H_{F}^\top -  \hat{f}_{t+1}^{d} \hat{f}_{t+1}^{d \top}   \right)} \\
& + \norm{\frac{1}{T} \sum_{t=1}^T (\hat{\sigma}_{t+1}^2 - \sigma_{t+1}^2) \varpi_t \left(H_{F} f^d_{t+1}f^{d \top}_{t+1} H_{F}^\top - \hat{f}_{t+1}^{d} \hat{f}_{t+1}^{d \top}   \right)} .
\end{align*}
By Lemma \ref{lem:variance_est_tech_debias}, the first term can be bounded like
\begin{align*}
 \norm{\frac{1}{T} \sum_{t=1}^T (\hat{\sigma}_{t+1}^2 - \sigma_{t+1}^2) \varpi_t f^d_{t+1}f^{d \top}_{t+1}} &\leq 
\max_t |\varpi_t| \left( \frac{1}{T} \sum_{t=1}^T (\hat{\sigma}_{t+1}^2 - \sigma_{t+1}^2)^2 \right)^{1/2} 
\left( \frac{1}{T} \sum_{t=1}^T \norm{f^d_{t+1}f^{d \top}_{t+1}}^2 \right)^{1/2}  \\
&= o_p(1).   
\end{align*}
In addition, the second term can be bound like 
\begin{align*}
&\norm{\frac{1}{T} \sum_{t=1}^T \sigma_{t+1}^2 \varpi_t \left(H_{F} f^d_{t+1}f^{d \top}_{t+1} H_{F}^\top -  \hat{f}_{t+1}^{d} \hat{f}_{t+1}^{d \top}   \right)} \\
&\ \ \lesssim \max_t\abs{\sigma_{t+1}^2 \varpi_t}  \frac{1}{T} \sum_{t=1}^T \norm{\hat{f}_{t+1}^{d} - H_{F} f^d_{t+1}}\norm{f^d_{t+1}} 
 = o_p(1).
\end{align*}
Moreover, the third term can be bounded like 
\begin{align*}
&\norm{\frac{1}{T} \sum_{t=1}^T (\hat{\sigma}_{t+1}^2 - \sigma_{t+1}^2) \varpi_t \left(H_{F} f^d_{t+1}f^{d \top}_{t+1} H_{F}^\top - \hat{f}_{t+1}^{d} \hat{f}_{t+1}^{d \top}   \right)} \\
&\leq \max_t\abs{ \varpi_t} \left( \frac{1}{T} \sum_{t=1}^T (\hat{\sigma}_{t+1}^2 - \sigma_{t+1}^2)^2 \right)^{1/2} \left( \frac{1}{T} \sum_{t=1}^T \norm{H_{F} f^d_{t+1}f^{d \top}_{t+1} H_{F}^\top - \hat{f}_{t+1}^{d} \hat{f}_{t+1}^{d \top}}^2 \right)^{1/2} = o_p(1)
\end{align*}
by Lemma \ref{lem:variance_est_tech_debias}. Here, $\frac{1}{T} \sum_{t=1}^T \norm{H_{F} f^d_{t+1}f^{d \top}_{t+1} H_{F}^\top - \hat{f}_{t+1}^{d} \hat{f}_{t+1}^{d \top}}^2 = o_p(1)$ because 
$$
\frac{1}{T} \sum_{t=1}^T \norm{H_{F} f^d_{t+1} - \hat{f}_{t+1}^{d} }^2\norm{f_{t+1}^d}^2 = o_p(1), \quad \frac{1}{T} \sum_{t=1}^T \norm{H_{F} f^d_{t+1} - \hat{f}_{t+1}^{d} }^4 = o_p(1) .
$$
Hence, we have 
\begin{align*}
\norm{\frac{1}{T} \sum_{t=1}^T \hat{\sigma}_{t+1}^2 \varpi_t \hat{f}_{t+1}^{d} \hat{f}_{t+1}^{d \top}  -   \frac{1}{T} \sum_{t=1}^T \sigma_{t+1}^2 \varpi_t H_{F} f^d_{t+1}f^{d \top}_{t+1} H_{F}^\top } = o_p(1).
\end{align*}
In addition, we have 
\begin{align*}
&\norm{\left( \frac{1}{T}\sum_t \hat{f}_{t+1}^d \hat{f}_{t+1}^{d \top}\right)^{-1} - \left( H_{F} \frac{1}{T} \sum_t f_{t+1}^{d} f_{t+1}^{d\top} H_{F}^\top \right)^{-1} } \\
&\leq
 \norm{\left( \frac{1}{T}\sum_t \hat{f}_{t+1}^d \hat{f}_{t+1}^{d \top}\right)^{-1}} \norm{\left( H_{F} \frac{1}{T} \sum_t f_{t+1}^{d} f_{t+1}^{d\top} H_{F}^\top \right)^{-1} } \\
&\ \ \times \norm{\frac{1}{T}\sum_t \hat{f}_{t+1}^d \hat{f}_{t+1}^{d \top} - H_{F} \frac{1}{T}\sum_t f_{t+1}^{d} f_{t+1}^{d\top} H_{F}^\top  } \\
& = o_p(1)
\end{align*}
as noted in the proof of Lemma \ref{lem:Gamma_bound}. Hence, we have
\begin{align*}
&\left( \frac{1}{T}\sum_t \hat{f}_{t+1}^d \hat{f}_{t+1}^{d \top}\right)^{-1} \frac{1}{T} \sum_{t=1}^T \hat{\sigma}_{t+1}^2 \varpi_t \hat{f}_{t+1}^{d} \hat{f}_{t+1}^{d \top} \left( \frac{1}{T}\sum_t \hat{f}_{t+1}^d \hat{f}_{t+1}^{d \top}\right)^{-1} \\
& \conP  \left( H_{F} \frac{1}{T} \sum_t f_{t+1}^{d} f_{t+1}^{d\top} H_{F}^\top \right)^{-1} \frac{1}{T} \sum_{t=1}^T \sigma_{t+1}^2 \varpi_t H_{F} f^d_{t+1}f^{d \top}_{t+1} H_{F}^\top  \left( H_{F} \frac{1}{T} \sum_t f_{t+1}^{d} f_{t+1}^{d\top} H_{F}^\top \right)^{-1} \\
& =  H_{F}^{- \top} \left( \frac{1}{T} \sum_t f_{t+1}^{d} f_{t+1}^{d\top}  \right)^{-1} \frac{1}{T} \sum_{t=1}^T \sigma_{t+1}^2 \varpi_t  f^d_{t+1}f^{d \top}_{t+1}   \left(  \frac{1}{T} \sum_t f_{t+1}^{d} f_{t+1}^{d\top} \right)^{-1}  H_{F}^{-1}.
\end{align*}
Since $H_{F}^{-1} \conP \calH$ by Lemma \ref{lem:rotation_matrix_tech}, we have the desired result.\\
(ii) Let $g_{js} = (a_s A - b_s B)x_{js}$ where $a_s = 1 - (\eta^\top Q_t \Gamma (Q_t^B)^{-1} + \bar{\breve{f}}^\top) (Q^f)^{-1} f_{s+1}^d$, $b_s = B_{it}^\top (Q_t^B)^{-1} (Q_f)^{-1} f_{s+1}^d$, $A = x_{it}^\top Q_t^{-1} - B_{it}^\top (Q_t^B)^{-1} \Gamma^\top $, and $B = (\eta^\top - \eta^\top Q_t \Gamma (Q_t^B)^{-1} \Gamma^\top$. Then, we can have
\begin{align*}
 \frac{1}{NTL} \sum_{j=1}^N \sum_{s=1}^T \sigma_{s+1}^2 g_{js}^2 &= \frac{1}{NTL} \sum_{j=1}^N \sum_{s=1}^T \sigma_{s+1}^2 (a_s A - b_s B) x_{js} x_{js}^\top (A^\top a_s - B^\top b_s)  \\
& = \frac{1}{TL} \sum_{s=1}^T \sigma_{s+1}^2 (a_s A - b_s B) Q_s (A^\top a_s - B^\top b_s).
\end{align*}
First, note that 
\begin{align}\label{eq:variance_est_1}
&\norm{ \frac{1}{TL} \sum_{s=1}^T\hat{\sigma}_{s+1}^2 \hat{a}_s^2 \hat{A} Q_s \hat{A}^\top - \frac{1}{TL} \sum_{s=1}^T\sigma_{s+1}^2 a_s^2 A Q_s A^\top }    \\
\nonumber&\lesssim \left( \frac{1}{T}\sum_{s=1}^T (\hat{\sigma}_{s+1}^2 - \sigma_{s+1}^2)^2 \right)^{1/2} \left( \frac{1}{TL^2}\sum_{s=1}^T \norm{a_s^2 A Q_s A^\top}^2 \right)^{1/2} \\
\nonumber& \ \ + \max_s \sigma_{s+1}^2  \frac{1}{TL} \sum_{s=1}^T \norm{ \hat{a}_s^2 \hat{A} Q_s \hat{A}^\top - a_s^2 A Q_s A^\top}\\
\nonumber& \ \ +\left( \frac{1}{T}\sum_{s=1}^T (\hat{\sigma}_{s+1}^2 - \sigma_{s+1}^2)^2 \right)^{1/2} \left( \frac{1}{TL^2}\sum_{s=1}^T \norm{ \hat{a}_s^2 \hat{A} Q_s \hat{A}^\top - a_s^2 A Q_s A^\top}^2 \right)^{1/2},
\end{align}
where $\hat{a}_s = 1 - (\hat{\eta}^\top Q_t \hat{\Gamma} (\hat{\Gamma}^\top Q_t \hat{\Gamma} )^{-1} + (\hat{\bar{\breve{f}}})^\top ) ( \frac{1}{T} \sum_{u=1}^T \hat{f}_{u+1}^d \hat{f}_{u+1}^{d \top})^{-1} \hat{f}_{s+1}^d$ and $\hat{A} = x_{it}^\top Q_t^{-1} - x_{it}^\top \hat{\Gamma} (\hat{\Gamma}^\top Q_t \hat{\Gamma} )^{-1} \hat{\Gamma}^\top $. Note that
\begin{align*}
\frac{1}{TL^2}\sum_{s=1}^T \norm{a_s^2 A Q_s A^\top}^2 
\leq \max_{s} \norm{Q_s}^2 \left(\frac{\norm{A}}{\sqrt{L}}\right)^4 \frac{1}{T} \sum_{s=1}^T a_s^2 = O_p(1)
\end{align*}
since $\frac{1}{T} \sum_{s=1}^T || f_{s+1}^d ||^2 = O_p(1)$. Hence, the first term converges to $0$ by Lemma \ref{lem:variance_est_tech_debias}. In addition, note that
\begin{align*}
&\frac{1}{TL^2}\sum_{s=1}^T \norm{ \hat{a}_s^2 \hat{A} Q_s \hat{A}^\top - a_s^2 A Q_s A^\top}^2 \\
&\lesssim \max_{s} \norm{Q_s}^2 \left(\frac{\norm{A}}{\sqrt{L}}\right)^4 \frac{1}{T}\sum_{s=1}^T \norm{\hat{a}_s^2 -  a_s^2}^2 
+ \max_{s} \norm{Q_s}^2 \left(\frac{\norm{\hat{A} - A}}{\sqrt{L}}\right)^2 \left(\frac{\norm{A}}{\sqrt{L}}\right)^2 \frac{1}{T}\sum_{s=1}^T a_s^4 .
\end{align*}
By using similar bounds as in Lemmas \ref{lem:gamma_f_final_bound}, \ref{lem:fhat}, and \ref{lem:etahat}, we have
\begin{align*}
\frac{1}{T}\sum_{s=1}^T \norm{\hat{a}_s^2 -  a_s^2}^2 \lesssim \frac{1}{T}\sum_{s=1}^T \norm{\hat{a}_s -  a_s}^2 \norm{a_s}^2 + \frac{1}{T}\sum_{s=1}^T \norm{\hat{a}_s -  a_s}^4 = o_p(1),
\end{align*}
because 
$$
\frac{1}{T} \sum_{t=1}^T \norm{\hat{\breve{f}}_{t+1} - H_{\Gamma}^{-1} \breve{f}_{t+1}}^2\norm{f_{t+1}^d}^2 = o_p(1), \quad \frac{1}{T} \sum_{t=1}^T \norm{\hat{\breve{f}}_{t+1} - H_{\Gamma}^{-1} \breve{f}_{t+1}}^4 = o_p(1) .
$$
Moreover, we can easily check that $||\hat{A} - A||/\sqrt{L} = o_p(1)$ by Lemmas \ref{lem:gamma_f_final_bound} and \ref{lem:alpha_I_1_tech}. Hence, we have 
$$
\frac{1}{TL^2}\sum_{s=1}^T \norm{ \hat{a}_s^2 \hat{A} Q_s \hat{A}^\top - a_s^2 A Q_s A^\top}^2  = o_p(1)
$$
and, similarly, we can show $ \frac{1}{TL} \sum_{s=1}^T \norm{ \hat{a}_s^2 \hat{A} Q_s \hat{A}^\top - a_s^2 A Q_s A^\top} = o_p(1)$. Then, with the aid of Lemma \ref{lem:variance_est_tech_debias}, we can show that the second third terms of \eqref{eq:variance_est_1} are $o_p(1)$. Using a similar way, we can also show that 
\begin{gather*}
\norm{ \frac{1}{TL} \sum_{s=1}^T\hat{\sigma}_{s+1}^2 \hat{a}_s \hat{b}_s \hat{A} Q_s \hat{B}^\top - \frac{1}{TL} \sum_{s=1}^T\sigma_{s+1}^2 a_s b_s A Q_s B^\top }  =o_p(1), \\
\norm{ \frac{1}{TL} \sum_{s=1}^T\hat{\sigma}_{s+1}^2 \hat{b}_s^2  \hat{B} Q_s \hat{B}^\top - \frac{1}{TL} \sum_{s=1}^T\sigma_{s+1}^2 b_s^2 B Q_s B^\top }  =o_p(1).
\end{gather*}
(iii) By Lemma \ref{lem:variance_est_tech_debias}, we have
$$
\frac{1}{T}\sum_{s=1}^T \hat{\sigma}^2_{s+1} - \frac{1}{T}\sum_{s=1}^T \sigma^2_{s+1}  = o_p(1).
$$
(iv) By Lemma \ref{lem:variance_est_tech_debias}, we have $\hat{\sigma}^2_{t+1} - \sigma^2_{t+1}  = o_p(1)$. $\square$

\subsection{Auxiliary lemmas}\label{sec:aux_lemma}

 Let $U_K D_K V_K^\top$ be the top-$K$ singular value decomposition of $\ddot{R}^d$.

\begin{lemma}[naive bound of $\tilde{\Gamma}$]\label{lem:naivegamma}
We have
$$
\norm{ \tilde{\Gamma} - \Gamma H_\Gamma }_F = O_p\left( \frac{\sqrt{L}}{\sqrt{N}} \right) .
$$
\end{lemma}

\noindent\textbf{Proof of Lemma \ref{lem:naivegamma}.} By the equation \eqref{eq:gamma_decomposition_naive}, we have 
\begin{align*}
\tilde{\Gamma} - \Gamma H_\Gamma = \ddot{E}^d F^{d \top}H_F^\top(\tilde{F}^d\tilde{F}^{d \top})^{-1}
+\ddot{E}^d\ddot{E}^{d \top}\tilde{\Gamma}(\tilde{F}^d\tilde{F}^{d \top})^{-1},
\end{align*}
since $\tilde{\Gamma}^\top \tilde{\Gamma} = U_K^\top U_K = I_K$. Note that $\norm{H_F} = \norm{\tilde{\Gamma}^\top \Gamma} \leq \norm{\Gamma}$ is bounded and w.h.p.,
\begin{align*}
\sqrt{\psi_{K} \left( \tilde{F}^d\tilde{F}^{d \top} \right)} = \psi_{K} (D_K) \geq \psi_{K} (\Gamma F^d) - \norm{\ddot{E}^d} \geq c \sqrt{T}
\end{align*}
for some constant $c>0$ since we have
$\norm{\ddot{R}^d  - \Gamma \calF^d} =  \norm{\ddot{E}^d} = O_p \left( \frac{\sqrt{LT}}{\sqrt{N}} \right)$ by Lemma \ref{lem:error_tech}. Hence, we have
\begin{align*}
\norm{\tilde{\Gamma} - \Gamma H_\Gamma} \leq \norm{\ddot{E}^d F^{d \top}H_F^\top(\tilde{F}^d\tilde{F}^{d \top})^{-1}} + \norm{\ddot{E}^d\ddot{E}^{d \top}\tilde{\Gamma}(\tilde{F}^d\tilde{F}^{d \top})^{-1}} = O_p\left( \frac{\sqrt{L}}{\sqrt{N}} + \frac{L}{N} \right).
\end{align*}
Because $K$ is finite, we have the desired result. $\square$
\smallskip

\begin{lemma}\label{lem:fd_bound}
Assume that $\norm{ \tilde{\Gamma} - \Gamma H_{\Gamma} }_F = O_p\left( b_{NT} \right)$ for some sequence $b_{NT} \rightarrow 0$. Then, we have
\begin{gather*}
\norm{ \tilde{f}_{t+1}^{d} - H_{F} f^d_{t+1}} = O_p\left( \frac{\sqrt{L}}{\sqrt{N}} b_{NT} + \frac{1}{\sqrt{N}} \right) ,\ \ 
\frac{1}{T}\sum_{t=1}^T \norm{ \tilde{f}_{t+1}^d - H_{F} f^d_{t+1}} = O_p\left( \frac{\sqrt{L}}{\sqrt{N}} b_{NT} + \frac{1}{\sqrt{N}} \right),\\
\frac{1}{T}\sum_{t=1}^T \norm{ \tilde{f}_{t+1}^d - H_{F} f^d_{t+1}}^2 = O_p\left( \left(\frac{\sqrt{L}}{\sqrt{N}} b_{NT} + \frac{1}{\sqrt{N}} \right)^2 \right).
\end{gather*}
\end{lemma}

\noindent\textbf{Proof of Lemma \ref{lem:fd_bound}.}
We have
\begin{equation}\label{eq:fd_expansion}
\tilde{f}^{d}_{t+1} - H_{F} f^d_{t+1}  = \left(\tilde{\Gamma} - \Gamma H_{\Gamma} \right)^\top   \ddot{E}_{t+1}^d +  H_{\Gamma}^\top \Gamma^{\top}   \ddot{E}_{t+1}^d .
\end{equation}
By Lemma \ref{lem:error_tech}, the order of the first term in \eqref{eq:fd_expansion} is $O_p\left(\frac{\sqrt{L}}{\sqrt{N}} b_{NT}\right)$ and that of the second term is $O_p\left(\frac{1}{\sqrt{N}}\right)$. Here, we use the fact that
$$
\norm{H_\Gamma} = \norm{F^d \tilde{F}^{d \top} (\tilde{F}^d\tilde{F}^{d \top})^{-1}} \leq \norm{F^d} \norm{V_K} \norm{D_K^{-1}} = O_p(1) .
$$
In addition, note that
\begin{align*}
\frac{1}{T} \sum_{t=1}^T \norm{\tilde{f}^{d}_{t+1} - H_{F}^{-1} f^d_{t+1}}  \leq  \norm{ \tilde{\Gamma} - \Gamma H_{\Gamma} }  \frac{1}{T} \sum_{t=1}^T \norm{\ddot{E}_{t+1}^d} +  \norm{ H_{\Gamma} } \frac{1}{T} \sum_{t=1}^T  \norm{\Gamma^{\top}   \ddot{E}_{t+1}^d}.
\end{align*}
Hence, using the same method as above with Lemma \ref{lem:error_tech}, we can have $\frac{1}{T}\sum_{t=1}^T \norm{ \tilde{f}_{t+1}^d - H_{F} f^d_{t+1}} = O_p\left( \frac{\sqrt{L}}{\sqrt{N}} b_{NT} + \frac{1}{\sqrt{N}} \right)$. In addition, by Lemma \ref{lem:error_tech}, we have
$$
\bbE\left[ \left. \frac{1}{T} \sum_{t=1}^T \norm{\ddot{E}_{t+1}^d}^2 \right| X \right] = O_p\left(\frac{L}{N} \right) , \quad \bbE\left[ \left.  \frac{1}{T} \sum_{t=1}^T  \norm{\Gamma^{\top}   \ddot{E}_{t+1}^d}^2 \right| X \right] = O_p\left(\frac{1}{N} \right) .
$$
Using this result, we have $\frac{1}{T}\sum_{t=1}^T \norm{ \tilde{f}_{t+1}^d - H_{F} f^d_{t+1}}^2 = O_p\left( \left(\frac{\sqrt{L}}{\sqrt{N}} b_{NT} + \frac{1}{\sqrt{N}} \right)^2 \right)$. $\square$
\smallskip

\begin{lemma}\label{lem:Gamma_bound}
Assume that there is a sequence $a_{NT} \rightarrow 0$ such that
\begin{gather*}
\norm{ \tilde{f}_{t+1}^d - H_{F} f^d_{t+1}} = O_p\left( a_{NT} \right) ,\ \ 
\frac{1}{T}\sum_{t=1}^T \norm{ \tilde{f}_{t+1}^d - H_{F} f^d_{t+1}} = O_p\left( a_{NT} \right),\\
\frac{1}{T}\sum_{t=1}^T \norm{ \tilde{f}_{t+1}^d - H_{F} f^d_{t+1}}^2 = O_p\left( a_{NT}^2 \right)
\end{gather*}
Then, we have
$$
\norm{ \tilde{\Gamma} - \Gamma H_{\Gamma} }_F =  O_p\left( \frac{\sqrt{L}}{\sqrt{N}} a_{NT} + \frac{\sqrt{L}}{\sqrt{NT}} \right).
$$
\end{lemma}

\noindent\textbf{Proof of Lemma \ref{lem:Gamma_bound}.}
We have
\begin{align}\label{eq:gamma_expansion}
\nonumber\vect\left( \tilde{\Gamma}^{ \top} \right) - \vect\left( H_{\Gamma}^{\top} \Gamma^\top \right) & = \left( I_L \otimes  \left(\frac{1}{T}\sum_t \tilde{f}_{t+1}^d \tilde{f}_{t+1}^{d \top} \right)^{-1}  \right)
\frac{1}{T} \sum_{t=1}^T \ddot{E}_{t+1}^d \otimes  \left(\tilde{f}_{t+1}^d -  H_{F} f_{t+1}^{d} \right)   \\
& \ \ +\left( I_L \otimes  \left(\frac{1}{T}\sum_t \tilde{f}_{t+1}^d \tilde{f}_{t+1}^{d \top} \right)^{-1}  \right)
\frac{1}{T} \sum_{t=1}^T \ddot{E}_{t+1}^d \otimes \left( H_{F} f_{t+1}^{d} \right).
\end{align}
First, Note that 
\begin{align*}
 \psi_{\min} \left( \frac{1}{T}\sum_t \tilde{f}_{t+1}^d \tilde{f}_{t+1}^{d \top}\right) &\geq 
 \psi_{\min} \left( H_{F} \frac{1}{T}\sum_t f_{t+1}^{d} f_{t+1}^{d\top} H_{F}^{\top}  \right) \\
 &- \norm{ \frac{1}{T}\sum_t \tilde{f}_{t+1}^d \tilde{f}_{t+1}^{d \top} - H_{F} \frac{1}{T}\sum_t f_{t+1}^{d} f_{t+1}^{d\top} H_{F}^\top  } \\
 &\geq c_1   , 
\end{align*}
for some constant $c_1>0$, w.p.c. to 1, because
\begin{align*}
&\norm{\frac{1}{T}\sum_t \tilde{f}_{t+1}^d \tilde{f}_{t+1}^{d \top} - H_{F} \frac{1}{T}\sum_t f_{t+1}^{d} f_{t+1}^{d\top} H_{F}^\top  }\\
&\lesssim \norm{H_{F} \frac{1}{T}\sum_t f_{t+1}^d (\tilde{f}_{t+1}^d - H_{F} f_{t+1}^{d} )^{\top}}\\
&\leq \norm{H_{F}} \left( \frac{1}{T}\sum_t \norm{f_{t+1}^d}^2 \right)^{1/2}\left( \frac{1}{T}\sum_t \norm{\tilde{f}_{t+1}^d - H_{F} f_{t+1}^{d}}^2 \right)^{1/2} = O_p(a_{NT}) = o_p(1),
\end{align*}
and $\psi_{\min} \left( H_{F} \frac{1}{T}\sum_t f_{t+1}^{d} f_{t+1}^{d\top} H_{F}^{\top}   \right) \geq \psi_{\min}^2 \left( H_{F} \right) \psi_{\min} \left( \frac{1}{T}\sum_t f_{t+1}^{d} f_{t+1}^{d\top} \right) \geq c_2$ for some constant $c_2 >0$ w.p.c. to 1. Here, $\psi_{\min} \left( H_{F} \right) = \psi_{\min} \left( \tilde{\Gamma}^\top \Gamma   \right) > c$ comes from a typical assertion such as Proposition 1 of \cite{bai2003inferential}. Hence, $\norm{\left(\frac{1}{T}\sum_t \tilde{f}_{t+1}^d \tilde{f}_{t+1}^{d \top} \right)^{-1} } = O_p(1)$. In addition, we have $\frac{1}{T} \sum_{t=1}^T \ddot{E}_{t+1}^d \otimes  \left(\tilde{f}_{t+1}^d -  H_{F} f_{t+1}^{d} \right) = \frac{1}{T} \sum_{t=1}^T \ddot{E}_{t+1} \otimes  \left(\tilde{f}_{t+1}^d -  H_{F} f_{t+1}^{d} \right)$ and 
\begin{align*}
\norm{ \frac{1}{T} \sum_{t=1}^T \ddot{E}_{t+1} \otimes  \left(\tilde{f}_{t+1}^d -  H_{F} f_{t+1}^{d} \right)}
&\leq \left(\frac{1}{T} \sum_{t=1}^T \norm{\ddot{E}_{t+1}}^2 \right)^{1/2} \left(\frac{1}{T} \sum_{t=1}^T \norm{\tilde{f}_{t+1}^d -  H_{F} f_{t+1}^{d} }^2 \right)^{1/2} \\
& = O_p\left( \frac{\sqrt{L}}{\sqrt{N}} a_{NT} \right),
\end{align*}
by Lemma \ref{lem:error_tech}. Moreover, $\frac{1}{T} \sum_{t=1}^T \ddot{E}_{t+1}^d \otimes \left( H_{F} f_{t+1}^{d} \right) = \left(I_L \otimes H_{F} \right) \frac{1}{T}\sum_{t=1}^T \ddot{E}_{t+1} \otimes  f_{t+1}^{d} $ and $\norm{\frac{1}{T}\sum_{t=1}^T \ddot{E}_{t+1} \otimes  f_{t+1}^{d} }= O_p\left( \frac{\sqrt{L}}{\sqrt{NT}} \right)$ because 
$$
\frac{1}{T}\sum_{t=1}^T \ddot{E}_{t+1} \otimes  f_{t+1}^{d} = \frac{1}{NT} \sum_{it} (Q_t^{-1} x_{it} \otimes f_{t+1}^{d}) \epsilon_{i,t+1}  = \frac{1}{NT} A^\top \vect(E)
$$ 
where
$A$ is the $NT \times LK$ matrix of $Q_t^{-1} x_{it} \otimes f_{t+1}^{d}$ and
$$
\bbE\norm{\frac{1}{NT}A^\top \vect(E)}_F^2 \lesssim \frac{1}{N^2T^2} \norm{\bbE[\vect(E)\vect(E)^\top]} \norm{A}_F^2 = O_p\left( \frac{L}{NT} \right)
$$
since $\norm{A}_F^2 \leq \max_t ||Q_t^{-1}||^2 \sum_t ||f_{t+1}^{d}||^2 \sum_i ||x_{it}||^2 = O_p(NTL)$. Then, applying these bounds to \eqref{eq:gamma_expansion}, we have 
$$
\norm{ \tilde{\Gamma} - \Gamma H_{\Gamma} }_F = \norm{\vect\left( \tilde{\Gamma}^{ \top} \right) - \vect\left( H_{\Gamma}^{\top} \Gamma^\top \right)} = O_p\left( \frac{\sqrt{L}}{\sqrt{N}} a_{NT} + \frac{\sqrt{L}}{\sqrt{NT}} \right). \ \ \square
$$

\begin{lemma}\label{lem:gamma_f_final_bound}
We have
\begin{gather*}
\norm{ \tilde{f}_{t+1}^{d} - H_{F} f^d_{t+1}} , \ \ \frac{1}{T}\sum_{t=1}^T \norm{ \tilde{f}_{t+1}^{d} - H_{F} f^d_{t+1}} = O_p\left( \left( \frac{L}{N} \right)^{10} + \frac{L}{N\sqrt{T}} + \frac{1}{\sqrt{N}} \right),\\
\frac{1}{T}\sum_{t=1}^T \norm{ \tilde{f}_{t+1}^{d} - H_{F} f^d_{t+1}}^2 = O_p\left( \left( \left( \frac{L}{N} \right)^{10} + \frac{L}{N\sqrt{T}} + \frac{1}{\sqrt{N}} \right)^2 \right),\\
\norm{\tilde{\Gamma} - \Gamma H_{\Gamma}}_F = O_p\left(  \frac{\sqrt{L}}{\sqrt{NT}} + \left( \frac{L}{N}  \right)^{10 + \frac{1}{2} } + \frac{\sqrt{L}}{N} \right).
\end{gather*}
\end{lemma}

\noindent\textbf{Proof of Lemma \ref{lem:gamma_f_final_bound}.}
Starting from the naive bound in Lemma \ref{lem:naivegamma}, by applying Lemmas \ref{lem:fd_bound} and \ref{lem:Gamma_bound} recursively, we can derive sharper and sharper bounds. But the number of recursions should be bounded. We apply Lemmas \ref{lem:fd_bound} and \ref{lem:Gamma_bound} 10 times here. $\square$

\begin{lemma}\label{lem:fhat}
Let $\tilde{\breve{f}}_{t+1} = (\tilde{\Gamma}^\top \tilde{\Gamma})^{-1} \tilde{\Gamma}^\top \ddot{R}_{t+1}$. We have
\begin{gather*}
\norm{\tilde{\breve{f}}_{t+1} - H_{\Gamma}^{-1} \breve{f}_{t+1}}  = O_p \left( \frac{1}{\sqrt{N}} + \frac{\sqrt{L}}{\sqrt{NT}} + \left( \frac{L}{N}  \right)^{10 + \frac{1}{2} } \right),\\
\frac{1}{T} \sum_{t=1}^T \norm{\tilde{\breve{f}}_{t+1} - H_{\Gamma}^{-1} \breve{f}_{t+1}}  = O_p \left( \frac{1}{\sqrt{N}} + \frac{\sqrt{L}}{\sqrt{NT}} + \left( \frac{L}{N}  \right)^{10 + \frac{1}{2} } \right),\\
\frac{1}{T} \sum_{t=1}^T \norm{\tilde{\breve{f}}_{t+1} - H_{\Gamma}^{-1} \breve{f}_{t+1}}^2  = O_p \left( \left[\frac{1}{\sqrt{N}} + \frac{\sqrt{L}}{\sqrt{NT}} + \left( \frac{L}{N}  \right)^{10 + \frac{1}{2} } \right]^2 \right).
\end{gather*}
\end{lemma}

\noindent\textbf{Proof of Lemma \ref{lem:fhat}.} Note that
$$
\tilde{\breve{f}}_{t+1} - H_{\Gamma}^{-1} \breve{f}_{t+1}  =  (\tilde{\Gamma} - \Gamma H_{\Gamma})^\top  \eta  - \tilde{\Gamma}^\top (\tilde{\Gamma} - \Gamma H_{\Gamma}) H_{\Gamma}^{-1}\breve{f}_{t+1}   + \tilde{\Gamma}^\top \ddot{E}_{t+1}.
$$
We have $\norm{\tilde{\Gamma} - \Gamma H_{\Gamma}}_F = O_p\left(  \frac{\sqrt{L}}{\sqrt{NT}} + \left( \frac{L}{N}  \right)^{10 + \frac{1}{2} } + \frac{\sqrt{L}}{N} \right)$ by Lemma \ref{lem:gamma_f_final_bound}. In addition, using the same token as in the proof of Lemma \ref{lem:fd_bound}, we have
$$
 \tilde{\Gamma}^\top \ddot{E}_{t+1} = O_p\left( \frac{\sqrt{L}}{\sqrt{N}} \left(  \frac{\sqrt{L}}{\sqrt{NT}} + \left( \frac{L}{N}  \right)^{10 + \frac{1}{2} } + \frac{\sqrt{L}}{N} \right) + \frac{1}{\sqrt{N}} \right).
$$
Hence, we have 
$$
\norm{\tilde{\breve{f}}_{t+1} - H_{\Gamma}^{-1} \breve{f}_{t+1}}  = O_p \left( \frac{1}{\sqrt{N}} + \frac{\sqrt{L}}{\sqrt{NT}} + \left( \frac{L}{N}  \right)^{10 + \frac{1}{2} } \right).
$$
Similarly, we can show the other bounds. $\square$
\smallskip

\begin{lemma}\label{lem:etahat}
We have (i) $\norm{\tilde{\eta} - \eta} = O_p\left( \frac{\sqrt{L}}{\sqrt{NT}} + \left( \frac{L}{N}\right)^{10 + \frac{1}{2}} + \frac{\sqrt{L}}{N}   \right)$; (ii) $\norm{x_{it}^\top \tilde{\eta} - x_{it}^\top \eta} = O_p\left( \frac{\sqrt{L}}{\sqrt{NT}} + \left( \frac{L}{N}\right)^{10 + \frac{1}{2}} + \frac{\sqrt{L}}{N} + \frac{L^{3/2}}{N^{3/2} \sqrt{T}}  \right)$.
\end{lemma}

\noindent\textbf{Proof of Lemma \ref{lem:etahat}.}
(i) Note that
$$
\tilde{\eta} - \eta =  \left(P_\Gamma - P_{\tilde{\Gamma}} \right) \left( \eta + \Gamma \bar{\breve{f}} + \bar{\ddot{E}} \right) + \left( I_L - P_\Gamma \right)\bar{\ddot{E}} 
$$
where $P_\Gamma = \Gamma \left( \Gamma^\top \Gamma\right)^{-1}\Gamma^\top$ and $P_{\tilde{\Gamma}} = \tilde{\Gamma} \left( \tilde{\Gamma}^\top \tilde{\Gamma}\right)^{-1} \tilde{\Gamma}^\top$. From Lemma \ref{lem:gamma_f_final_bound}, we have $\norm{\tilde{\Gamma} - H_{\Gamma}\Gamma}_F = O_p\left( \frac{\sqrt{L}}{\sqrt{NT}} + \left( \frac{L}{N}\right)^{ 10 + \frac{1}{2}} + \frac{\sqrt{L}}{N}   \right)$. So, a simple calculation using Lemma \ref{lem:projection_tech} shows that $\norm{P_\Gamma - P_{\tilde{\Gamma}}} =O_p\left( \frac{\sqrt{L}}{\sqrt{NT}} + \left( \frac{L}{N}\right)^{10 + \frac{1}{2}} + \frac{\sqrt{L}}{N}   \right)$. In addition, we have $\norm{\bar{\ddot{E}} } = \norm{\frac{1}{T} \sum_{t=1}^T \ddot{E}_{t+1}} = O_p\left( \frac{\sqrt{L}}{\sqrt{NT}}\right) $ by Lemma \ref{lem:error_tech}. Then, since $\norm{P_\Gamma} \leq 1$, we have
$$
\norm{\tilde{\eta} - \eta} = O_p\left( \frac{\sqrt{L}}{\sqrt{NT}} + \left( \frac{L}{N}\right)^{10 + \frac{1}{2}} + \frac{\sqrt{L}}{N}   \right).
$$
(ii) The proof is similar to (i). Here, we use Lemma \ref{lem:alpha_I_1_tech} (i). $\square$
\smallskip

\begin{comment}

\begin{lemma}\label{lem:delta}
We have
$$
\norm{\hat{\delta} - \delta} = O_p\left( \frac{1}{\sqrt{T}} \right).
$$
\end{lemma}

\noindent\textbf{Proof of Lemma \ref{lem:delta}.}
Note that
$$
\hat{\delta} - \delta = \frac{1}{NT} \sum_{s=1}^T B_s^{o\top} E_{s+1} = \frac{1}{NT} \sum_{s=1}^T \sum_{j=1}^N B_{s,j}^{o} \epsilon_{j,s+1}
$$
where $B_{s,j}^{o} = B_{s}^{o\top}e_j$. In addition, we have
\begin{align*}
\bbE\left[ \left.  \norm{ \sum_{s=1}^T \sum_{j=1}^N B_{s,j}^{o} \epsilon_{j,s+1}}^2  \right| X \right] &= \sum_{q=1}^{N-L} \bbE\left[ \left.  \left( \sum_{s=1}^T \sum_{j=1}^N B_{s,jq}^{o} \epsilon_{j,s+1} \right)^2  \right| X \right] \\
& =  \sum_{q=1}^{N-L} \calB_q^\top  \bbE\left[ \left. \vect(E) \vect(E)^\top  \right| X \right] \calB_q \\
& \leq \sum_{q=1}^{N-L} \norm{\calB_q}^2 \norm{\bbE\left[ \left. \vect(E) \vect(E)^\top  \right| X \right]} \\
&= \sum_{s=1}^T \norm{B_s^o}_F^2 \norm{\bbE\left[ \left. \vect(E) \vect(E)^\top  \right| X \right]} \\
& \lesssim N(N-L)T
\end{align*}
since $\norm{B_s^o}_F^2 = N(N-L)$ by Lemma \ref{lem:B_tech} and $\norm{\bbE\left[ \left. \vect(E) \vect(E)^\top  \right| X \right]}$ is bounded. Hence, we have 
$$
\norm{\hat{\delta} - \delta} = O_p\left( \frac{1}{\sqrt{T}} \right). \ \ \square
$$

\end{comment}

\begin{lemma}\label{lem:variance_est_tech}
Let $\hat{\sigma}^2_{t+1} = \frac{1}{N} \sum_{i=1}^N \hat{\epsilon}_{i,t+1}^2$, where $m_{i,t+1} = \alpha_{O,it} + x_{it}^\top \eta + x_{it}^\top \Gamma \breve{f}_{t+1}$, $\tilde{m}_{i,t+1}= \hat{\alpha}_{O,it} + x_{it}^\top \tilde{\eta} + x_{it}^\top \tilde{\Gamma} \tilde{\breve{f}}_{t+1}$, and $\hat{\epsilon}_{i,t+1} = r_{i,t+1} - \tilde{m}_{i,t+1}$. Then, we have (i) $\abs{\hat{\sigma}^2_{t+1} - \sigma^2_{t+1}} = o_p(1)$, (ii) $\frac{1}{T}\sum_{t=1}^T \abs{\hat{\sigma}^2_{t+1} - \sigma^2_{t+1}} = o_p(1)$, (iii) $\frac{1}{T}\sum_{t=1}^T \abs{\hat{\sigma}^2_{t+1} - \sigma^2_{t+1}}^2 = o_p(1)$.
\end{lemma}

\noindent\textbf{Proof of Lemma \ref{lem:variance_est_tech}.}
(i) Note that
\begin{align*}
\hat{\sigma}_{t+1}^2 - \bbE[\epsilon_{j,t+1}^2] &= \frac{1}{N}\sum_{j=1}^N \left( \epsilon_{j,t+1}^2 - \bbE[\epsilon_{j,t+1}^2] \right) + \frac{1}{N}\sum_{j=1}^N \left(\hat{\epsilon}_{j,t+1}^2 - \epsilon_{j,t+1}^2 \right) \\
&=  \frac{1}{N}\sum_{j=1}^N \left( \epsilon_{j,t+1}^2 - \bbE[\epsilon_{j,t+1}^2] \right) + 2 \frac{1}{N}\sum_{j=1}^N \epsilon_{j,t+1} \left( \tilde{m}_{j,t+1} - m_{j,t+1} \right) \\
& \ \ + \frac{1}{N}\sum_{j=1}^N \left( \tilde{m}_{j,t+1} - m_{j,t+1} \right)^2  .
\end{align*}
The first term is $o_p(1)$ by the concentration inequality. For the second term, we can derive
$$
\frac{1}{N}\sum_{j=1}^N \epsilon_{j,t+1} \left( \hat{\alpha}_{O,jt} - \alpha_{O,jt} \right) = o_p(1) 
$$
by the same token as in the proof of Lemma \ref{lem:debiasing_tech}. In addition, we have
$$
\frac{1}{N}\sum_{j=1}^N \epsilon_{j,t+1} \left( x_{jt}^\top \eta - x_{jt}^\top \tilde{\eta} \right) = \frac{1}{N}\sum_{j=1}^N \epsilon_{j,t+1} x_{jt}^\top \left( \eta - \tilde{\eta}\right) = o_p(1)
$$
by Lemma \ref{lem:etahat} since $\norm{\frac{1}{N}\sum_{j=1}^N \epsilon_{j,t+1} x_{jt}^{\top }} = O_p(\frac{\sqrt{L}}{\sqrt{N}})$. Moreover, we have
$$
\frac{1}{N} \sum_{j=1}^N \epsilon_{j,t+1} x_{it}^\top (\tilde{\Gamma} - \Gamma H_{\Gamma}) H_{\Gamma}^{-1} \breve{f}_{t+1} = o_p(1), \quad 
\frac{1}{N} \sum_{j=1}^N \epsilon_{j,t+1} B_{jt}^\top H_{\Gamma} ( \tilde{\breve{f}}_{t+1} - H_{\Gamma}^{-1} \breve{f}_{t+1}) = o_p(1)
$$
by Lemmas \ref{lem:gamma_f_final_bound} and \ref{lem:fhat}. Hence, we have $\frac{1}{N}\sum_{j=1}^N \epsilon_{j,t+1} \left( \tilde{m}_{j,t+1} - m_{j,t+1} \right) = o_p(1)$. For the third term, we have $\frac{1}{N}\sum_{j=1}^N \left( \hat{\alpha}_{O,jt} - \alpha_{O,jt} \right)^2 = o_p(1)$ by using the bound of $\bbE \left[ \left. (\hat{\alpha}_{O,jt} - \alpha_{O,jt})^2  \right| X \right]$ as in the proof of Lemma \ref{lem:debiasing_tech}. In addition, we have
$$
\frac{1}{N}\sum_{j=1}^N \left( x_{jt}^\top \eta - x_{jt}^\top \tilde{\eta} \right)^2 =  \left( \eta - \tilde{\eta}\right)^\top Q_t \left( \eta - \tilde{\eta}\right) = o_p(1).
$$
Moreover, we have
$$
\frac{1}{N} \sum_{j=1}^N  (x_{it}^\top (\tilde{\Gamma} - \Gamma H_{\Gamma}) H_{\Gamma}^{-1} \breve{f}_{t+1})^2 = \breve{f}_{t+1}^\top H_{\Gamma}^{-\top} (\tilde{\Gamma} - \Gamma H_{\Gamma})^\top Q_t (\tilde{\Gamma} - \Gamma H_{\Gamma}) H_{\Gamma}^{-1} \breve{f}_{t+1} = o_p(1).
$$
Lastly, we have 
$$
\frac{1}{N} \sum_{j=1}^N (B_{jt}^\top H_{\Gamma} ( \tilde{\breve{f}}_{t+1} - H_{\Gamma}^{-1} \breve{f}_{t+1}))^2 = ( \tilde{\breve{f}}_{t+1} - H_{\Gamma}^{-1} \breve{f}_{t+1})^\top H_{\Gamma}^{\top} Q_B H_{\Gamma} ( \tilde{\breve{f}}_{t+1} - H_{\Gamma}^{-1} \breve{f}_{t+1}) = o_p(1).
$$
Hence, we have $\frac{1}{N}\sum_{j=1}^N \left( \tilde{m}_{j,t+1} - m_{j,t+1} \right)^2 = o_p(1)$. \\
(ii) The proof is similar to that of (iii).\\
(iii) First, note that
$$
\frac{1}{T} \sum_{t=1 }^T \norm{\frac{1}{N}\sum_{j=1}^N \left( \epsilon_{j,t+1}^2 - \bbE[\epsilon_{j,t+1}^2] \right)}^2 = o_p(1)
$$
because $\bbE \norm{\frac{1}{N}\sum_{j=1}^N \left( \epsilon_{j,t+1}^2 - \bbE[\epsilon_{j,t+1}^2] \right)}^2 = O \left( \frac{1}{N}\right)$. In addition, we can show that
$$
\frac{1}{T} \sum_{t=1}^T \norm{\frac{1}{N}\sum_{j=1}^N \epsilon_{j,t+1} \left( \hat{\alpha}_{O,jt} - \alpha_{O,jt} \right)}^2 = o_p(1)
$$
by using the similar method as in the proof of Lemma \ref{lem:debiasing_tech}. Moreover, we have 
$$
\frac{1}{T} \sum_{t=1}^T \norm{\frac{1}{N}\sum_{j=1}^N \epsilon_{j,t+1} \left( x_{jt}^\top \eta - x_{jt}^\top \tilde{\eta} \right)}^2 \leq  \frac{1}{T} \sum_{t=1}^T \norm{\frac{1}{N}\sum_{j=1}^N \epsilon_{j,t+1} x_{jt}}^2 \norm{ \eta - \tilde{\eta} }^2 = o_p(1)
$$
since $\bbE[||\frac{1}{N}\sum_{j=1}^N \epsilon_{j,t+1} x_{jt}||^2|X] \lesssim \frac{L}{N}$. In addition, we can bound
$$
\frac{1}{T} \sum_{t=1}^T \norm{\frac{1}{N} \sum_{j=1}^N \epsilon_{j,t+1} x_{it}^\top (\tilde{\Gamma} - \Gamma H_{\Gamma}) H_{\Gamma}^{-1} \breve{f}_{t+1}}^2 \leq \frac{1}{T} \sum_{t=1}^T \norm{\frac{1}{N}\sum_{j=1}^N \epsilon_{j,t+1} x_{jt}}^2 \norm{\breve{f}_{t+1}}^2 \norm{\tilde{\Gamma} - \Gamma H_{\Gamma}}^2 = o_p(1)
$$
since $\frac{1}{T} \sum_{t=1}^T \bbE[||\frac{1}{N}\sum_{j=1}^N \epsilon_{j,t+1} x_{jt}||^2||\breve{f}_{t+1}||^2|X,F] \lesssim \frac{L}{N}$. Besides, we have
\begin{align*}
&\frac{1}{T} \sum_{t=1}^T \norm{\frac{1}{N} \sum_{j=1}^N \epsilon_{j,t+1} B_{jt}^\top H_{\Gamma} ( \hat{f}_{t+1} - H_{\Gamma}^{-1} f_{t+1})} ^2 \\
&\leq 
\left( \frac{1}{T} \sum_{t=1}^T \norm{\frac{1}{N} \sum_{j=1}^N \epsilon_{j,t+1} B_{jt} }^4 \right)^{1/2} \left( \frac{1}{T} \sum_{t=1}^T \norm{ \tilde{\breve{f}}_{t+1} - H_{\Gamma}^{-1} \breve{f}_{t+1}}^4 \right)^{1/2} = o_p(1)
\end{align*}
since $\bbE[||\frac{1}{N} \sum_{j=1}^N \epsilon_{j,t+1} B_{jt}||^4|X] \lesssim \frac{1}{N}$. Here, we can show that $ \frac{1}{T} \sum_{t=1}^T \norm{ \tilde{\breve{f}}_{t+1} - H_{\Gamma}^{-1} \breve{f}_{t+1}}^4 = o_p(1)$ in a similar way to the proof of Lemmas \ref{lem:fhat} and \ref{lem:fd_bound} with a weak dependence of noises across $i$. Hence, we have 
$$
\frac{1}{T} \sum_{t=1}^T \norm{\frac{1}{N}\sum_{j=1}^N \epsilon_{j,t+1} \left( \tilde{m}_{j,t+1} - m_{j,t+1} \right) }^2 = o_p(1) .
$$
Moreover, we can show that $\frac{1}{T} \sum_{t=1}^T \norm{\frac{1}{N}\sum_{j=1}^N \left( \hat{\alpha}_{O,jt} - \alpha_{O,jt} \right)^2}^2  = o_p(1)$ by using a concentration inequality for the sub-Gaussian random variable like Lemma \ref{lem:uniformbound_1}. In addition, we have $\frac{1}{T} \sum_{t=1}^T \norm{\frac{1}{N}\sum_{j=1}^N (x_{jt}^\top (\tilde{\eta} - \eta))^2}^2 \lesssim \norm{\tilde{\eta} - \eta}^4 = o_p(1)$. Similarly, we have 
$\frac{1}{T} \sum_{t=1}^T \norm{\frac{1}{N}\sum_{j=1}^N (x_{jt}^\top (\tilde{\Gamma} - \Gamma H_{\Gamma})\breve{f}_{t+1})^2}^2 \lesssim \frac{1}{T} \sum_{t=1}^T \norm{\breve{f}_{t+1}}^4 \norm{\tilde{\Gamma} - \Gamma H_{\Gamma}}^4 = o_p(1)$.
Lastly, we have 
$$ 
\frac{1}{T} \sum_{t=1}^T \norm{\frac{1}{N}\sum_{j=1}^N (x_{jt}^\top \Gamma H_{\Gamma} (\tilde{\breve{f}}_{t+1} - H_{\Gamma}^{-1} \breve{f}_{t+1}))^2}^2 \lesssim \frac{1}{T} \sum_{t=1}^T \norm{\tilde{\breve{f}}_{t+1} - H_{\Gamma}^{-1} \breve{f}_{t+1})}^4  = o_p(1).
$$
Hence, we have 
$$
\frac{1}{T} \sum_{t=1}^T \norm{\frac{1}{N}\sum_{j=1}^N \left( \tilde{m}_{j,t+1} - m_{j,t+1} \right)^2 }^2 = o_p(1) . \ \ \square
$$

\begin{lemma}\label{lem:variance_est_tech_debias}
Let $\hat{\sigma}^2_{t+1} = \frac{1}{N} \sum_{i=1}^N \hat{\epsilon}_{i,t+1}^2$, where $m_{i,t+1} = \alpha_{O,it} + x_{it}^\top \eta + x_{it}^\top \Gamma \breve{f}_{t+1}$, $\hat{m}_{i,t+1}= \hat{\alpha}_{O,it} + x_{it}^\top \hat{\eta} + x_{it}^\top \hat{\Gamma} \hat{\breve{f}}_{t+1}$, and $\hat{\epsilon}_{i,t+1} = r_{i,t+1} - \hat{m}_{i,t+1}$. Then, we have (i) $\abs{\hat{\sigma}^2_{t+1} - \sigma^2_{t+1}} = o_p(1)$, (ii) $\frac{1}{T}\sum_{t=1}^T \abs{\hat{\sigma}^2_{t+1} - \sigma^2_{t+1}} = o_p(1)$, (iii) $\frac{1}{T}\sum_{t=1}^T \abs{\hat{\sigma}^2_{t+1} - \sigma^2_{t+1}}^2 = o_p(1)$.
\end{lemma}

\noindent\textbf{Proof of Lemma \ref{lem:variance_est_tech_debias}.} The proof is the same as that of Lemma \ref{lem:variance_est_tech} and we omit it here. $\square$

\subsection{Technical lemmas}

\begin{lemma}\label{lem:error_tech}
(i) $\norm{(X_t^\top X_t)^{-1} X_t^\top E_{t+1}} = O_p\left( \frac{\sqrt{L}}{\sqrt{N}}\right)$; (ii) $\frac{1}{T} \sum_{t=1}^T  \norm{(X_t^\top X_t)^{-1} X_t^\top E_{t+1}}^2  = O_p\left( \frac{L}{N} \right)$; (iii) $\frac{1}{T} \sum_{t=1}^T  \norm{(X_t^\top X_t)^{-1} X_t^\top E_{t+1}}  = O_p\left( \frac{\sqrt{L}}{\sqrt{N}} \right)$; (iv) $\norm{\frac{1}{T} \sum_{t} (X_t^\top X_t)^{-1} X_t^\top E_{t+1}} = O_p\left( \frac{\sqrt{L}}{\sqrt{NT}} \right)$; (v) $\norm{ \Gamma^\top (X_t^\top X_t)^{-1} X_t^\top E_{t+1}}  = O_p\left(\frac{1}{\sqrt{N}} \right)$, (vi) $\frac{1}{T} \sum_{t=1}^T \norm{ \Gamma^\top (X_t^\top X_t)^{-1} X_t^\top E_{t+1}}  = O_p\left(\frac{1}{\sqrt{N}} \right)$; (vii) $\frac{1}{T} \sum_{t=1}^T \norm{ \Gamma^\top (X_t^\top X_t)^{-1} X_t^\top E_{t+1}}^2  = O_p\left(\frac{1}{N} \right)$;\\
(viii) $\max_t \bbE[(e_l^\top (X_t^\top X_t)^{-1} X_t^\top E_{t+1})^2|X] = O_p(\frac{1}{N})$; (ix) $||\frac{1}{T} \sum_{t=1}^T (e_l^\top \ddot{E}_{t+1})|| = O_p\left(\frac{1}{\sqrt{NT}}\right)$.
\end{lemma}

\noindent\textbf{Proof of Lemma \ref{lem:error_tech}.} (i) Note that
\begin{align*}
\bbE\left[ \left. \norm{(X_t^\top X_t)^{-1} X_t^\top E_t}^2 \right| X \right] 
&= \sum_{l=1}^L e_l^\top (X_t^\top X_t)^{-1} X_t^\top \bbE\left[  \left. E_{t+1} E_{t+1}^\top \right| X \right] X_t (X_t^\top X_t)^{-1}  e_l \\
&\leq \sum_{l=1}^L \norm{ e_l^\top (X_t^\top X_t)^{-1} X_t^\top }^2 \norm{\bbE\left[  \left. E_{t+1} E_{t+1}^\top \right| X \right]} \\
& =  \norm{(X_t^\top X_t)^{-1} X_t^\top }_F^2 \norm{\bbE\left[  \left. E_{t+1} E_{t+1}^\top \right| X \right]} 
 = O_p\left( \frac{L}{N} \right).
\end{align*}
by Assumptions \ref{asp:characteristics} and \ref{asp:noise}. Hence, we have $\norm{(X_t^\top X_t)^{-1} X_t^\top E_{t+1}} = O_p\left( \frac{\sqrt{L}}{\sqrt{N}}\right)$.\\
(ii) Because
\begin{align*}
\bbE\left[ \left. \sum_{t=1}^T  \norm{(X_t^\top X_t)^{-1} X_t^\top E_{t+1}}^2 \right| X \right] &= \sum_{t=1}^T \bbE\left[ \left. \norm{(X_t^\top X_t)^{-1} X_t^\top E_{t+1}}^2 \right| X \right] \\
&\leq \max_t \norm{(X_t^\top X_t)^{-1}}^2 \max_t \norm{\bbE\left[  \left. E_{t+1} E_{t+1}^\top \right| X \right]} \sum_{t=1}^T \norm{X_t}_F^2 \\ 
& = O_p\left( \frac{LT}{N} \right) ,
\end{align*}
we have $\frac{1}{T} \sum_{t=1}^T  \norm{(X_t^\top X_t)^{-1} X_t^\top E_{t+1}}^2  = O_p\left( \frac{L}{N} \right)$.\\
(iii) Note that
\begin{align*}
\frac{1}{T} \sum_{t=1}^T \bbE\left[ \left. \norm{(X_t^\top X_t)^{-1} X_t^\top E_{t+1}} \right| X \right] &\leq \frac{1}{T} \sum_{t=1}^T  \left( \bbE\left[ \left. \norm{(X_t^\top X_t)^{-1} X_t^\top E_{t+1}}^2 \right| X \right] \right)^{1/2} \\
&\leq \max_t \norm{(X_t^\top X_t)^{-1}} \max_t \norm{\bbE\left[  \left. E_{t+1} E_{t+1}^\top \right| X \right]}^{1/2} \frac{1}{T} \sum_{t=1}^T \norm{X_t}_F  \\
 &= O_p\left( \frac{\sqrt{L}}{\sqrt{N}} \right).
\end{align*}
(iv) Note that
\begin{align*}
\norm{\frac{1}{T} \sum_{t=1}^T (X_t^\top X_t)^{-1} X_t^\top E_{t+1}}^2 &= \sum_{l=1}^L \left(\frac{1}{T} \sum_{t=1}^T e_l^\top (X_t^\top X_t)^{-1} X_t^\top E_{t+1} \right)^2 \\
&=\frac{1}{N^2T^2} \sum_{l=1}^L \left( \sum_{i=1}^N \sum_{t=1}^T e_l^\top (X_t^\top X_t/N)^{-1} x_{it} \epsilon_{i,t+1} \right)^2 \\
& = \frac{1}{N^2 T^2} \sum_{l=1}^L \left( A_l^\top \vect(E) \right)^2,
\end{align*}
where $A_l$ is the $NT \times 1$ vector whose $(i,t)$-th element is $e_l^\top (X_t^\top X_t/N)^{-1} x_{it}$. Hence, we have
$$
\bbE\norm{\frac{1}{T} \sum_{t=1}^T (X_t^\top X_t)^{-1} X_t^\top E_{t+1}}^2 \leq \frac{1}{N^2T^2} \sum_{l=1}^L \norm{A_l}^2 \norm{\bbE\left[ \vect(E) \vect(E)^\top | X  \right]} = O_p\left( \frac{L}{NT} \right)
$$
because $\max_t e_l^\top (X_t^\top X_t/N)^{-1} e_l \leq \max_t \norm{(X_t^\top X_t/N)^{-1}} \norm{e_l}^2 \leq C$ for some constant $C > 0$ and
$$
\sum_{l=1}^L \norm{A_l}^2 = \sum_{l=1}^L \sum_{i=1}^N \sum_{t=1}^T e_l^\top (X_t^\top X_t/N)^{-1} x_{it} x_{it}^\top (X_t^\top X_t/N)^{-1} e_l 
= N \sum_{l=1}^L \sum_{t=1}^T e_l^\top (X_t^\top X_t/N)^{-1} e_l .
$$
(v) Note that
\begin{align*}
\bbE\left[ \left. \norm{ \Gamma^\top (X_t^\top X_t)^{-1} X_t^\top E_{t+1}}^2 \right| X \right]
&= \sum_{k=1}^K e_k^\top  \Gamma^\top (X_t^\top X_t)^{-1} X_t^\top \bbE\left[  \left. E_{t+1} E_{t+1}^\top \right| X \right] X_t (X_t^\top X_t)^{-1} \Gamma e_k \\
&\leq  \norm{\bbE\left[  \left. E_{t+1} E_{t+1}^\top \right| X \right]} \norm{ \Gamma^\top (X_t^\top X_t)^{-1} X_t^\top}_F^2 \\
&\lesssim \norm{\bbE\left[  \left. E_{t+1} E_{t+1}^\top \right| X \right]}\norm{ \Gamma^\top (X_t^\top X_t)^{-1} X_t^\top}^2 \\
& \lesssim \norm{\Gamma}^2 \norm{(X_t^\top X_t)^{-1} X_t^\top}^2 = O_p\left(\frac{1}{N} \right).
\end{align*}
(vi), (vii) trivially follow from (v).\\
(viii) We have
\begin{align*}
\max_t \bbE[(e_l^\top (X_t^\top X_t)^{-1} X_t^\top E_{t+1})^2|X] 
& = \max_t e_l^\top (X_t^\top X_t)^{-1} X_t^\top \bbE[ E_{t+1}E_{t+1}^\top] X_t (X_t^\top X_t)^{-1} e_l \\    
& \leq \max_t \norm{\bbE[ E_{t+1}E_{t+1}^\top]} \max_t \norm{X_t (X_t^\top X_t)^{-1}}^2 \\
& = O_p\left( \frac{1}{N} \right).
\end{align*}
(ix) Note that
\begin{align*}
\norm{\frac{1}{T} \sum_{t=1}^T e_l^\top (X_t^\top X_t)^{-1} X_t^\top E_{t+1}}^2 &=\frac{1}{N^2T^2} \left( \sum_{i=1}^N \sum_{t=1}^T e_l^\top (X_t^\top X_t/N)^{-1} x_{it} \epsilon_{i,t+1} \right)^2 \\
& = \frac{1}{N^2 T^2} \left( A_l^\top \vect(E) \right)^2,
\end{align*}
where $A_l$ is the $NT \times 1$ vector whose $(i,t)$-th element is $e_l^\top (X_t^\top X_t/N)^{-1} x_{it}$. Hence, we have
$$
\bbE\norm{\frac{1}{T} \sum_{t=1}^T e_l^\top (X_t^\top X_t)^{-1} X_t^\top E_{t+1}}^2 \leq \frac{1}{N^2T^2} \norm{A_l}^2 \norm{\bbE\left[ \vect(E) \vect(E)^\top | X  \right]} = O_p\left( \frac{1}{NT} \right) ,
$$
because $\max_t e_l^\top (X_t^\top X_t/N)^{-1} e_l \leq \max_t \norm{(X_t^\top X_t/N)^{-1}} \norm{e_l}^2 \leq C$ for some constant $C > 0$ and $\norm{A_l}^2 = \sum_{i=1}^N \sum_{t=1}^T e_l^\top (X_t^\top X_t/N)^{-1} x_{it} x_{it}^\top (X_t^\top X_t/N)^{-1} e_l 
= N  \sum_{t=1}^T e_l^\top (X_t^\top X_t/N)^{-1} e_l $. $\square$
\smallskip

\begin{lemma}\label{lem:rotation_matrix_tech}
(i) W.h.p., $\norm{H_F}$ and $\norm{H_F^{-1}}$ are bounded.
(ii) W.h.p., $\norm{H_\Gamma}$ and $\norm{H_\Gamma^{-1}}$ are bounded. 
(iii) $\norm{H_F - H_\Gamma^{-1}} = o_p(1)$ and $\norm{H_\Gamma - H_F^{-1}} = o_p(1)$. (iv) $H_F \conP I_{sgn} \boldsymbol{G}^{-1} \left( \Gamma^{\top} \Gamma \right)^{1/2}$ and $H_F^{-1} \conP \left( \Gamma^{\top} \Gamma \right)^{-1/2} \boldsymbol{G} I_{sgn}$.
\end{lemma}

\noindent\textbf{Proof of Lemma \ref{lem:rotation_matrix_tech}.}
(i) First, $\norm{H_F} = \norm{\tilde{\Gamma}^\top \Gamma} \leq \norm{\Gamma}$ is bounded. In addition, by a typical assertion for the spectral method such as Proposition 1 of \cite{bai2003inferential}, we have $\psi_{\min} \left( H_{F} \right) = \psi_{\min} \left( \tilde{\Gamma}^\top \Gamma   \right) > c$. Hence, $\norm{H_F^{-1}}$ is bounded.\\
(ii) Simple calculation shows that
\begin{gather*}
H_{\Gamma}^{\top} 
= \left[ \left(\frac{1}{T}\sum_t \tilde{f}_{t+1}^{d} \tilde{f}_{t+1}^{d \top} \right)^{-1} \left( \frac{1}{T}\sum_t \tilde{f}_{t+1}^{d} ( H_{F}f_{t+1}^{d} - \tilde{f}_{t+1}^{d} )^{\top}  \right) + I_K \right] H_{F}^{- \top }. 
\end{gather*}
Let $U_K D_K V_K^\top$ be the top-$K$ singular value decomposition of $\ddot{R}^d$. Note that w.h.p.,
\begin{align*}
\sqrt{\psi_{K} \left( \tilde{F}^d\tilde{F}^{d \top} \right)} = \psi_{K} (D_K) \geq \psi_{K} (\Gamma F^d) - \norm{\ddot{E}^d} \geq c \sqrt{T}
\end{align*}
for some constant $c>0$ since we have
$\norm{\ddot{R}^d  - \Gamma \calF^d} =  \norm{\ddot{E}^d} = O_p \left( \frac{\sqrt{LT}}{\sqrt{N}} \right)$ by Lemma \ref{lem:error_tech}. Hence,
$$
\psi_{\min} \left( \frac{1}{T}\sum_t \tilde{f}_{t+1}^{d} \tilde{f}_{t+1}^{d \top} \right) = \frac{1}{T} \psi_{\min} \left( \tilde{F}^d \tilde{F}^{d \top} \right) = \frac{1}{T} \psi_{\min} \left( D_K \right)^2 > c,
$$
for some constant $c >0$ and $\norm{\left(\frac{1}{T}\sum_t \tilde{f}_{t+1}^{d} \tilde{f}_{t+1}^{d \top} \right)^{-1} }$ is bounded. In addition, 
$$
\norm{ \frac{1}{T}\sum_t \tilde{f}_{t+1}^{d} ( H_{F}f_{t+1}^{d} - \tilde{f}_{t+1}^{d} )^{\top}  } = o_p(1)
$$
by Lemma \ref{lem:gamma_f_final_bound}. Hence, w.h.p.,
\begin{align*}
&\psi_{\min} \left[ \left(\frac{1}{T}\sum_t \tilde{f}_{t+1}^{d} \tilde{f}_{t+1}^{d \top} \right)^{-1} \left( \frac{1}{T}\sum_t \tilde{f}_{t+1}^{d} ( H_{F}f_{t+1}^{d} - \tilde{f}_{t+1}^{d} )^{\top}  \right) + I_K \right] \\
&\geq 1 - \norm{\left(\frac{1}{T}\sum_t \tilde{f}_{t+1}^{d} \tilde{f}_{t+1}^{d \top} \right)^{-1} \left( \frac{1}{T}\sum_t \tilde{f}_{t+1}^{d} ( H_{F}f_{t+1}^{d} - \tilde{f}_{t+1}^{d} )^{\top} \right) }\\
&\geq \frac{1}{2}.
\end{align*}
Then, $\psi_{\min}(H_\Gamma) \geq \frac{1}{2}\psi_{\min}(H_F^{-1}) > c$ for some $c>0$. Hence, $\norm{H_\Gamma^{-1}}$ is bounded. In addition, because $\norm{\left(\frac{1}{T}\sum_t \tilde{f}_{t+1}^{d} \tilde{f}_{t+1}^{d \top} \right)^{-1} \left( \frac{1}{T}\sum_t \tilde{f}_{t+1}^{d} ( H_{F}f_{t+1}^{d} - \tilde{f}_{t+1}^{d} )^{\top} \right) } = o_p(1)$, $\norm{H_\Gamma - H_F^{-1}} = o_p(1)$ and $\norm{H_\Gamma}$ is also bounded.\\
(iii) We prove $\norm{H_\Gamma - H_F^{-1}} = o_p(1)$ in (ii). In addition, we have
$$
\norm{H_{\Gamma}^{-1} -H_{F}}  \leq \norm{H_{F}}\norm{H_{\Gamma} -H_{F}^{-1}} \norm{H_{\Gamma}^{-1}} = o_p(1).
$$
(iv) Let $\Omega = \left( \Gamma^{\top} \Gamma \right)^{1/2} \left( \frac{1}{T}  F^{d \top} F^{d} \right)  \left( \Gamma^{\top} \Gamma \right)^{1/2}$ and $G$ be a $K \times K$ matrix whose columns are the eigenvectors of $\Omega$ such that $\Lambda = G^\top \Omega G$ is the descending order diagonal matrix of the eigenvalues of $\Omega$. Define $H = \left( \Gamma^{\top}\Gamma \right)^{-1/2} G$. Then, we have
\begin{align*}
(\Gamma F^{d\top} F^{d} \Gamma^{\top})\Gamma H
&= \Gamma \left( \Gamma^{\top}\Gamma \right)^{-1/2} \left( \Gamma^{\top} \Gamma \right)^{1/2} F^{d\top} F^{d} \left( \Gamma^{\top} \Gamma \right)^{1/2} \left( \Gamma^{\top} \Gamma \right)^{1/2} H \\
& = \Gamma \left( \Gamma^{\top} \Gamma \right)^{-1/2}  \left[ \left( \Gamma^{\top} \Gamma \right)^{1/2} F^{d\top} F^{d} \left( \Gamma^{\top} \Gamma \right)^{1/2} 
 G \right] \\
& = \Gamma \left( \Gamma^{\top} \Gamma \right)^{-1/2} T \Omega G
 = \Gamma \left( \Gamma^{\top} \Gamma \right)^{-1/2} G  T  \Lambda \\
 & = \Gamma H  T \Lambda .
\end{align*}
In addition, note that $\left( \Gamma H \right)^\top \left( \Gamma H \right) = H^{\top} \Gamma^{ \top} \Gamma H = G^\top G = I_K$. Therefore, $\Gamma H$ is the eigenvector of $\Gamma F^{d\top} F^{d} \Gamma^{\top}$ and the left singular vector of $\Gamma F^{d\top}$. Let $U$ be the left singular vector of $\Gamma F^{d\top}$. Then, we have $H_F = U_K^\top U H^{-1}$ since $\Gamma = U H^{-1}$. Since $\norm{\Omega - \left( \Gamma^{\top} \Gamma \right)^{1/2} \Sigma_f \left( \Gamma^{\top} \Gamma \right)^{1/2}} = o_p(1)$ and the eigenvalues of $\left( \Gamma^{\top} \Gamma \right)^{1/2} \Sigma_f \left( \Gamma^{\top} \Gamma \right)^{1/2}$ are distinct, by the eigenvector perturbation theory, there is a unique eigenvector of $\left( \Gamma^{\top} \Gamma \right)^{1/2} \Sigma_f \left( \Gamma^{\top} \Gamma \right)^{1/2}$, says, $\boldsymbol{G}$, such that $\norm{G - \boldsymbol{G}} = o_p(1)$. Therefore, $\norm{H - \left( \Gamma^{\top}\Gamma \right)^{-1/2} \boldsymbol{G}} \conP 0$. Moreover, because $\norm{G^{-1} - \boldsymbol{G}^{-1}} = \norm{G^{\top} - \boldsymbol{G}^{\top}} = o_p(1)$, we also have $\norm{H^{-1} - \boldsymbol{G}^{-1} \left( \Gamma^{\top} \Gamma \right)^{1/2} } = o_p(1)$. By the same method in Claim E.1 of \cite{choi2024high}, we know $U_K^\top U \conP I_{sgn}$ where $I_{sgn}$ is the $K \times K$ diagonal matrix consisting of the diagonal elements of $\pm 1$ and the sign of these are determined by the sign alignment between $U_K$ and $U$. Hence, we have $H_F \conP I_{sgn} \boldsymbol{G}^{-1} \left( \Gamma^{\top} \Gamma \right)^{1/2}$ and $H_F^{-1} \conP \left( \Gamma^{\top} \Gamma \right)^{-1/2} \boldsymbol{G} I_{sgn}$. $\square$
\smallskip

\begin{lemma}\label{lem:debiasing_tech}
Let $a_{ijt} = (e_l^\top Q_t^{-1} x_{it}) Q_t^{-1} x_{jt}$ where $Q_t = X_t^\top X_t / N$ and $u_{ijt} = \epsilon_{it} \epsilon_{jt}$. Then, we have (i) $\frac{1}{T N^2} \sum_{i=1}^N \sum_{j=1}^N \sum_{t=1}^T a_{ijt} (u_{ij,t+1} - \bbE[u_{ij,t+1}]) = O_p\left( \frac{\sqrt{L}}{N \sqrt{T} } \right)$;\\
(ii) $\frac{1}{T N^2} \sum_{i=1}^N \sum_{t=1}^T a_{iit} \left( \bbE[\epsilon_{i,t+1}^2] - \hat{\sigma}_{t+1}^2 \right)  = o_p\left( \frac{1}{\sqrt{NT}} \right)$.
\end{lemma}

\noindent\textbf{Proof of Lemma \ref{lem:debiasing_tech}.}
(i) Let $A$ be the $N^2T \times L$ matrix whose $(i,j,t)$-th row is $a_{ijt}^\top$ and $U$ be the $N^2T \times 1$ vector whose $(i,j,t)$-th element is $u_{ijt} - \bbE[u_{ijt}]$. Then, we have
\begin{gather*}
\frac{1}{T N^2} \sum_{i=1}^N \sum_{j=1}^N \sum_{t=1}^T a_{ijt} (u_{ijt} - \bbE[u_{ijt}]) = \frac{1}{T N^2} A^\top U
\end{gather*}
and
\begin{align*}
\bbE\left[ \left. \left( \frac{1}{T N^2} \norm{A^\top U} \right)^2 \right| X \right] 
&= \frac{1}{T^2 N^4}\sum_{r=1}^{L} A_r^\top \bbE\left[\left.  U U^\top \right| X \right] A_r \leq \frac{1}{T^2 N^4} \norm{A}_F^2 \norm{\bbE\left[\left.  U U^\top \right| X \right]} \\
&= O_p\left( \frac{1}{N^2 T} \right),
\end{align*}
where $A_r = A e_r$, because $\norm{\bbE \left[\left.  U U^\top \right| X \right]} $ is bounded and
\begin{align*}
\norm{A}_F^2 &= \sum_{r,i,j,t} (e_l^\top Q_t^{-1} x_{it})^2 (e_r^\top Q_t^{-1} x_{jt})^2 
= N^2 \sum_t e_l^\top Q_t^{-1} e_l \sum_r e_r^\top Q_t^{-1} e_r
\lesssim LN^2 \sum_t \abs{e_l^\top Q_t^{-1} e_l}  \\
&= O_p \left( L N^2 T \right),
\end{align*}
since $\norm{e_l} =\norm{e_r} = 1$ where $1 \leq l,r \leq L$. Note that, when $\epsilon_{it}$ is independent across $i$, $\Cov(\epsilon_{i,t+1}\epsilon_{j,t+1},\epsilon_{i',s+1}\epsilon_{j',s+1} )$ is nonzero only when $i\neq j$, $i=i'$, $j=j'$ or $i\neq j$, $i=j'$, $j=i'$ except for the case $i=j=i'=j'$. Hence, the condition $\max_{i,t} \sum_{s=1}^T \abs{\Cov(\epsilon_{i,t+1}, \epsilon_{i,s+1})} \leq C_1$ and $\max_{i,t} \sum_{s=1}^T \abs{\Cov(\epsilon_{i,t+1}^2, \epsilon_{i,s+1}^2)} \leq C_1$ for some constant $C_1>0$ is enough to have $\norm{\bbE \left[\left.  U U^\top \right| X \right]} < C_2$ for some constant $C_2>0$ because $\max_{ijt} \sum_{i',j',s} \abs{\Cov(\epsilon_{i,t+1}\epsilon_{j,t+1},\epsilon_{i',s+1}\epsilon_{j',s+1})}$ is bounded.\\
(ii) Let $m_{i,t+1} = \alpha_{O,it} + x_{it}^\top \eta + x_{it}^\top \Gamma \breve{f}_{t+1}$, $\tilde{m}_{i,t+1}= \hat{\alpha}_{O,it} + x_{it}^\top \tilde{\eta} + x_{it}^\top \tilde{\Gamma} \tilde{\breve{f}}_{t+1}$, and $\hat{\epsilon}_{i,t+1} = r_{i,t+1} - \tilde{m}_{i,t+1}$. Note that
\begin{gather*}
\hat{\sigma}_{t+1}^2 - \bbE[\epsilon_{j,t+1}^2] = \frac{1}{N}\sum_{j=1}^N \left(\hat{\epsilon}_{j,t+1}^2 - \epsilon_{j,t+1}^2 \right) + \frac{1}{N}\sum_{j=1}^N \left( \epsilon_{j,t+1}^2 - \bbE[\epsilon_{j,t+1}^2] \right).
\end{gather*}
First of all, by the concentration inequality with the weak dependent error condition, the part related to the second term is bounded as
$$
 \frac{1}{T N^2} \sum_{i=1}^N \sum_{t=1}^T a_{iit} \frac{1}{N}\sum_{j=1}^N \left( \epsilon_{j,t+1}^2 - \bbE[\epsilon_{j,t+1}^2] \right) = \frac{1}{TN^2} \sum_{j=1}^N\sum_{t=1}^T \bar{a}_t \left( \epsilon_{j,t+1}^2 - \bbE[\epsilon_{j,t+1}^2] \right) = O_p\left( \frac{1}{\sqrt{T} N^{3/2}}\right),
$$
where $\bar{a}_t = \frac{1}{N}\sum_{i=1}^N a_{iit} = Q_t^{-1} e_l$. Here, we use the bounds $\max_t||\bar{a}_t|| = O_p(1)$ and $\frac{1}{T} \sum_{t=1}^T ||\bar{a}_t ||^2 = O_p(1) $. For the part related to the first term, note that 
$$
\frac{1}{N}\sum_{j=1}^N \left(\hat{\epsilon}_{j,t+1}^2 - \epsilon_{j,t+1}^2 \right) = 2 \frac{1}{N}\sum_{j=1}^N \epsilon_{j,t+1} \left( \tilde{m}_{j,t+1} - m_{j,t+1} \right) + \frac{1}{N}\sum_{j=1}^N \left( \tilde{m}_{j,t+1} - m_{j,t+1} \right)^2  ,
$$ 
and 
\begin{align*}
&\frac{1}{N}\sum_{j=1}^N \epsilon_{j,t+1} \left( \tilde{m}_{j,t+1} - m_{j,t+1} \right) \\
& \ \ = \frac{1}{N}\sum_{j=1}^N \epsilon_{j,t+1}(\hat{\alpha}_{O,jt} - \alpha_{O,jt}) + \frac{1}{N}\sum_{j=1}^N \epsilon_{j,t+1} x_{jt}^\top (\tilde{\eta} - \eta) + \frac{1}{N}\sum_{j=1}^N \epsilon_{j,t+1} x_{jt}^\top (\tilde{\Gamma} \tilde{\breve{f}}_{t+1} - \Gamma f_{t+1}).    
\end{align*}
Note that the dominating terms of $\hat{\alpha}_{O,jt} - \alpha_{O,jt}$ are
\begin{align*}
\frac{1}{NT} \sum_{k=1}^N \sum_{s=1}^T B_{t,j}^{o \top} B_{s,k}^{o} \epsilon_{k,s+1}, \ \  \frac{1}{N} \sum_{k=1}^N \sum_{q \in D_t} B_{t,jq}^{o \top} B_{t,kq}^{o} \epsilon_{k,t+1} , \ \ \sum_{q \notin D_t} B_{t,jq}^{o} \left( \tilde{\bar{\xi}}_q - \bar{\xi}_q \right).
\end{align*}
The part related to the first term is bounded as
\begin{align*}
&\norm{\frac{1}{T N^2} \sum_{i=1}^N \sum_{t=1}^T a_{iit} \frac{1}{N} \sum_{j=1}^N \epsilon_{j,t+1}[\hat{\alpha}_{O,jt} - \alpha_{O,jt}]_{part_1}} \\
&\leq \frac{1}{T N^2} \norm{\sum_{t=1}^T \sum_{j=1}^N \epsilon_{j,t+1} \bar{a}_t B^{o\top}_{t,j}} \norm{\frac{1}{NT} \sum_{k=1}^N \sum_{s=1}^T B_{s,k}^{o} \epsilon_{k,s+1}} = O_P\left( \frac{1}{NT} \right).
\end{align*}
Here we use the relation that
\begin{align*}
\bbE\left[ \left.  \norm{ \sum_{s=1}^T \sum_{k=1}^N B_{s,k}^{o} \epsilon_{k,s+1}}^2  \right| X \right] &= \sum_{q=1}^{N-L} \bbE\left[ \left.  \left( \sum_{s=1}^T \sum_{k=1}^N B_{s,kq}^{o} \epsilon_{k,s+1} \right)^2  \right| X \right] \\
& =  \sum_{q=1}^{N-L} \calB_q^\top  \bbE\left[ \left. \vect(E) \vect(E)^\top  \right| X \right] \calB_q \\
& \leq \sum_{q=1}^{N-L} \norm{\calB_q}^2 \norm{\bbE\left[ \left. \vect(E) \vect(E)^\top  \right| X \right]} \\
&= \sum_{s=1}^T \norm{B_s^o}_F^2 \norm{\bbE\left[ \left. \vect(E) \vect(E)^\top  \right| X \right]} \\
& \lesssim N(N-L)T
\end{align*}
since $\norm{B_s^o}_F^2 = N(N-L)$ by Lemma \ref{lem:B_tech} and $\norm{\bbE\left[ \left. \vect(E) \vect(E)^\top  \right| X \right]}$ is bounded. In addition, we use the relation that
\begin{align*}
\bbE \left[ \left.  \norm{\sum_{t=1}^T \sum_{j=1}^N \epsilon_{j,t+1} \bar{a}_t B^{o\top}_{t,j}}_F^2  \right| X \right] &= \sum_{r=1}^{L}\sum_{k=1}^{N-L} \bbE \left[ \left. \left( \sum_{t=1}^T \sum_{j=1}^N \epsilon_{j,t+1} \bar{a}_{t,r} B^{o}_{t,jk} \right)^2 \right| X \right]\\
& = \sum_{r=1}^{L}\sum_{k=1}^{N-L} A_{rk}^\top \bbE \left[ \left. E E^\top \right| X \right] A_{rk}\\
& \leq \norm{\bbE \left[ \left. E E^\top \right| X \right]} \sum_{t=1}^T \sum_{j=1}^N \sum_{r=1}^{L} \sum_{k=1}^{N-L} \bar{a}_{t,r}^2 B_{t,jk}^{o2}\\
& = \norm{\bbE \left[ \left. E E^\top \right| X \right]} \sum_{t=1}^T \norm{\bar{a}_{t}}^2 \norm{B^{o}_t}_F^2 = O_p\left( TN(N-L) \right)
\end{align*}
where $A_{rk}$ is the $NT \times 1$ vector whose $(i,t)$-th element is $\bar{a}_{t,r} B^{o}_{t,ik}$, because $\norm{\bbE \left[ \left. E E^\top \right| X \right]}$ is bounded, $\sum_{t=1}^T \norm{\bar{a}_{t}}^2 = O_p(T)$, and $\norm{B^{o}_t}_F^2 = N(N-L)$. In addition, for the second part, we have
\begin{align*}
&\norm{\frac{1}{T N^2} \sum_{i=1}^N \sum_{t=1}^T a_{iit} \frac{1}{N} \sum_{j=1}^N \epsilon_{j,t+1}[\hat{\alpha}_{O,jt} - \alpha_{O,jt}]_{part_2}} \\
&\leq \frac{1}{N} \left( \frac{1}{T} \sum_{t=1}^T \norm{\bar{a}_t}^2 \right)^{1/2} \left( \frac{1}{T} \sum_{t=1}^T  \left( \frac{1}{N^2} \sum_{j=1}^N \sum_{k=1}^N \epsilon_{j,t+1} \epsilon_{k,t+1} \left( \sum_{q \in D_t} B_{t,jq}^o B^o_{t,kq} \right) \right)^{2} \right)^{1/2} \\
& = o_p\left( \frac{1}{\sqrt{NT}} \right)
\end{align*}
since by Lemma \ref{lem:uniformbound_1}, w.h.p., for all $t$,
\begin{align*}
 \frac{1}{N^2} \sum_{j=1}^N \sum_{k=1}^N \epsilon_{j,t+1} \epsilon_{k,t+1} \left( \sum_{q \in D_t} B_{t,jq}^o B_{t,kq} \right) 
& =\sum_{q \in D_t} \left( \frac{1}{N} \sum_{j=1}^N \epsilon_{j,t+1} B_{t,jq}^o \right) \left( \frac{1}{N} \sum_{k=1}^N \epsilon_{k,t+1} B_{t,kq}^o \right) \\
&\lesssim \frac{\log N}{N} | D_t| \ll \frac{\sqrt{N}}{\sqrt{T}} .   
\end{align*}
For the third term, the dominating parts of $[\hat{\alpha}_{O,jt} - \alpha_{O,jt}]_{part_3}$ are
$$
 \frac{1}{T} \sum_{s=1}^T \sum_{q \in D_s / D_t} B_{t,jq}^o \frac{1}{N} \sum_{k=1}^N B_{s,kq}^o \epsilon_{k,s+1}, \ \  \frac{1}{T} \sum_{s=1}^T \sum_{q \in D_s / D_t} B_{t,jq}^o \frac{1}{T} \sum_{s'=1}^T \xi_{s',q}  .
$$
For the first part, we have
\begin{align*}
&\frac{1}{NT} \sum_{t=1}^T \bar{a}_t \frac{1}{N} \sum_{j=1}^N \epsilon_{j,t+1}  \frac{1}{T} \sum_{s=1}^T \sum_{q \in D_s / D_t} B_{t,jq}^o \frac{1}{N} \sum_{k=1}^N B_{s,kq}^o \epsilon_{k,s+1}\\
&\leq \max_t \norm{\bar{a}_t} \frac{1}{N^3 T^2} \sum_{t=1}^T \norm{ \sum_{j=1}^N \sum_{k=1}^N \sum_{s=1}^T \epsilon_{j,t+1}   \epsilon_{k,s+1} \sum_{q \in D_s / D_t} B_{t,jq}^o  B_{s,kq}^o   }\\
& = o_p\left( \frac{1}{\sqrt{NT}} \right)
\end{align*}
since by Lemma \ref{lem:uniformbound_1}, w.h.p., for all $t$,
\begin{align*}
\sum_{j=1}^N \sum_{k=1}^N \sum_{s=1}^T \epsilon_{j,t+1}   \epsilon_{k,s+1} \sum_{q \in D_s / D_t} B_{t,jq}^o  B_{s,kq}^o 
& = \sum_{s=1}^T \sum_{q \in D_s / D_t} \left( \sum_{k=1}^N \epsilon_{k,s+1} B_{s,kq}^o \right) \left( \sum_{j=1}^N \epsilon_{j,t+1} B_{t,jq}^o \right) \\
& \lesssim T |\bar{D}| N \log N
\end{align*}
where $|\bar{D}| = \frac{1}{T} \sum_{s=1}^T \bar{D_s}$. In addition, we have
\begin{align*}
&\frac{1}{NT} \sum_{t=1}^T \bar{a}_t \frac{1}{N} \sum_{j=1}^N \epsilon_{j,t+1} \frac{1}{T} \sum_{s=1}^T \sum_{q \in D_s / D_t} B_{t,jq}^o \frac{1}{T} \sum_{s'=1}^T \xi_{s',q} \\
& \leq \frac{1}{N} \max_t \norm{\bar{a}_t} \max_{t,q} \norm{\frac{1}{N} \sum_{j =1}^N \epsilon_{j,t+1} B_{t,jq}^o } \frac{1}{T^2} \sum_{t=1}^T \sum_{s=1}^T \sum_{q \in D_s / D_t} \norm{\bar{\xi}_{q}}     \\
& = o_p \left( \frac{1}{\sqrt{NT}} \right)
\end{align*}
by Lemma \ref{lem:uniformbound_1} and the sparsity condition. Moreover, the part related to $\tilde{\eta} - \eta$ is bounded as
\begin{align*}
\norm{\frac{1}{T N^2} \sum_{i=1}^N \sum_{t=1}^T a_{iit} \frac{1}{N} \sum_{j=1}^N \epsilon_{j,t+1} x_{jt}^\top  (\tilde{\eta} - \eta)}
&\leq \frac{1}{T N^2} \norm{\sum_{t=1}^T \sum_{j=1}^N \epsilon_{j,t+1} \bar{a}_t  x_{jt}^\top} \norm{\tilde{\eta} - \eta}\\
&= O_P\left( \frac{1}{\sqrt{NT}} \frac{\sqrt{L}}{N} \left( \frac{\sqrt{L}}{\sqrt{NT}} + \left(\frac{L}{N} \right)^{10 + \frac{1}{2}} + \frac{\sqrt{L}}{N} \right) \right)
\end{align*}
by Lemma \ref{lem:etahat} and the fact that $\norm{\sum_{t=1}^T \sum_{j=1}^N \epsilon_{j,t+1} \bar{a}_t  x_{jt}} = O_p\left( \sqrt{NTL} \right)$. For the part related to $\tilde{\Gamma} \tilde{\breve{f}}_{t+1} - \Gamma \breve{f}_{t+1}$, note that
\begin{align*}
&\norm{\frac{1}{T N^2} \sum_{i=1}^N \sum_{t=1}^T a_{iit} \frac{1}{N} \sum_{j=1}^N \epsilon_{j,t+1} x_{jt}^\top ( \tilde{\Gamma} \tilde{\breve{f}}_{t+1} - \Gamma \breve{f}_{t+1})}\\
&\lesssim 
\norm{\frac{1}{T N^2}  \sum_{t=1}^T\sum_{j=1}^N  \bar{a}_{t}   \epsilon_{j,t+1} x_{jt}^\top (\tilde{\Gamma} - \Gamma H_{\Gamma})  H_{\Gamma}^{-1} \breve{f}_{t+1}}
+
\norm{\frac{1}{T N^2}  \sum_{t=1}^T\sum_{j=1}^N  \bar{a}_{t}   \epsilon_{j,t+1} x_{jt}^\top \Gamma H_{\Gamma} (\tilde{\breve{f}}_{t+1} - H_{\Gamma}^{-1} \breve{f}_{t+1})} .
\end{align*}
The first term can be bounded as
\begin{align*}
&\norm{\frac{1}{T N^2}  \sum_{t=1}^T \sum_{j=1}^N  \bar{a}_{t}   \epsilon_{j,t+1} x_{jt}^\top (\tilde{\Gamma} - \Gamma H_{\Gamma})  H_{\Gamma}^{-1} \breve{f}_{t+1}}\\
&\leq \frac{1}{N} \left( \frac{1}{T} \sum_{t=1}^T \norm{\frac{1}{N} \sum_{j=1}^N   \epsilon_{j,t+1} x_{jt}^\top}^2 \right)^{1/2} \left( \frac{1}{T} \sum_{t=1}^T \norm{\bar{a}_{t}}^2 \norm{\breve{f}_{t+1}}^2 \right)^{1/2} \norm{\tilde{\Gamma} - \Gamma H_{\Gamma}}\\
& = O_p\left( \frac{\sqrt{L}}{N^{3/2}} \left( \frac{\sqrt{L}}{\sqrt{NT}} + \left( \frac{L}{N} \right)^{10 + \frac{1}{2}}  + \frac{\sqrt{L}}{N}  \right)  \right)
\end{align*}
using the bound $\norm{\tilde{\Gamma} - \Gamma H_{\Gamma}} = O_p\left(\frac{\sqrt{L}}{\sqrt{NT}} + \left( \frac{L}{N} \right)^{10 + \frac{1}{2}}  + \frac{\sqrt{L}}{N}  \right) $ from Lemma \ref{lem:gamma_f_final_bound}. In addition, the second term is bounded as
\begin{align*}
&\norm{\frac{1}{T N^2}  \sum_{t=1}^T\sum_{j=1}^N  \bar{a}_{t}   \epsilon_{j,t+1} x_{jt}^\top \Gamma H_{\Gamma} (\tilde{\breve{f}}_{t+1} - H_{\Gamma}^{-1} \breve{f}_{t+1})}\\
&\leq \frac{1}{N} \left( \frac{1}{T} \sum_{t=1}^T \norm{\frac{1}{N} \sum_{j=1}^N   \epsilon_{j,t+1} \bar{a}_{t} B_{jt}^\top }^2 \right)^{1/2} \left( \frac{1}{T} \sum_{t=1}^T \norm{\tilde{\breve{f}}_{t+1} - H_{\Gamma}^{-1} \breve{f}_{t+1}}^2 \right)^{1/2} \\
&= O_p\left( \frac{1}{N^{3/2}}  \left( \frac{1}{\sqrt{N}} + \frac{\sqrt{L}}{\sqrt{NT}} + \left( \frac{L}{N} \right)^{10 + \frac{1}{2}}  \right)  \right),
\end{align*}
by Lemma \ref{lem:fhat}. Next, we bound the parts related to the following term:
\begin{align*}
&\frac{1}{N}\sum_{j=1}^N \left( \tilde{m}_{j,t+1} - m_{j,t+1} \right)^2 \\
&\lesssim \frac{1}{N}\sum_{j=1}^N \left[ (\hat{\alpha}_{O,jt} - \alpha_{O,jt})^2 +  (x_{jt}^\top (\tilde{\eta} - \eta))^2 + (x_{jt}^\top (\tilde{\Gamma} - \Gamma H_{\Gamma})  H_{\Gamma}^{-1} \breve{f}_{t+1})^2 + (x_{jt}^\top \Gamma H_{\Gamma} (\tilde{\breve{f}}_{t+1} - H_{\Gamma}^{-1} \breve{f}_{t+1})^2
\right].
\end{align*}
The first term is bounded as
\begin{align*}
\frac{1}{T N^2} \sum_{i=1}^N \sum_{t=1}^T a_{iit} \frac{1}{N} \sum_{j=1}^N (\hat{\alpha}_{O,jt} - \alpha_{O,jt})^2 = \frac{1}{T N^2} \sum_{t=1}^T \sum_{j=1}^N \bar{a}_t (\hat{\alpha}_{O,jt} - \alpha_{O,jt})^2 = o_p\left( \frac{1}{\sqrt{NT}} \right),
\end{align*}
by using the bound from the independent sub-Gaussian assumption like Lemma \ref{lem:uniformbound_1} as above. The second term is bounded like 
\begin{align*}
&\frac{1}{T N^2} \sum_{i=1}^N \sum_{t=1}^T a_{iit} \frac{1}{N} \sum_{j=1}^N (x_{jt}^\top (\tilde{\eta} - \eta))^2 = \frac{1}{T N} \sum_{t=1}^T  \bar{a}_t (\tilde{\eta} - \eta)^\top Q_t (\tilde{\eta} - \eta)  \\
& \ \ = O_p\left( \frac{1}{N} \left( \frac{L}{NT} + \left( \frac{L}{N}\right)^{20 + 1} + \frac{L}{N^2} \right)  \right).
\end{align*}
by Lemma \ref{lem:etahat}. The third term is bounded like
\begin{align*}
& \frac{1}{T N^2} \sum_{i=1}^N \sum_{t=1}^T a_{iit} \frac{1}{N} \sum_{j=1}^N (x_{jt}^\top (\tilde{\Gamma} - \Gamma H_{\Gamma})  H_{\Gamma}^{-1} \breve{f}_{t+1})^2 \\
& \ \ =  \frac{1}{T N} \sum_{t=1}^T  \bar{a}_t  \breve{f}_{t+1}^\top H_{\Gamma}^{-\top} (\tilde{\Gamma} - \Gamma H_{\Gamma})^\top Q_t (\tilde{\Gamma} - \Gamma H_{\Gamma})  H_{\Gamma}^{-1} \breve{f}_{t+1}  \\
& \ \ = O_p\left( \frac{1}{N} \left( \frac{L}{NT} + \left( \frac{L}{N}\right)^{20 + 1} + \frac{L}{N^2} \right)  \right).
\end{align*}
using the bound $\norm{\tilde{\Gamma} - \Gamma H_{\Gamma}} = O_p\left(\frac{\sqrt{L}}{\sqrt{NT}} + \left( \frac{L}{N} \right)^{10 + \frac{1}{2}}  + \frac{\sqrt{L}}{N}  \right) $ from Lemma \ref{lem:gamma_f_final_bound}. Lastly, the fourth term is bounded like
\begin{align*}
&\norm{\frac{1}{T N^2} \sum_{i=1}^N \sum_{t=1}^T a_{iit} \frac{1}{N} \sum_{j=1}^N (x_{jt}^\top \Gamma H_{\Gamma} (\tilde{\breve{f}}_{t+1} - H_{\Gamma}^{-1} \breve{f}_{t+1})^2}\\
&\ \  = \norm{\frac{1}{T N} \sum_{t=1}^T  \bar{a}_t  (\tilde{\breve{f}}_{t+1} - H_{\Gamma}^{-1} \breve{f}_{t+1})^\top H_{\Gamma}^\top \Gamma^\top Q_t \Gamma H_{\Gamma} (\tilde{\breve{f}}_{t+1} - H_{\Gamma}^{-1} \breve{f}_{t+1})} \\
&\ \  \lesssim \frac{1}{T N} \max_{t}\norm{\bar{a}_t} \sum_{t=1}^T \norm{\tilde{\breve{f}}_{t+1} - H_{\Gamma}^{-1} \breve{f}_{t+1} }^2 = O_p\left( \frac{1}{N} \left( \frac{L}{NT} + \left( \frac{L}{N}\right)^{20 + 1} + \frac{L}{N^2} \right)  \right),
\end{align*}
since a simple calculation with Lemma \ref{lem:fhat}. Here, we use the fact that $\max_t||\bar{a}_t|| = O_p(1)$ To sum up, we have
 $$
 \frac{1}{T N^2} \sum_{i=1}^N \sum_{t=1}^T a_{iit} \left( \bbE[\epsilon_{i,t+1}^2] - \hat{\sigma}_{t+1}^2 \right)  = o_p\left( \frac{1}{\sqrt{NT}} \right). \ \ \square
 $$

\begin{lemma}\label{lem:alpha_I_1_tech}
(i) $\norm{x_{it}^\top \Gamma H_{\Gamma} - x_{it}^\top \tilde{\Gamma}} = O_p \left( \frac{L^{3/2}}{N^{3/2} \sqrt{T}} + \frac{\sqrt{L}}{N} + \left( \frac{L}{N} \right)^{10 + \frac{1}{2}} 
 + \frac{\sqrt{L}}{\sqrt{NT}} \right)$;\\
(ii) $\norm{ \left( H_{\Gamma}^\top \Gamma^\top Q_t \Gamma  H_{\Gamma} \right)^{-1} - \left( \tilde{\Gamma}^\top Q_t \tilde{\Gamma} \right)^{-1} } = O_p\left(\frac{\sqrt{L}}{\sqrt{NT}} + \left( \frac{L}{N} \right)^{10 + \frac{1}{2}}  + \frac{\sqrt{L}}{N}  \right) $; (iii) $\norm{X_{t}^\top \Gamma H_{\Gamma} - X_{t}^\top \tilde{\Gamma}} = O_p\left( \sqrt{N}\left(\frac{\sqrt{L}}{\sqrt{NT}} + \left( \frac{L}{N} \right)^{10 + \frac{1}{2}}  + \frac{\sqrt{L}}{N} \right) \right)$; (iv) $\norm{x_{it}^\top \Gamma H_{\Gamma} - x_{it}^\top \hat{\Gamma}} = O_p\left( \frac{\sqrt{L}}{\sqrt{NT}} \right)$; (v) $\norm{\Gamma H_{\Gamma} - \hat{\Gamma}} = O_p\left( \frac{\sqrt{L}}{\sqrt{NT}} \right)$; (vi) $\norm{ \left( H_{\Gamma}^\top \Gamma^\top Q_t \Gamma  H_{\Gamma} \right)^{-1} - \left( \hat{\Gamma}^{\top} Q_t \hat{\Gamma} \right)^{-1} } = O_p\left(\frac{\sqrt{L}}{\sqrt{NT}}  \right) $.
\end{lemma}

\noindent\textbf{Proof of Lemma \ref{lem:alpha_I_1_tech}.} (i) Note that
\begin{align*}
\tilde{\Gamma}^\top x_{it} - H_{\Gamma}^\top \Gamma^\top x_{it} 
&= \left( x_{it}^\top \otimes I_K \right)   \vect \left( \tilde{\Gamma}^\top - H_{\Gamma}^\top \Gamma^\top \right) \\
&= \left( \frac{1}{T} \sum_{t=1}^T \tilde{f}_{t+1}^{d} \tilde{f}_{t+1}^{d \top} \right)^{-1}
H_F \frac{1}{NT} \sum_{js}  x_{it}^\top Q_s^{-1} x_{js} f_{s+1}^d \epsilon_{j,s+1} \\
&+ \left( \frac{1}{T} \sum_{t=1}^T \tilde{f}_{t+1}^{d} \tilde{f}_{t+1}^{d \top} \right)^{-1}
\frac{1}{NT} \sum_{s}  x_{it}^\top Q_s^{-1} X_s^\top E_{s+1} \left( \tilde{f}_{s+1}^{d} - H_{F} f_{s+1}^d \right) .
\end{align*}
By the weak dependence assumption, we have $\frac{1}{NT} \sum_{js}  x_{it}^\top Q_s^{-1} x_{js} f_{s+1}^d \epsilon_{j,s+1} = O_p\left( \frac{\sqrt{L}}{\sqrt{NT}}\right)$. In addition, we have
\begin{align*}
&\frac{1}{NT} \sum_{s}  x_{it}^\top Q_s^{-1} X_s^\top E_{s+1} \left( \tilde{f}_{s+1}^{d} - H_{F} f_{s+1}^d \right) \\
&\leq \left( \frac{1}{T} \sum_s \norm{\frac{1}{N} \sum_j x_{it}^\top Q_s^{-1} x_{js} \epsilon_{j,s+1} }^2   \right)^{\frac{1}{2}}
\left( \frac{1}{T} \sum_s \norm{\tilde{f}_{s+1}^{d} - H_{F} f_{s+1}^d}^2  \right)^{\frac{1}{2}} \\
& = O_p \left( \frac{\sqrt{L}}{\sqrt{N}} \left( \frac{1}{\sqrt{
N}} + \frac{L}{N\sqrt{T}} + \left( \frac{L}{N} \right)^{10} \right)  \right),
\end{align*}
since $\frac{1}{T} \sum_s \norm{\tilde{f}_{s+1}^{d} - H_{F} f_{s+1}^d}^2 = O_p\left( \left( \frac{1}{\sqrt{
N}} + \frac{L}{N\sqrt{T}} + \left( \frac{L}{N} \right)^{10} \right)^2 \right)$ by Lemma \ref{lem:gamma_f_final_bound} and\\ $\frac{1}{T} \sum_s \bbE[||\frac{1}{N} \sum_j x_{it}^\top Q_s^{-1} x_{js} \epsilon_{j,s+1} ||^2 | X] = O_p\left( \frac{L}{N} \right)$.\\
(ii) First, note that
\begin{align*}
\norm{ H_{\Gamma}^\top \Gamma^\top Q_t \Gamma  H_{\Gamma}  -  \tilde{\Gamma}^\top Q_t \tilde{\Gamma}  }  
 \lesssim \norm{\tilde{\Gamma}  - \Gamma  H_{\Gamma} }\norm{Q_t} \norm{\Gamma  H_{\Gamma}} 
 = O_p\left(\frac{\sqrt{L}}{\sqrt{NT}} + \left( \frac{L}{N} \right)^{10 + \frac{1}{2}}  + \frac{\sqrt{L}}{N}  \right) = o_p(1)
\end{align*}
by using the bound $\norm{\tilde{\Gamma} - \Gamma H_{\Gamma}} = O_p\left(\frac{\sqrt{L}}{\sqrt{NT}} + \left( \frac{L}{N} \right)^{10 + \frac{1}{2}}  + \frac{\sqrt{L}}{N}  \right) $ from Lemma \ref{lem:gamma_f_final_bound}. Then, since $\psi_{\min} \left(H_{\Gamma}^\top \Gamma^\top Q_t \Gamma  H_{\Gamma} \right) > c$ for some constant $c>0$ we have $\norm{\left( \tilde{\Gamma}^\top Q_t \tilde{\Gamma} \right)^{-1}} = O_p(1)$. Then, we have
\begin{align*}
\norm{ \left( H_{\Gamma}^\top \Gamma^\top Q_t \Gamma  H_{\Gamma} \right)^{-1} - \left( \tilde{\Gamma}^\top Q_t \tilde{\Gamma} \right)^{-1} } 
&\leq \norm{\left( H_{\Gamma}^\top \Gamma^\top Q_t \Gamma  H_{\Gamma} \right)^{-1}} \norm{\left( \tilde{\Gamma}^\top Q_t \tilde{\Gamma} \right)^{-1}} \norm{ H_{\Gamma}^\top \Gamma^\top Q_t \Gamma  H_{\Gamma}  -  \tilde{\Gamma}^\top Q_t \tilde{\Gamma}  }\\
&= O_p\left(\frac{\sqrt{L}}{\sqrt{NT}} + \left( \frac{L}{N} \right)^{10 + \frac{1}{2}}  + \frac{\sqrt{L}}{N}  \right) .
\end{align*}
(iii) It follows from
$$
\norm{X_t \tilde{\Gamma} - X_t \Gamma H_{\Gamma}} \leq \norm{X_t} \norm{\tilde{\Gamma} - \Gamma H_{\Gamma}}
= O_p\left( \sqrt{N}\left(\frac{\sqrt{L}}{\sqrt{NT}} + \left( \frac{L}{N} \right)^{10 + \frac{1}{2}}  + \frac{\sqrt{L}}{N} \right) \right).
$$
(iv) As noted in the proof of Theorem \ref{thm:gamma_1} (b), we need to bound the following term:
\begin{align*}
&\tilde{\Gamma}^{\top} \frac{1}{T} \sum_{s=1}^T (x_{it}^\top \ddot{E}_{t+1}) \times \ddot{E}_{t+1}  - \tilde{\Gamma}^{\top} \frac{1}{T} \sum_{t=1}^T  \hat{\sigma}_{t+1}^2  (X_t^\top X_t)^{-1} x_{it}\\
&= \tilde{\Gamma}^{\top} \frac{1}{N^2 T} \sum_{j,j',t}(x_{it}^\top Q_t^{-1} x_{jt}) Q_t^{-1} x_{j't} \epsilon_{j,t+1} \epsilon_{j',t+1} 
- \tilde{\Gamma}^{\top} \frac{1}{N^2 T} \sum_{j,t}(x_{it}^\top Q_t^{-1} x_{jt}) Q_t^{-1} x_{jt}  \hat{\sigma}_{t+1}^2 \\
&=\tilde{\Gamma}^{\top} \frac{1}{N^2 T} \sum_{j,j',t}a_{jj't} (u_{jj',t+1} - \bbE[u_{jj',t+1}])  +
\tilde{\Gamma}^{\top} \frac{1}{N^2 T} \sum_{j,t}a_{jjt} (\bbE[\epsilon_{j,t+1}^2] - \hat{\sigma}_{t+1}^2),
\end{align*}
where $a_{jj't} = (x_{it}^\top Q_t^{-1} x_{jt}) Q_t^{-1} x_{j't}$. Using a similar method as in the proof of Lemma \ref{lem:debiasing_tech}, we can show that it is $o_p\left( \frac{\sqrt{L}}{\sqrt{NT}} \right)$. In addition, we have
$$
\frac{1}{T}\sum_{t=1}^T x_{it}^\top \ddot{E}_{t+1} \otimes  f_{t+1}^{d} = \frac{1}{NT} \sum_{j=1}^N \sum_{t=1}^T (x_{it}^\top Q_t^{-1} x_{jt})  f_{t+1}^{d} \epsilon_{j,t+1}  = O_p\left( \frac{\sqrt{L}}{\sqrt{NT}} \right).
$$ 
Hence, we have $\norm{x_{it}^\top \Gamma H_{\Gamma} - x_{it}^\top \hat{\Gamma}} =  O_p\left( \frac{\sqrt{L}}{\sqrt{NT}} \right)$.\\
(v) By the similar method as in the proof of Theorem \ref{thm:gamma_1} (b), we need to bound the following term:
\begin{align*}
& \frac{1}{T} \sum_{s=1}^T  \ddot{E}_{t+1} \otimes \ddot{E}_{t+1}  -  \frac{1}{TN^2} \sum_{i=1}^N \sum_{t=1}^T  a_{iit} \hat{\sigma}_{t+1}^2  \\
&=\frac{1}{N^2 T} \sum_{i,j,t} a_{ijt} (u_{ij,t+1} - \bbE[u_{ij,t+1}])  +
\frac{1}{N^2 T} \sum_{i,t} a_{iit} (\bbE[\epsilon_{i,t+1}^2] - \hat{\sigma}_{t+1}^2),
\end{align*}
where $a_{ijt} = (Q_t^{-1} \otimes Q_t^{-1}) (x_{it} \otimes x_{jt})$. Here, we have 
$$
\max_t \norm{\bar{a}_t} = \max_t \norm{\vect(Q_t^{-1})} = \max_t \norm{Q_t^{-1}}_F \leq \sqrt{L} \max_t \norm{Q_t^{-1}} = O_p \left( \sqrt{L} \right)
$$
since $\bar{a}_t = \frac{1}{N} \sum_i a_{iit} = \vect(Q_t^{-1} \frac{1}{N} \sum_i x_{it} x_{it}^\top Q_t^{-1} ) = \vect(Q_t^{-1})$. Then, using a similar method as in the proof of Lemma \ref{lem:debiasing_tech}, we can show that it is $o_p\left( \frac{\sqrt{L}}{\sqrt{NT}} \right)$. In addition, we have
$$
\frac{1}{T}\sum_{t=1}^T  \ddot{E}_{t+1} \otimes  f_{t+1}^{d} = \frac{1}{NT} \sum_{j=1}^N \sum_{t=1}^T ( Q_t^{-1} x_{jt} \otimes f_{t+1}^{d}) \epsilon_{j,t+1}  = O_p\left( \frac{\sqrt{L}}{\sqrt{NT}} \right).
$$ 
Hence, we have $\norm{\Gamma H_{\Gamma} -  \hat{\Gamma}} =  O_p\left( \frac{\sqrt{L}}{\sqrt{NT}} \right)$.\\
(vi) With the aid of (v), we can proof it in the same way as that of (ii). $\square$
\smallskip

\begin{lemma}\label{lem:projection_tech}
For an $a \times b$ matrix $\Phi$ and its estimator $\hat{\Phi}$, we have the following decomposition:
\begin{align*}
P_{\hat{\Phi}} - P_{\Phi} &= \hat{\Phi} \left( \hat{\Phi}^\top \hat{\Phi} \right)^{-1} \hat{\Phi}^\top - \Phi \left( \Phi^\top \Phi \right)^{-1} \Phi^\top \\
&= \Phi \left( \Phi^\top \Phi \right)^{-1} \left( \hat{\Phi} - \Phi \right)^\top (I - P_{\Phi}) 
+ (I - P_{\Phi}) \left( \hat{\Phi} - \Phi \right)   \left( \Phi^\top \Phi \right)^{-1} \Phi^\top + \text{higher order terms},
\end{align*}
where
\begin{align*}
\text{higher order terms} &= (\hat{\Phi} - \Phi) \left[ \left( \hat{\Phi}^\top \hat{\Phi} \right)^{-1} - \left( \Phi^\top \Phi \right)^{-1} \right] \Phi^\top 
+ (\hat{\Phi} - \Phi) \left( \Phi^\top \Phi \right)^{-1}(\hat{\Phi} - \Phi)^\top  \\
&+ \Phi \left[ \left( \hat{\Phi}^\top \hat{\Phi} \right)^{-1} - \left( \Phi^\top \Phi \right)^{-1} \right] (\hat{\Phi} - \Phi)^\top
+ (\hat{\Phi} - \Phi) \left[ \left( \hat{\Phi}^\top \hat{\Phi} \right)^{-1} - \left( \Phi^\top \Phi \right)^{-1} \right] (\hat{\Phi} - \Phi)^\top \\
& - \Phi \left( \Phi^\top \Phi \right)^{-1} \left( \hat{\Phi} - \Phi \right)^\top \Phi \left[ \left( \hat{\Phi}^\top \hat{\Phi} \right)^{-1} - \left( \Phi^\top \Phi \right)^{-1} \right] \Phi^\top \\
& - \Phi \left( \Phi^\top \Phi \right)^{-1} \left( \hat{\Phi} - \Phi \right)^\top \left( \hat{\Phi} - \Phi \right)\left( \Phi^\top \Phi \right)^{-1} \Phi^\top \\
& -  \Phi \left( \Phi^\top \Phi \right)^{-1}  \Phi^\top \left( \hat{\Phi} - \Phi \right)\left[ \left( \hat{\Phi}^\top \hat{\Phi} \right)^{-1} - \left( \Phi^\top \Phi \right)^{-1} \right] \Phi^\top\\
&- \Phi \left( \Phi^\top \Phi \right)^{-1}  \left( \hat{\Phi} - \Phi \right)^\top \left( \hat{\Phi} - \Phi \right)\left[ \left( \hat{\Phi}^\top \hat{\Phi} \right)^{-1} - \left( \Phi^\top \Phi \right)^{-1} \right] \Phi^\top.
\end{align*}
\end{lemma}

\noindent\textbf{Proof of Lemma \ref{lem:projection_tech}.}
First, a simple calculation shows that
\begin{align*}
P_{\hat{\Phi}} - P_{\Phi} & = (\hat{\Phi} - \Phi)  \left( \Phi^\top \Phi \right)^{-1} \Phi^\top + \Phi \left[ \left( \hat{\Phi}^\top \hat{\Phi} \right)^{-1} - \left( \Phi^\top \Phi \right)^{-1} \right] \Phi^\top +  \Phi   \left( \Phi^\top \Phi \right)^{-1} (\hat{\Phi} - \Phi)^\top  \\
 &+ (\hat{\Phi} - \Phi) \left[ \left( \hat{\Phi}^\top \hat{\Phi} \right)^{-1} - \left( \Phi^\top \Phi \right)^{-1} \right] \Phi^\top 
+ (\hat{\Phi} - \Phi) \left( \Phi^\top \Phi \right)^{-1}(\hat{\Phi} - \Phi)^\top  \\
&+ \Phi \left[ \left( \hat{\Phi}^\top \hat{\Phi} \right)^{-1} - \left( \Phi^\top \Phi \right)^{-1} \right] (\hat{\Phi} - \Phi)^\top
+ (\hat{\Phi} - \Phi) \left[ \left( \hat{\Phi}^\top \hat{\Phi} \right)^{-1} - \left( \Phi^\top \Phi \right)^{-1} \right] (\hat{\Phi} - \Phi)^\top.
\end{align*}
In addition, a simple calculation shows that
\begin{align*}
\Phi \left[ \left( \hat{\Phi}^\top \hat{\Phi} \right)^{-1} - \left( \Phi^\top \Phi \right)^{-1} \right] \Phi^\top 
&= \Phi \left( \Phi^\top \Phi \right)^{-1} \left[\Phi^\top \Phi -  \hat{\Phi}^\top \hat{\Phi} \right] \left( \hat{\Phi}^\top \hat{\Phi} \right)^{-1} \Phi^\top \\
&= - \Phi \left( \Phi^\top \Phi \right)^{-1} \left( \hat{\Phi} - \Phi \right)^\top P_{\Phi}
-  P_{\Phi} \left( \hat{\Phi} - \Phi \right)   \left( \Phi^\top \Phi \right)^{-1} \Phi^\top \\
& - \Phi \left( \Phi^\top \Phi \right)^{-1} \left( \hat{\Phi} - \Phi \right)^\top \Phi \left[ \left( \hat{\Phi}^\top \hat{\Phi} \right)^{-1} - \left( \Phi^\top \Phi \right)^{-1} \right] \Phi^\top \\
& - \Phi \left( \Phi^\top \Phi \right)^{-1} \left( \hat{\Phi} - \Phi \right)^\top \left( \hat{\Phi} - \Phi \right)\left( \Phi^\top \Phi \right)^{-1} \Phi^\top \\
& -  \Phi \left( \Phi^\top \Phi \right)^{-1}  \Phi^\top \left( \hat{\Phi} - \Phi \right)\left[ \left( \hat{\Phi}^\top \hat{\Phi} \right)^{-1} - \left( \Phi^\top \Phi \right)^{-1} \right] \Phi^\top\\
&- \Phi \left( \Phi^\top \Phi \right)^{-1}  \left( \hat{\Phi} - \Phi \right)^\top \left( \hat{\Phi} - \Phi \right)\left[ \left( \hat{\Phi}^\top \hat{\Phi} \right)^{-1} - \left( \Phi^\top \Phi \right)^{-1} \right] \Phi^\top.
\end{align*}
Therefore, we have
\begin{align*}
P_{\hat{\Phi}} - P_{\Phi} = \Phi \left( \Phi^\top \Phi \right)^{-1} \left( \hat{\Phi} - \Phi \right)^\top (I - P_{\Phi}) 
+ (I - P_{\Phi}) \left( \hat{\Phi} - \Phi \right)   \left( \Phi^\top \Phi \right)^{-1} \Phi^\top + \text{higher order terms}. \ \ \square
\end{align*}

\begin{lemma}\label{lem:B_tech}
(i) $B_t^{o\top} B_t^o = N \cdot I_{N-L}$ and $\norm{B_t^o} = \sqrt{N}$; (ii) For all $1\leq q \leq N-L$, we have $\norm{B^o_{t,q}}^2 = N$ where $B^o_{t,q} = B^o_t e_q$ and $e_q$ is an $(N-L) \times 1$ unit vector.  That is, $\sum_{i=1}^N (B^{o}_{t,iq})^2 = N$ where $B^o_{t,iq} = e_i^\top B_t^o e_r$; (iii) $\norm{B_t^o}_F^2 = \sum_{i=1}^{N} \norm{B^o_{t,i}}^2 = N(N-L)$ where $B^o_{t,i} = B^{o \top}_t e_i$ and $e_i$ is an $N \times 1$ unit vector.
\end{lemma}

\noindent\textbf{Proof of Lemma \ref{lem:B_tech}.}
(i) $B_t^{o\top} B_t^o = (X_t^{o\top}X_t^o/N)^{-1/2}X_t^{o\top} X_t^o(X_t^{o\top}X_t^o/N)^{-1/2} = N \cdot I_{N-L}$.
(ii) $\norm{B^o_{t,q}}^2 = e_q^\top B_t^{o\top} B_t^o e_q = N \cdot e_q^\top e_q = N$. (iii) $\norm{B_t^o}_F^2 = \sum_{q=1}^{N-L} \norm{B_{t,q}^o}^2 = N(N-L)$ by (i). $\square$
\smallskip

\begin{lemma}\label{lem:uniformbound_1}
We have w.p.c. to 1, for all $t$,
$$
\max_{1 \leq q \leq N-L} \abs{\frac{1}{N} \sum_{j=1}^{N} B_{t,jq}^{o} \epsilon_{j,t+1}} \leq C_u \sigma_{t+1} \frac{\sqrt{\log NT}}{\sqrt{N}} 
$$
for some constant $C_u>0$. In addition, if $T / N^a$ is bounded for some $a \geq 1$, then we can have the same result with $\sqrt{\log N}$ in place of $\sqrt{\log NT}$.
\end{lemma}

\noindent\textbf{Proof of Lemma \ref{lem:uniformbound_1}.}
For each $1 \leq q \leq N - L$, we have
$\sum_{j=1}^{N} \left( B_{t,jq}^{o} \right)^2 = \norm{B_{t,q}^{o}}^2 = N$ by Lemma \ref{lem:B_tech}. Hence, by Hoeffding's inequality (e.g., Theorem 2.6.3 of \cite{vershynin2018high}), for each $q$ and $t$, we have with probability exceeding $1 - O((NT)^{-9})$ that
$$
\frac{1}{N} \sum_{j=1}^{N} B_{t,jq}^{o} \epsilon_{j,t+1} \leq C_u \sigma_{t+1} \frac{\sqrt{\log NT}}{\sqrt{N}} 
$$
for some universal constants $C_u >0$. Hence, we have w.p.c. to 1, for all $t$,
$$
\max_{1 \leq q \leq N - L}\abs{\frac{1}{N} \sum_{j=1}^{N} B_{t,jq}^{o} \epsilon_{j,t+1}} \leq C_u \sigma_{t+1} \frac{\sqrt{\log NT}}{\sqrt{N}} . \ \ \square
$$

\begin{lemma}\label{lem:uniformbound_2}
We have w.p.c. to 1,
$$
\max_{1 \leq q \leq N - L} \abs{\frac{1}{NT} \sum_{s=1}^T \sum_{j=1}^{N} B_{s,jq}^{o} \epsilon_{j,s+1}} \ll \sigma_{t+1} \frac{\sqrt{\log NT}}{\sqrt{N}} 
$$
for all $1 \leq t \leq T$. In addition, if $T / N^a$ is bounded for some $a \geq 1$, then we can have the same result with $\sqrt{\log N}$ in place of $\sqrt{\log NT}$.
\end{lemma}

\noindent\textbf{Proof of Lemma \ref{lem:uniformbound_2}.}
We want to show that w.p.c. to 1,
$$
\max_{1 \leq q \leq N - L} \abs{\frac{1}{NT} \sum_{s=1}^T \sum_{j=1}^{N} B_{s,jq}^{o} \epsilon_{j,s+1}} \leq \sigma_{\min} \frac{(\log NT)^{1/4}}{\sqrt{N}} 
$$
where $\sigma_{\min}$ is some constant such that $\sigma_{\min} \leq \sigma_t$ for all $t$. Then, Lemma \ref{lem:uniformbound_2} is followed from it. By the Markov's inequality, we have
$$
\Pr\left( \abs{\frac{1}{NT} \sum_{s=1}^T \sum_{j=1}^{N} B_{s,jq}^{o} \epsilon_{j,s+1}} \geq  \sigma_{\min} \frac{(\log NT)^{1/4}}{\sqrt{N}}   \right) \leq \frac{\bbE \left[ \abs{\frac{1}{NT} \sum_{s=1}^T \sum_{j=1}^{N} B_{s,jq}^{o} \epsilon_{j,s+1}}^\alpha   \right]}{ \left(  \sigma_{\min} \frac{(\log NT)^{1/4}}{\sqrt{N}}\right)^\alpha}
$$
for some integer $\alpha \geq 1$. Then, because
\begin{align*}
&\Pr\left( \abs{\frac{1}{NT} \sum_{s=1}^T \sum_{j=1}^{N} B_{s,jq}^{o} \epsilon_{j,s+1}} \geq  \sigma_{\min} \frac{(\log NT)^{1/4}}{\sqrt{N}}  \text{  at least one  } q \right)  \\
&\leq
\Pr\left( \bigcup_{1\leq q \leq N-L} \left\{ \abs{\frac{1}{NT} \sum_{s=1}^T \sum_{j=1}^{N} B_{s,jq}^{o} \epsilon_{j,s+1}} \geq  \sigma_{\min} \frac{(\log NT)^{1/4}}{\sqrt{N}} \right\}  \right) \\
& \leq (N-L) \times \frac{ \frac{1}{(NT)^{\alpha/2}} \bbE \left[ \abs{\frac{1}{\sqrt{NT}} \sum_{s=1}^T \sum_{j=1}^{N} B_{s,jq}^{o} \epsilon_{j,s+1}}^\alpha   \right]}{ \left( \sigma_{\min} \frac{(\log NT)^{1/4}}{\sqrt{N}}\right)^\alpha} \longrightarrow 0
\end{align*}
under our assumptions, we can say that w.p.c. to 1,
$$
\max_{1 \leq q \leq N - L} \abs{\frac{1}{NT} \sum_{s=1}^T \sum_{j=1}^{N} B_{s,jq}^{o} \epsilon_{j,s+1}} \leq  \sigma_{\min} \frac{(\log NT)^{1/4}}{\sqrt{N}} . \ \ \square
$$

\end{document}